\documentclass[floatfix,twocolumn,prd,showpacs,preprintnumbers,amsmath,nofootinbib,amssymb,superscriptaddress]{revtex4-1}

\usepackage{color} 
\newcommand{\blue}[1]{{\color{blue} #1}}
\definecolor{green}{rgb}{0,.5,0}

\definecolor{red}{rgb}{1,0,0}
\newcommand{\red}[1]{{ #1}}
\usepackage{graphicx}% Include figure files
\usepackage[subfigure]{graphfig}
\usepackage{epsfig}
\usepackage{dcolumn}% Align table columns on decimal point
\usepackage{amsmath}
\usepackage{multirow}
\usepackage{booktabs}   %%表格横线宏包
\usepackage{diagbox}
\usepackage{colortbl}
\usepackage{soul}
\usepackage{slashed}

\usepackage[colorlinks,linkcolor=blue,citecolor=blue,urlcolor=blue]{hyperref}

\def\bea{\begin{eqnarray}}

\def\eea{\end{eqnarray}}

\def\bal#1\eal{\begin{align}#1\end{align}}
\newcommand{\MSbar}{{\overline{\text{MS}}}}
\newcommand{\MOM}{\text{RI/MOM}}

\immediate\write18{texcount -inc -incbib 
-sum borra.tex > /tmp/wordcount.tex}

\begin{document}

\title{\vspace{1.0in}Quark masses and low energy constants in the continuum from the tadpole improved clover ensembles}

\author{\includegraphics[scale=0.30]{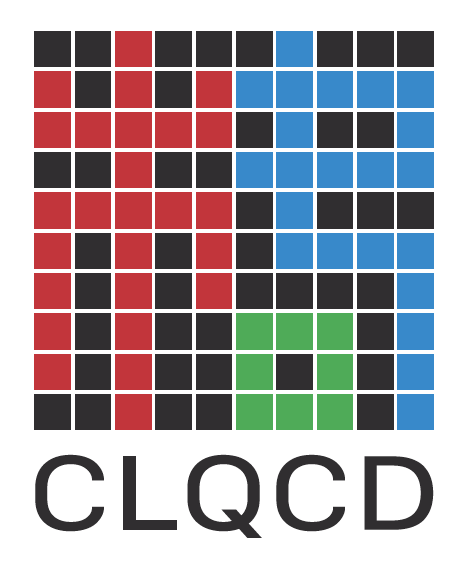}\\
Zhi-Cheng Hu}
\affiliation{Institute of Modern Physics, Chinese Academy of Sciences, Lanzhou, 730000, China}
\affiliation{University of Chinese Academy of Sciences, School of Physical Sciences, Beijing 100049, China}

\author{Bo-Lun Hu}
\affiliation{CAS Key Laboratory of Theoretical Physics, Institute of Theoretical Physics, Chinese Academy of Sciences, Beijing 100190, China}

\author{Ji-Hao Wang}
\affiliation{CAS Key Laboratory of Theoretical Physics, Institute of Theoretical Physics, Chinese Academy of Sciences, Beijing 100190, China}
\affiliation{University of Chinese Academy of Sciences, School of Physical Sciences, Beijing 100049, China}

\author{Ming Gong}
\affiliation{Institute of High Energy Physics, Chinese Academy of Sciences, Beijing 100049, China}

\author{Guoming Liu}
\affiliation{Key Laboratory of Atomic and Subatomic Structure and Quantum Control (MOE), Guangdong Basic Research Center of Excellence for Structure and Fundamental Interactions of Matter, Institute of Quantum Matter, South China Normal University, Guangzhou 510006, China}
\affiliation{Guangdong-Hong Kong Joint Laboratory of Quantum Matter, Guangdong Provincial Key Laboratory of Nuclear Science, Southern Nuclear Science Computing Center, South China Normal University, Guangzhou 510006, China}

\author{Liuming Liu}
\email[Corresponding author: ]{liuming@impcas.ac.cn}
\affiliation{Institute of Modern Physics, Chinese Academy of Sciences, Lanzhou, 730000, China}
\affiliation{University of Chinese Academy of Sciences, Beijing 100049, China. }

\author{Peng Sun}
\email[Corresponding author: ]{pengsun@impcas.ac.cn}
\affiliation{Institute of Modern Physics, Chinese Academy of Sciences, Lanzhou, 730000, China}

\author{Wei Sun}
\affiliation{Institute of High Energy Physics, Chinese Academy of Sciences, Beijing 100049, China}

\author{Wei Wang}
\affiliation{INPAC, Shanghai Key Laboratory for Particle Physics and Cosmology, Key Laboratory for Particle Astrophysics and Cosmology (MOE), School of Physics and Astronomy, Shanghai Jiao Tong University, Shanghai 200240, China}
\affiliation{Southern Center for Nuclear-Science Theory (SCNT), Institute of Modern Physics, Chinese Academy of Sciences, Huizhou 516000, Guangdong Province, China}

\author{Yi-Bo Yang}
\email[Corresponding author: ]{ybyang@itp.ac.cn}
\affiliation{University of Chinese Academy of Sciences, School of Physical Sciences, Beijing 100049, China}
\affiliation{CAS Key Laboratory of Theoretical Physics, Institute of Theoretical Physics, Chinese Academy of Sciences, Beijing 100190, China}
\affiliation{School of Fundamental Physics and Mathematical Sciences, Hangzhou Institute for Advanced Study, UCAS, Hangzhou 310024, China}
\affiliation{International Centre for Theoretical Physics Asia-Pacific, Beijing/Hangzhou, China}

\author{Dian-Jun Zhao}
\affiliation{University of Chinese Academy of Sciences, School of Physical Sciences, Beijing 100049, China}
\affiliation{CAS Key Laboratory of Theoretical Physics, Institute of Theoretical Physics, Chinese Academy of Sciences, Beijing 100190, China}

\begin{abstract}
We present the light-flavor quark masses and low energy constants using the 2+1 flavor full-QCD ensembles with stout smeared clover fermion action and Symanzik gauge actions. Both the fermion and gauge actions are tadpole  improved  self-consistently. The simulations are performed on 11 ensembles at 3 lattice spacings $a\in[0.05,0.11]$ fm, 4 spatial sizes $L\in[2.5, 5.1]$ fm, 7 pion masses $m_{\pi}\in[135,350]$ MeV, and several values of the strange quark mass. The quark mass is defined through the partially conserved axial current (PCAC) relation and renormalized to $\overline{\mathrm{MS}}$ 2 GeV through the intermediate regularization independent momentum subtraction (RI/MOM) scheme. The systematic uncertainty of using the symmetric momentum subtraction (SMOM) scheme is also included. Eventually, we predict $m_u=2.45(22)(20)$ MeV, $m_d=4.74(11)(09)$ MeV, and $m_s=98.8(2.9)(4.7)$ MeV with the systematic uncertainties from lattice spacing determination, continuum extrapolation and renormalization constant included. We also obtain the chiral condensate $\Sigma^{1/3}=268.6(3.6)(0.7)$ MeV and the pion decay constant $F=86.6(7)(1.4) $ MeV in the $N_f=2$ chiral limit, and the next-to-leading order low energy constants $\ell_3=2.43(54)(05)$ and $\ell_4=4.322(75)(96)$.
 \end{abstract}

\maketitle

\section{Introduction}

As fundamental parameters of the standard model which are not directly measurable in experiments, the mass of the lightest three flavors can only be determined accurately using lattice quantum chromodynamics (QCD). Lattice QCD offers a non-perturbative approach to solve QCD, the underlying theory of the strong interactions, but a set of complete and accurate ensembles is essential to ensure reliable results. 

Due to the infamous fermion doubling problem which prevents a straightforward discretization of the continuum Dirac fermion action, an accurate determination of the light quark masses is highly non-trivial. Since the widely used clover fermions suffer the additional chiral symmetry breaking which induces power divergence with loop corrections, most of the light quark mass determinations are made with either the staggered fermion (or its improved versions)~\cite{MILC:2004qnl, HPQCD:2004hdp, Mason:2005bj, MILC:2009ltw, Davies:2009ih, MILC:2009mpl, Laiho:2011np, McNeile:2010ji, MILC:2015vfd, Maezawa:2016vgv,Bazavov:2017lyh,FermilabLattice:2018est,MILC:2018ddw} which suffers the mixing between four equivalent ``tastes" of a given flavor, or Ginsburg-Wilson fermion actions like  Domain Wall~\cite{RBC-UKQCD:2008mhs,RBC:2010qam,Blum:2010ym,RBC:2012cbl,RBC:2014ntl} or Overlap~\cite{Yang:2014sea} fermion which requires ${\cal O}(10–100)$ times more cost of computational resources than the clover fermions.

Thus a natural question is, whether it is possible for the clover fermion to reach a high accuracy determination of the light quark masses. In 2007, Ref.~\cite{JLQCD:2007xff} proposed an alternative approach to define light quark masses from the PCAC relation and renormalizes it with tadpole improved 1-loop matching. By utilizing the Schrodinger functional (SF) scheme~\cite{Sint:2010eh}, the PCAC quark mass can be renormalized non-perturbatively, and the calculation with physical pion mass but a single lattice spacing $a=$0.09 fm and $m_{\pi}L\sim$ 2 gives $m_{u,d}=3.12(24)(08)$ MeV ~\cite{PACS-CS:2008bkb,PACS-CS:2009sof,PACS-CS:2010gyf,Aoki:2012st} at $\MSbar\rm{(2\ GeV)}$ which is 10\% lower than the present lattice average value $m_{u,d}=3.381(40)$MeV~\cite{FlavourLatticeAveragingGroupFLAG:2021npn} with large uncertainty.

A more systematic study using the SF scheme was conducted by the ALPHA collaboration with multiple lattice spacing $a\in[0.05,0.086]$ fm but relatively heavy quark masses $m_{\pi}\ge$ 200 MeV, and their determination resulted in $m_{u,d}=3.54(12)(9)$MeV~\cite{Bruno:2019vup}. So far, the most precise determination of $m_{u,d}=3.469(47)(48)$\cite{BMW:2010skj,Bruno:2019vup} with clover fermion comes from the BMW collaboration, which utilized multiple lattice spacings $a\in[0.05,0.012]$ fm with the lightest pion mass $m_{\pi}=131(2)$ MeV and renormalized the quark mass using the widely-used RI/MOM scheme~\cite{Martinelli:1994ty}. 

But the systematic uncertainty of using the RI/MOM scheme could be underestimated, as the RI/MOM scheme exhibits poor perturbative convergence for the scalar/pseudoscalar current, leading to sensitivity in the final result due to the estimate of the missing higher-order corrections. Thus, the SMOM scheme~\cite{Aoki:2007xm,Sturm:2009kb} was proposed to suppress this uncertainty and has been employed in most recent quark mass determinations using chiral fermions. Nevertheless, a recent study~\cite{Hasan:2019noy} at multiple lattice spacings shows that using either the RI/MOM or SMOM intermediate scheme can result in the renormalized scalar current under $\overline{\textrm{MS}}$ scheme differing by 30\% at $a\sim 0.1$ fm for the clover fermion.

Additionally, it is worth mentioning that the renormalized quark mass using the RI/MOM scheme with the Twisted-mass fermion~\cite{EuropeanTwistedMass:2014osg, Giusti:2017dmp,ExtendedTwistedMass:2021gbo} is $m_{u,d}$= 3.64(7)(6) MeV, which is approximately 5\% higher than the results obtained with chiral fermions that predominantly use the SMOM scheme.

In this work, we conduct a detailed comparison of the renormalization constants (RCs) using the RI/MOM and SMOM schemes. It turns out that the sensitivity of the intermediate schemes can be suppressed to $\sim 5$\% level, which allows us to provide a relatively precise prediction of the quark mass. Based on the Kaon masses with the QED effect subtracted, we also obtain the up, down, and strange quark masses separately, along with other related quantities. We expect that further improvement in the prediction accuracy can be achieved through calculations on more lattice spacings.

\section{Simulation setup}\label{sec:setup}

%% basic informations
The results in this work, are based on the 2+1 flavor full QCD ensembles using the tadpole improved tree level Symanzik (TITLS) gauge action and the tadpole improved tree level Clover (TITLC) fermion action. 

The TITLS gauge action, denoted as $S_g$, is defined in the following,
\bal\label{eq:gauge_action}
    S_g = \frac{1}{N_c} \mathrm{Re} \sum_{x,\mu<\nu}\mathrm{Tr} 
    \left[ 
        1- \hat{\beta} 
        \left(
            \mathcal{P}^U_{\mu,\nu}(x)+\frac{c_1\mathcal{R}^U_{\mu,\nu}(x)}{1-8c_1^0}
        \right)
    \right],
\eal
where $N_c=3$, and  
\begin{align}
    \mathcal{P}^{U}_{\mu,\nu}(x)& = U_\mu(x)U_\nu(x+a\hat{\mu})U^{\dagger}_\mu(x+a\hat{\nu})U^{\dagger}_\nu(x), \notag
    \\
    \mathcal{R}^{U}_{\mu,\nu}(x)& = U_\mu(x) U_\mu(x+a\hat{\mu}) U_\nu(x+2a\hat{\mu})  \notag
    \\
    & ~~~\times  U^{\dagger}_\mu(x+a\hat{\mu}+a\hat{\nu}) U^{\dagger}_\mu(x+a\hat{\nu}) U^{\dagger}_\nu(x), \notag
    \\
    U_{\mu}(x) &= P
    \left[ 
        \mathrm{exp} 
        \left( 
            {\rm i} g_0 \int_{x}^{x+\hat{\mu}a} \mathrm{d}y A_{\mu}(y)
        \right) 
    \right], \notag
\end{align}
$\hat{\beta}=(1-8c_1^0)\frac{6}{g_0^2u_0^4}\equiv 10/(g_0^2u_0^4)$ with $c_1^0=-\frac{1}{12}$, $c_1=\frac{c_1^0}{ u_0^2}$, $u_0=\langle \frac{\mathrm{Re}\mathrm{Tr}\sum_{x,\mu<\nu}\mathcal{P}^{U}_{\mu\nu}(x)}{6N_c\tilde{V}} \rangle^{1/4}$ is the tadpole improvement factor, $\tilde{V}=\tilde{L}^3\times \tilde{T}$ is the dimensionless 4-D volume of the lattice, \red{and we use $\tilde{O}$ for the dimensionless value of any quantity $O$}.

The TITLC fermion action uses 1-step stout smeared link $V$ with smearing parameter $\rho=0.125$,

\begin{align}
    \label{eq:quark_action}
    &S_q(m)=\sum_{x,\mu=1,...,4,\eta=\pm}\bar{\psi}(x)\sum\frac{1+\eta\gamma_{\mu}}{2}V_{\eta\mu}(x)\psi(x{+\eta\hat{\mu}a})\nonumber
    \\ 
    &\quad +\sum_x\psi(x)
    \left[
        -(4+ma) \delta_{y,x} + c_{\rm sw} \sigma^{\mu\nu} g_0 F^V_{\mu\nu}
    \right]
    \psi(x),
\end{align}

where $c_{\rm sw}=\frac{1}{v^3_{0}}$ with $v_0=\langle \frac{\mathrm{Re}\mathrm{Tr}\sum_{x,\mu<\nu}\mathcal{P}^{V}_{\mu\nu}(x)}{6N_c\tilde{V}} \rangle^{1/4}$, and 
\begin{eqnarray}
F^V_{\mu\nu} &=& \frac{i}{8a^2g_0} (\mathcal{P}^V_{\mu,\nu}-\mathcal{P}^V_{\nu,\mu}+\mathcal{P}^V_{\nu,-\mu}-\mathcal{P}^V_{-\mu,\nu} \nonumber \\
&& + \mathcal{P}^V_{-\mu,-\nu} -\mathcal{P}^V_{-\nu,-\mu} + \mathcal{P}^V_{-\nu,\mu} -\mathcal{P}^V_{\mu,-\nu}).
\end{eqnarray}

\begin{table*}[t]
	\centering
	\caption{\label{tab:lattice}	Lattice size $\tilde{L}^3\times \tilde{T}$, gauge coupling $\hat{\beta}=10/(g^2u_0^4)$, dimensionless bare quark mass parameters $\tilde{m}^{\rm b}_{l,s}$, renormalized quark masses $m^R_{l,s}$ and the corresponding pseudoscalar mass $m_{\pi, K}$, and the statistics information.}
	%\begin{ruledtabular}
	 \resizebox{2.1\columnwidth}{!}{
		\begin{tabular}{l | rrrrrr | rrrr | r}
		    & C24P34 & C24P29 &  C32P29 & C32P23 & C48P23 & C48P14 & F32P30  & F48P30  & F32P21  & F48P21 & H48P32 \\   
		\hline    
 		$\tilde{L}^3\times \tilde{T}$ & $24^3\times 64$ & $24^3\times 72$ & $32^3\times 64$ & $32^3\times 64$ & $48^3\times 96$ & $48^3\times 96$ & $32^3\times 96$ & $48^3\times 96$ & $32^3\times 64$ & $48^3\times 96$ & $48^3\times 144$ \\
		$\hat{\beta}$ & \multicolumn{6}{c|}{6.20} & \multicolumn{4}{c|}{6.41}  & \multicolumn{1}{c}{6.72} \\
        $a$ (fm) & \multicolumn{6}{c|}{0.10530(18)} & \multicolumn{4}{c|}{0.07746(18)}  & \multicolumn{1}{c}{0.05187(26)} \\
		\hline 		
		$\tilde{m}^{b}_l$ & -0.2770  & -0.2770 & -0.2770 & -0.2790 & -0.2790 & -0.2825 & -0.2295 &  -0.2295& -0.2320 & -0.2320 & -0.1850\\	  
		$\tilde{m}^{b}_s$  & -0.2310 & -0.2400 & -0.2400 & -0.2400 & -0.2400 & -0.2310 & -0.2050 & -0.2050 & -0.2050 & -0.2050 & -0.1700\\	
        \hline 	
        $m^R_{l}$ (MeV) & 22.90(19)   & 16.94(12)       &  17.35(11)   &	 10.55(11)  &	10.27(10)	&   3.638(83)   &	18.54(12)	 & 18.511(92) &	8.58(16)   & 8.59(08)   &	19.42(05)\\
        $m^R_{s}$ (MeV) & 111.41(16) & 87.46(10)
 & 88.16(10) & 84.48(07) & 84.79(04)
 & 103.15(05) & 93.23(11) & 93.05(08) & 89.75(10)
 & 90.43(08)& 95.61(04)\\       
        \hline 	
        $m_{\pi}$\ \ (MeV) &341.1(1.8)  & 292.7(1.2)	  &  292.4(1.1)	 &   228.0(1.2)	&    225.6(0.9) &	135.5(1.6)  &   303.2(1.3)	 & 303.4(0.9) & 210.9(2.2) & 207.2(1.1)	 &  317.2(0.9)\\
        $m_{K}$ (MeV) & 582.7(1.6) &509.4(1.1)   &509.0(1.1)    & 484.1(1.0)   &484.1(1.3)    & 510.0(1.0)  & 524.6(1.8)   &523.6(1.4)   &  492.0(1.7)   & 493.0(1.4)  &536.1(3.0) \\
%          $m_{\eta_s}$ (MeV) & 750.7(0.9)  & 659.13(0.6)
% & 659.7(0.7) & 645.1(0.5)
%&645.1(0.3) & 708.1(0.3)
%  & 677.7(0.9) & 676.0(0.6) & 661.8(0.7) & 663.9(0.6) & 700.0(0.5)\\
		 \hline	
        $n_{\rm cfg}$  &200	 &476	&198	&400	&62	&203	&206	&99	&194	&98	  &176
\\ 
        $n_{\rm src}$ &32	&3	&3	&3	&3	&48	&3	&3	&3	&12	&12
\\
		\end{tabular}
	}
\end{table*}

The parameters utilized for the simulation, encompassing the lattice size ($\tilde{L}^3\times \tilde{T}$), gauge coupling ($\hat{\beta}$), and the lattice spacing ($a$) determined through the gradient flow~\cite{Luscher:2010iy} with $w_0$~\cite{BMW:2012hcm} using the Symanzik action, are outlined in Table~\ref{tab:lattice}. The dimensionless bare degenerated light and strange quark mass ($\tilde{m}^{\rm b}_{l,s}$), renormalized quark masses ($m_{l,s}^{R}$) at $\overline{\textrm{MS}}$ 2GeV, and the respective pion and kaon masses ($m_{\pi, K}$) are also included in the table. The details of the pseudoscalar meson mass and lattice spacing extraction can be found in the supplemental materials~\cite{ref_sm}. The impact of the mistuning effect of the tadpole improvement factors $u_0$ and $v_0$ can also be found there.

The ensemble set used in this work is designed to control the variables in the systematic uncertainty estimation. For example, the spatial size $L$ of the C24P29, F32P30, and H48P32 ensembles are all within 1-2\% of each other, and the unitary pion masses are also similar with a 10\% difference. Thus, they are very suitable for investigating the discretization error of the hadron structure with non-zero given momentum. The pion mass and volume of C32P23 are close to those of F48P21 within 10\%, and remaining differences can be further suppressed by interpolation with the other ensembles or by generating a new ensemble C36P21 using interpolated parameters. The other ensembles with larger dimensionless volume, such as F64P14 and/or H64P22, should also be helpful in achieving better control over the discretization error, and will be generated in the future.

For the Clover fermion action, defining the renormalized quark mass  $m_q^R$ from the bare quark mass parameter $\tilde{m}^b_q$ can be subtle since the critical quark mass $\tilde{m}_{\rm crti}$ vanishing the pion mass is non-zero.  A more practical solution defines it through the PCAC relation~\cite{JLQCD:2007xff}:
\begin{equation}
Z_A\partial_\mu A_\mu=2m^R_qZ_PP,    
\end{equation}
where $A_{\mu}=\bar{\psi}\gamma_{5}\gamma_{\mu}\psi$ and $P=\bar{\psi}\gamma_{5}\psi$. The PCAC quark mass $m_q^{\rm PC}$ is then defined through the pion correlation functions:
\bal
m_q^{\rm PC}&=\frac{m_{\rm PS}\sum_{\vec{x}}\langle A_4(\vec{x},t)P^\dag(
\vec{0},0) \rangle}{2\sum_{\vec{x}}\langle P(\vec{x},t)P^\dag(\vec{0},0) \rangle}|_{t\rightarrow \infty}\label{eq:mass_v1},
\eal
where $m_{\mathrm{PS}}$ is the pseudoscalar meson mass. The renormalized quark mass is subsequently defined as $m^R_q=Z_A/Z_P m^{\rm PC}_q$. 

\begin{figure}[thb]
\includegraphics[width=0.5\textwidth]{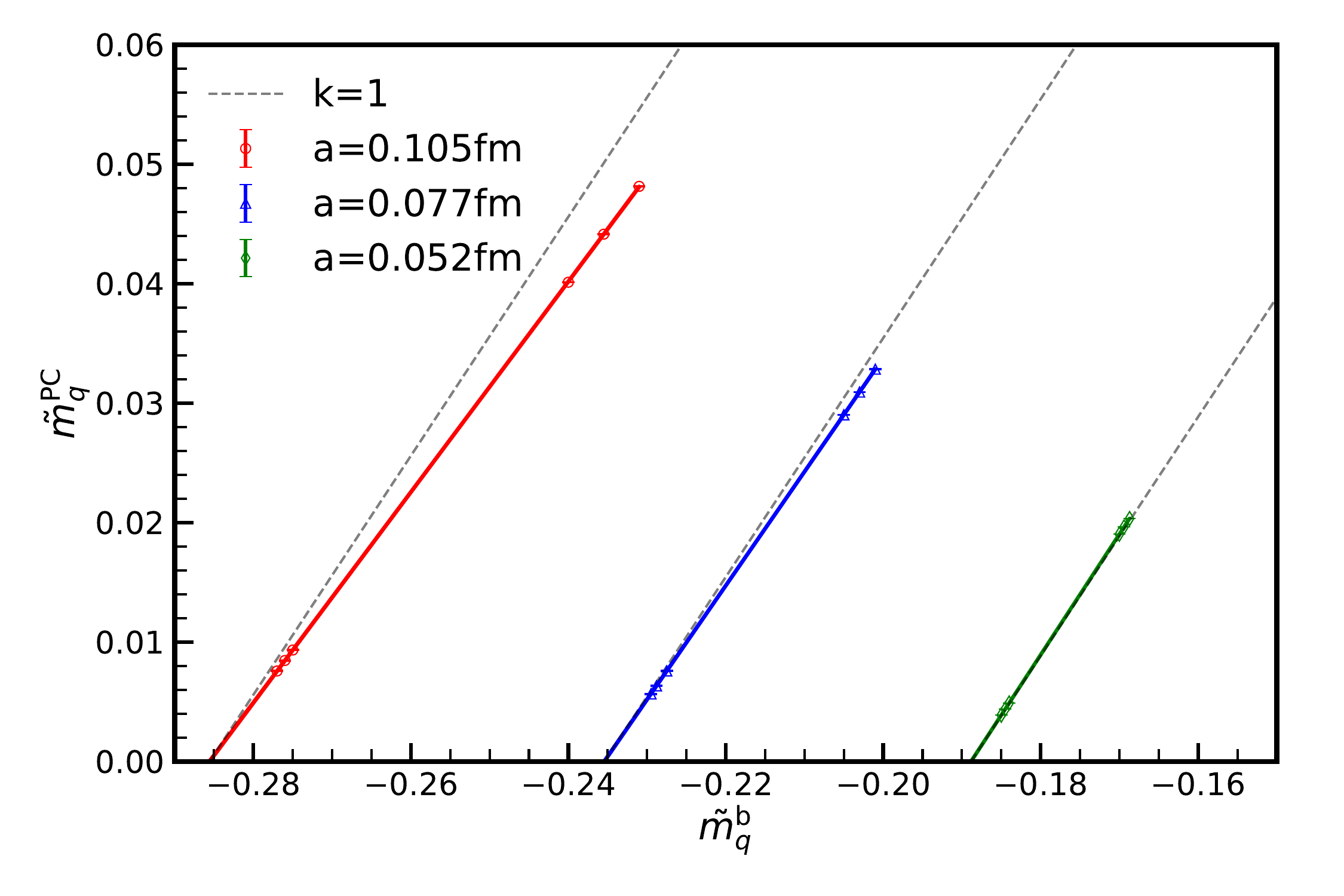}
\caption{The dimensionless PCAC quark mass $\tilde{m}_q^{\rm PC}=m_q^{\rm PC}a$ v.s. the bare quark mass $\tilde{m}_q^{\rm b}=m_q^{\rm b}a$ at three lattice spacings. The slope should approach 1 in the continuum limit. The data points correspond to six valence quark masses around the unitary light and strange quark masses in those ensembles.
}
\label{fig:quark_mass_def2}
\end{figure}

In Fig.~\ref{fig:quark_mass_def2}, we plot the dimensionless PCAC quark mass $\tilde{m}_q^{\rm PC} \equiv m_q^{\rm PC}a $  as a function of the dimensionless input bare quark mass parameters $\tilde{m}_q^b \equiv m_q^{\rm b}a$, at three lattice spacings with $m_{\pi}\sim 300$ MeV. The figure also includes linear fits using the following form:
\bal
\tilde{m}^{\rm PC}_q=k_m (\tilde{m}^{b}_q-\tilde{m}_{\rm crti}),
\eal
where $\tilde{m}_{\rm crti}$ corresponds to the critical pion mass that makes the pion mass and $\tilde{m}^{\rm PC}$ vanish. The parameter $k_m=1+{\cal O}(a^2,\alpha_s,{a\alpha_s})$ approaches $1/Z_A$ determined by non-perturbative RI/MOM renormalization (due to the relation $Z_mZ_P=1$) in the continuum limit, while it is affected by the ${\cal O}(a^2)$ discretization error and ${\cal O}(\alpha_s)$ loop effects at finite lattice spacing. 

Unlike the hadron mass, the determination of physical quark mass on the lattice using discretized actions requires additional renormalization. The RCs defined under the $\MSbar$ scheme, can only be obtained through regularization-independent (RI) schemes such as RI/MOM~\cite{Martinelli:1994ty} or SMOM~\cite{Aoki:2007xm,Sturm:2009kb}. These RCs should be independent of intermediate schemes.

For the overlap fermion action, which possesses strict chiral symmetry, the relations $Z_V=Z_A$ and $Z_P=Z_S=1/Z_m$ are satisfied strictly. The scheme dependence of RI/MOM or SMOM can be ignored {compared} to other systematic uncertainties ~\cite{He:2022lse}.

However, in the case of the clover fermion action, the ratio $Z_A/Z_V$ can deviate from unity due to the additive chiral symmetry breaking effect generated by additional terms in the action
, and it is sensitive to the choice of using RI/MOM or SMOM scheme. Based on the $f_{\pi}$ at three lattice spacing with $m_{\pi}=317$ MeV, the scheme sensitivity is approximately 1\% after a linear $a^2$ continuum extrapolation, with the discretization error through RI/MOM being 25\% smaller than that of SMOM.

\begin{figure}[thb]
\includegraphics[width=0.45\textwidth]{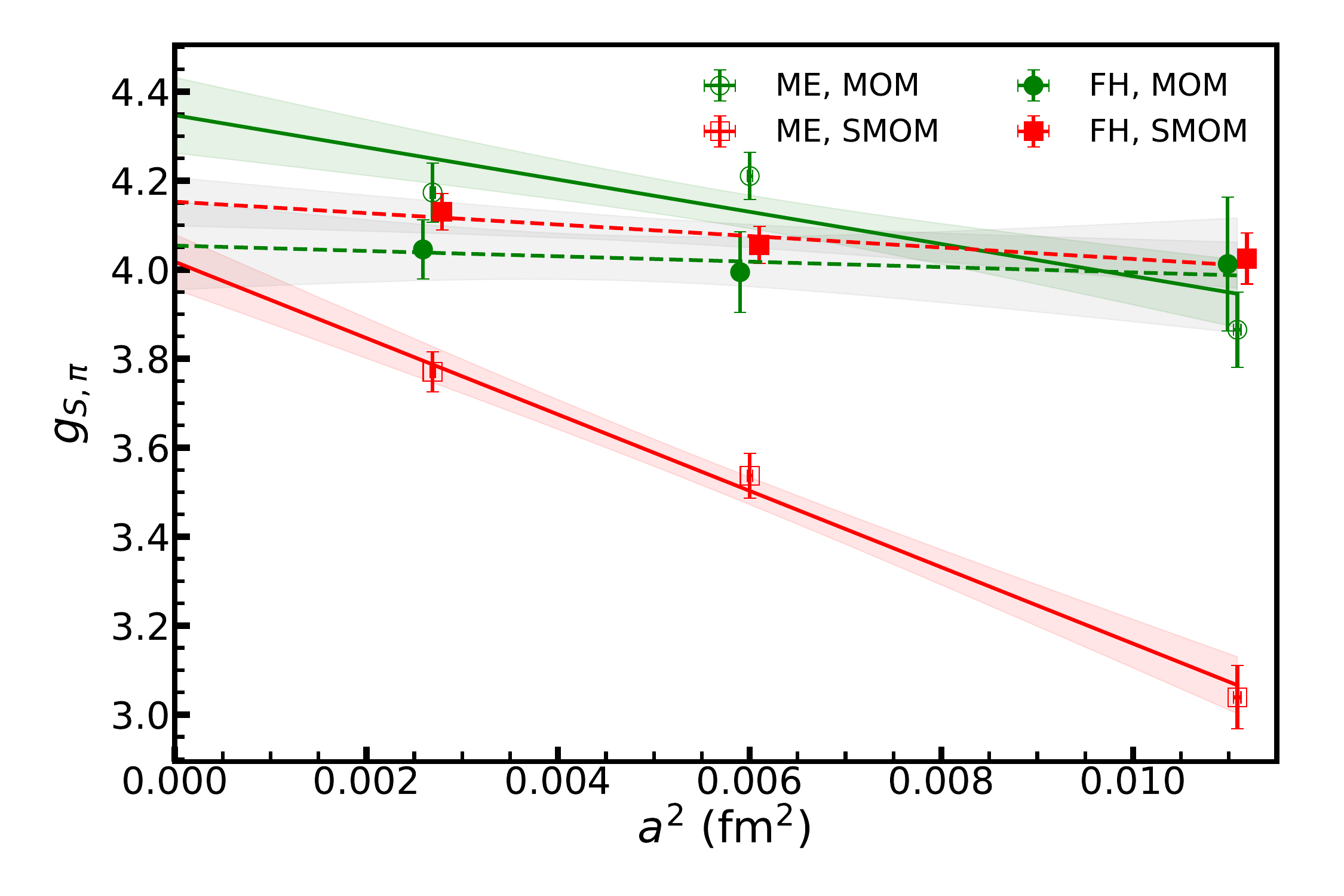}
\caption{
Renormalized scalar matrix element $g_{S,\pi}$ with $m_{\pi}$=317 MeV at three lattice spacing. $g_{S,\pi, {\rm ME}}$ used $Z^{\MSbar\rm{(2\ GeV)}}_S$, $g_{S,\pi, {\rm FH}}$ used $Z^{\MSbar\rm{(2\ GeV)}}_P$, through either RI/MOM or SMOM scheme. The extrapolated values deviate by {$\sim 7$\%}.
}
\label{fig:gs_lat}
\end{figure}

For the chiral symmetry breaking effect between $Z_S$ and $Z_P$, it is valuable to consider the scalar matrix element with the valence quark contribution only,
\bal
g_{S, \pi, {\rm ME}}={\langle \pi|S|\pi\rangle}_{\rm val}/{\langle \pi|\pi\rangle},
\eal
where $S=\bar{\psi} \psi$. We show the renormalized $g_{S,\pi, {\rm ME}}^{\MSbar\rm{(2\ GeV)}}=Z^{\MSbar\rm{(2\ GeV)}}_S g_{S,\pi}$ obtained from RI/MOM and SMOM schemes in Fig.~\ref{fig:gs_lat}, for $m_{\pi}=317$ MeV at three different lattice spacings. It can be observed that the RI/MOM scheme exhibits a smaller discretization error than that of the SMOM scheme, and the continuum extrapolated values differ from each other by approximately {7.6(2.3)\%}. 

Using the Feynman-Hellman theorem, one can also extract $g_{S,\pi}$ from the quark mass dependence of $m_{\pi}$, as
\bal
g_{S, \pi, {\rm FH}}=\frac{1}{2}\frac{\partial m_{\pi}(m_q)}{\partial m_q}\simeq \frac{m_{\pi}}{4m_q}+{\cal O}(m_q,a^2),
\eal
where the factor $\frac{1}{2}$ in front of $\frac{\partial m_{\pi}(m_q)}{\partial m_q}$ is used to average the contribution from two propagators in the pion correlator. Using the renormalized quark mass $m_q^{\MSbar \rm{(2\ GeV)}}$ extracted with RI/MOM or SMOM scheme, $g_{S,\pi, {\rm FH}}^{\MSbar\rm{(2\ GeV)}}$ (filled green dots for MOM and red boxes for SMOM) are in the range of 3.9 to 4.2 and then slightly smaller than the linear $a^2$ extrapolated value $g_{S, \pi, {\rm ME}}$={4.35(9)} (green band) using the RI/MOM scheme but consistent with the SMOM value 4.02(6) (red band) . Even more, $g_{S,\pi, {\rm FH}}$ is consistent with the $g_{S,\pi, {\rm ME}}$ using the RI/MOM scheme at each lattice spacing within two sigma, but have significant difference from the $g_{S,\pi, {\rm ME}}$ using the SMOM scheme. Thus the deviation between the $g_{S,\pi, {\rm FH}}$ and $g_{S,\pi, {\rm ME}}$ using the RI/MOM scheme would be only a systematic uncertainty due to the linear $a^2$ extrapolation. Thus the renormalized $g_S$ shall have about {7\%} systematic uncertainty with present data, and more reliable continuum extrapolation with data at more lattice spacing is essential to obtain accurate prediction on $g_S$. 

The renormalization constants for various quark bilinear operators are detailed in the supplemental materials~\cite{ref_sm}, along with a discussion on the discretization error from different renormalization methods of quark field and mass. 

\section{Results}\label{sec:result}

Using the lattice spacing shown in Table~\ref{tab:lattice}, we find that the unitary pion mass on the ensemble C48P14 at $a=0.1053(2)$ fm is 135.5(1.6) MeV, which perfectly agrees with the physical neutral pion mass $m_{\pi^0}$ of 134.98 MeV within 1\% statistical uncertainty. The charged pion mass $m_{\pi^{\pm}}$ 139.57 MeV receives the QED correction 4.53(6) MeV~\cite{Feng:2021zek} and then the subtracted pure QCD $m_{\pi^{\pm}}$ is consistent with that of $m_{\pi^0}$ within the uncertainty.

The corresponding renormalized light quark mass and pion decay constants can also be determined as:
\bal\label{eq:result_a0.105}
m_{l}^{\MSbar\rm{(2\ GeV)}}(a=0.105~\mathrm{fm})&=3.64(8)(11)~\mathrm{MeV},\nonumber\\
f_{\pi}(a=0.105~\mathrm{fm})&=121.9(5)~\mathrm{MeV},
\eal
where the second uncertainty of $m_{q}$ comes from that of the renormalization constant. Based on the continuum extrapolation with a 317 MeV pion mass, the pion decay constant can change by approximately 7\%, and then agree with the present PDG value $ 130.4(0.2){\rm MeV}$ \cite{ParticleDataGroup:2020ssz} after the continuum extrapolation. 

In order to process this continuum extrapolation systematically, we calculate the quark propagators with unitary light quark mass and also 2 partially quenched quark masses with the constraint $m_{\pi}L>3.5$, on each of the 11 ensembles. Then we use the NLO partially quenched $\chi$PT form~\cite{Sharpe:1997by} to describe the pion masses and decay constants with different valence and sea quark masses, in addition to extra parameters $c_{m/f,a/l}$ for the finite lattice spacing/volume corrections.

Since the statistics on each ensemble are different, we perform 4000 bootstrap re-samplings on each ensemble and conduct the correlated global fit based on these bootstrap samples. In such a strategy, the correlation between different data points in the same ensemble is included automatically, and that between different ensembles vanishes within the statistical uncertainty of the re-sampling. The lattice spacing and renormalization constants are sampled for each bootstrap sample using a Gaussian distribution with their uncertainties as the width of the distribution. 

\begin{table*}[ht!]                   
    \caption{Summary of our determination on quark masses at $\MSbar{\rm (2 GeV)}$ and the other quantities\red{, through the intermediate RI/MOM or SMOM schemes}, with comparison with FLAG~\cite{FlavourLatticeAveragingGroupFLAG:2021npn} and/or PDG~\cite{ParticleDataGroup:2020ssz}. \red{The difference between two schemes is considered as systematic uncertainty in the combined determination.}}  

    \begin{tabular}{l| l l l l l | l l }       
    \hline
    \hline
                &$m_l$ (MeV)        & $m_u$  (MeV)  & $m_d$  (MeV)  & $m_s$  (MeV)      &$\Sigma^{1/3}$(MeV)& $m_s/m_l$        &   $m_u/m_d$     \\
    RI/MOM      &3.60(11)           & 2.45(22)      & 4.74(11)      & 98.8(2.9)         &268.6(3.6)         & 27.47(30)        &   0.519(51)    \\ 
    SMOM        &3.45(05)           & 2.25(10)      & 4.65(08)      & 94.1(1.2)         &269.3(1.8)         & 27.28(22)        &   0.485(26)     \\ 
    Combined    &3.60(11)(15)       & 2.45(22)(20)  & 4.74(11)(09)  & 98.8(2.9)(4.7)    &268.6(3.6)(0.7)    & 27.47(30)(13)    &   0.519(51)(34) \\
    FLAG/PDG    &3.381(40)          & 2.27(09)      & 4.67(09)      & 92.2(1.0)         &272(5)             & 27.42(12)        &   0.485(19)     \\
    \hline
                &$F $ (MeV)        & $F_{\pi}/F$       & $f_{\pi}$(MeV)    &  $f_{K^{\pm}}$(MeV)   &  $f_{K^{\pm}}/f_{\pi}$ & $\ell_3$   &   $\ell_4$      \\
    RI/MOM      & 86.6(7)          & 1.0675(19)        & 130.7(0.9)        & 155.6(0.8)            &1.1907(76)              &2.43(54)    &   4.322(75)     \\
    SMOM        & 85.1(6)          & 1.0683(15)        & 128.6(0.8)        & 152.9(0.7)            &1.1890(74)              &2.49(23)    &   4.226(48)     \\
    Combined    & 86.6(7)(1.4)     & 1.0675(19)(08)     & 130.7(0.9)(2.1)   & 155.6(0.8)(2.7)       &1.1907(76)(03)          &2.43(54)(05)&   4.322(75)(96)  \\
    FLAG/PDG    & 86.8(6)          & 1.062(7)          & 130.2(0.8)        &  155.7(0.7)           &1.1917(37)              &3.07(64)    &   4.02(45)       \\
    \hline
    \hline
    \end{tabular} 
    \label{tab:final}
\end{table*}

\begin{figure}[thb]
    \centering
    \includegraphics[width=0.45\textwidth]{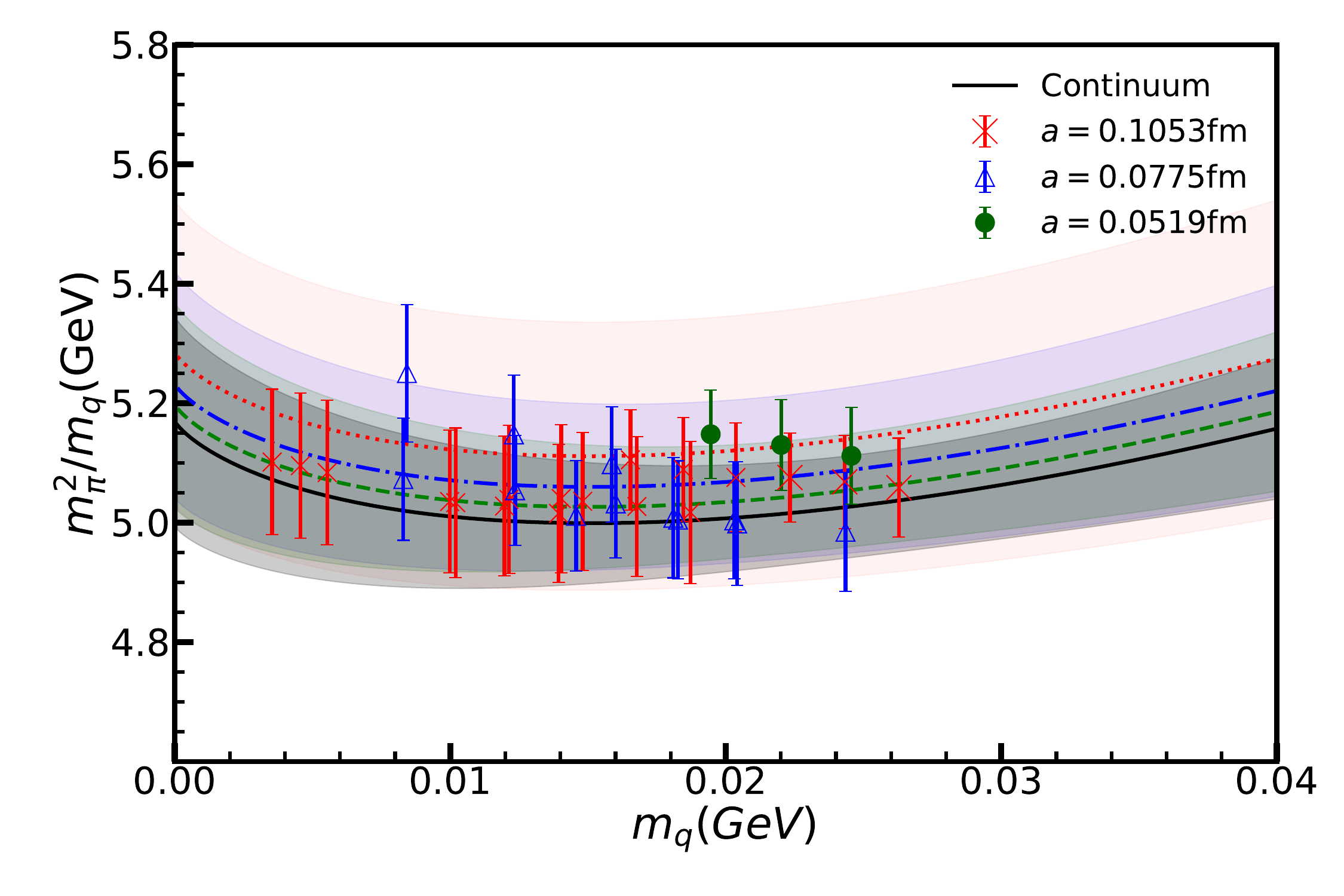}
    \includegraphics[width=0.45\textwidth]{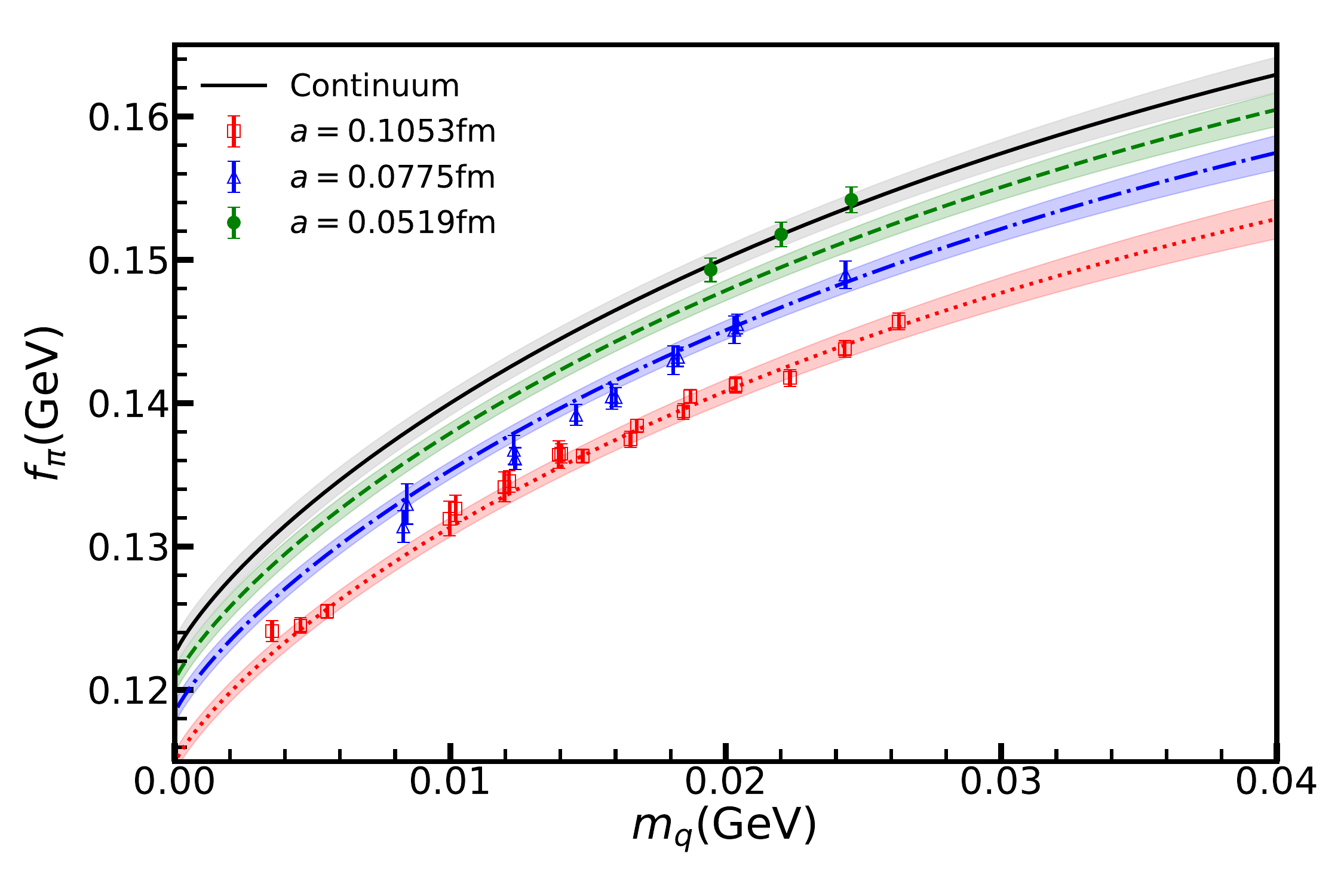}
    \caption{The corrected unitary $(m^{\rm uni}_{\pi})^2/m_q$ and the decay constant $f^{\rm uni}_\pi$ varies with the quark mass, at three lattice spacing (colored data points and {corresponding} bands {with dashed line for 0.0519 fm, dash-dotted line for 0.0775 fm, and dotted line for 0.1053 fm}) and also continuum (gray band).}
    \label{fig:pion_final}
\end{figure}

To illustrate the lattice spacing dependence and the unitary quark mass dependence, we subtract the partially quenching effect using bootstrap samples of the fit parameters from the original data points, and show the ratio $(m_{\pi})^2/m^{\MSbar\rm{(2\ GeV)}}_q$ 
(upper panel) and also $f_{\pi}$ (lower panel) at different quark masses $m_q$. The corrected data points at different lattice spacings use different symbols: red crosses for $a$=0.105 fm, blue triangles for $a$=0.077 fm, and filled green dots for $a$=0.052 fm. The bands with similar color represent the fitted band at the corresponding lattice spacing, and the gray band shows the final prediction in the continuum and infinite volume limit. It is observed that the continuum extrapolation pushes $f_{\pi}$ to be obviously higher, while the impact on the $(m_{\pi}^2/m_q)$ ratio and, consequently, $m_q$ is much weaker.

The $m_{\pi}$ and $f_{\pi}$ with unitary valence and sea quark masses have the following parameterization,
\begin{align}
    m_\pi^2&=\Lambda_{\chi}^2  2 y
    \left[
        1+y 
        \left( 
            \ln\frac{2y\Lambda^2_{\chi}}{m^2_{\pi, {\rm phys}}}-{\ell}_3
        \right)
        +\mathcal{O}(y^2)
    \right],
    \\
    F_\pi&=F
    \left[ 
        1-2y
        \left( 
            \ln\frac{2y\Lambda^2_{\chi}}{m^2_{\pi, {\rm phys}}}-\ell_4
        \right)
        +\mathcal{O}(y^2)
    \right],
\end{align}
where $\Lambda_{\chi}=4\pi F$, $y=\frac{\Sigma m_l}{F^2 \Lambda_{\chi}^2}$. $\Sigma$, $F$ and $\ell_{3,4}$ are low energy constants. Our determination of those constants are also collected in Table~\ref{tab:final}, consistent with the current $N_f=2+1$ FLAG average but have smaller uncertainties except $F$.

In this work, we use the $m_{K^{\pm}}$ and $m_{K^0}$ with the constraint $m^{\rm phys}_u+m^{\rm phys}_d=2m^{\rm phys}_l$, to determine the up, down and strange quark masses $m_{u,d,s}$. The QED correction on the kaon mass is subtracted based on the literature~\cite{Giusti:2017dmp} under the renormalization scheme $m^{\overline{\textrm{MS}}}_{q, \rm QCD+QED}(2\mathrm{GeV})=m^{\overline{\textrm{MS}}}_{q, \rm QCD}(2\mathrm{GeV})$. On each ensemble, we calculate the strange quark propagators with a unitary strange quark mass $m^{\rm v}_s=m^{\rm s}_s$, and also 2 partially quenched quark masses $m^{\rm v}_s\sim 100$ MeV. We construct the Kaon correlation functions with three strange quark masses and three light quark masses used in the pion case. The $3\times 3$ partially quenched Kaon masses on all the ensembles are fitted with the following form proposed in a recent work~\cite{ExtendedTwistedMass:2021gbo},
\bal
&m_K^2(m^{\rm v}_l,m^{\rm s}_l,m^{\rm v}_s,m^{\rm s}_s,a) 
 =(b_s^{\rm v} m^{\rm v}_s+b_s^{\rm s} m^{\rm s}_s+b_l^{\rm v} m^{\rm v}_l+b_l^{\rm s} m^{\rm s}_l) \notag \\ 
&~~~~~~ \times\left[ 1+c^K_l m^{\rm v}_l+c^K_m a^2 + c_{L}^K \exp{(-m_{{\pi}} L)} \right].
\eal

Based on the fit of $m_K^2$, the total strange quark mass dependence \red{$b_s^{\rm v}+b_s^{\rm s}= 2.37(08)$} is consistent with the leading order light quark mass dependence \red{$b_l^{\rm v}+b_l^{\rm s}= {2.59(95)}$}, and the coefficient of the non-linear quark mass dependence \red{$c_l^K= { 0.07(13) } $} can not be determined based on current statistics. 

\begin{figure}[thb]
    \centering
    \includegraphics[width=0.45\textwidth]{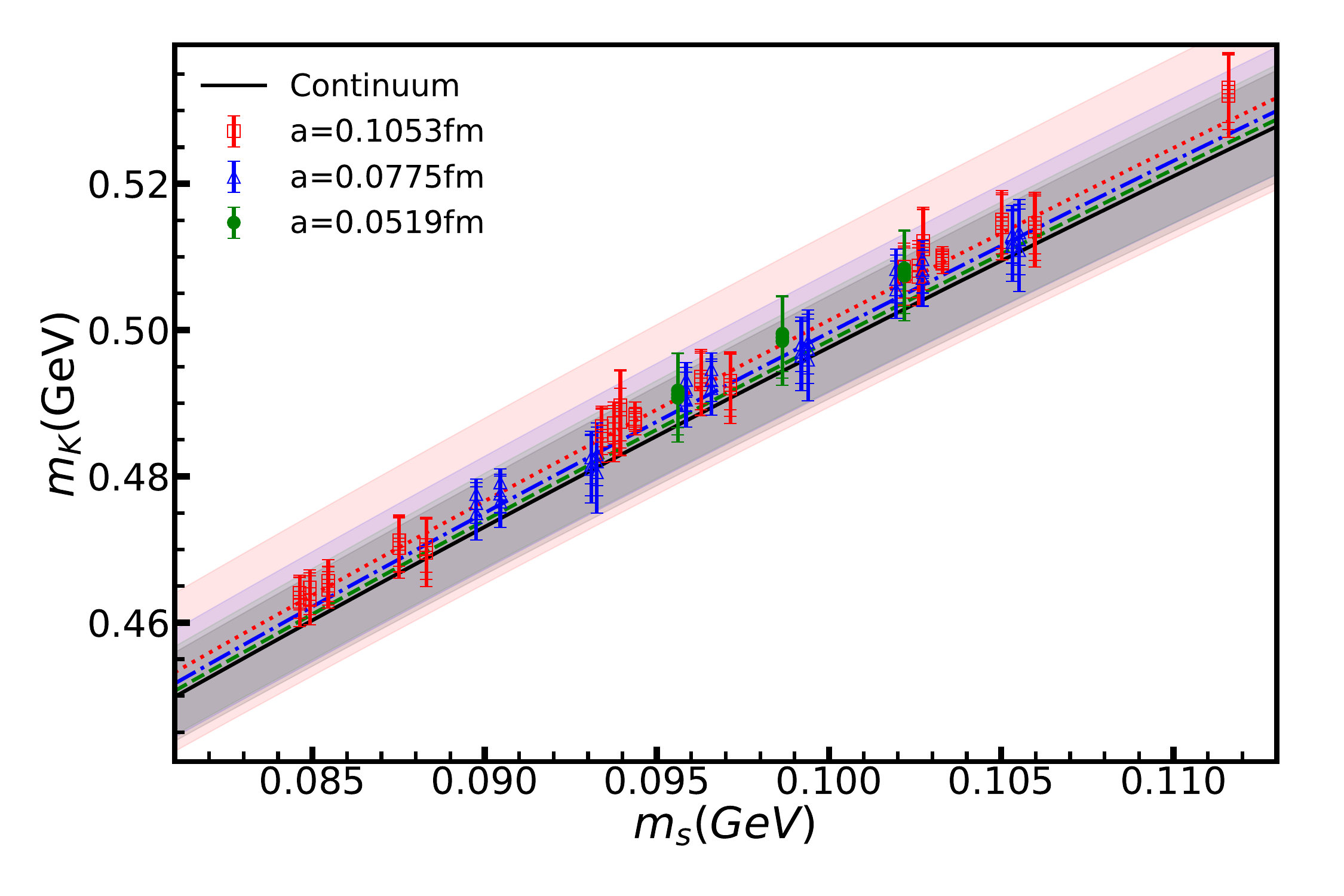}
    \includegraphics[width=0.45\textwidth]{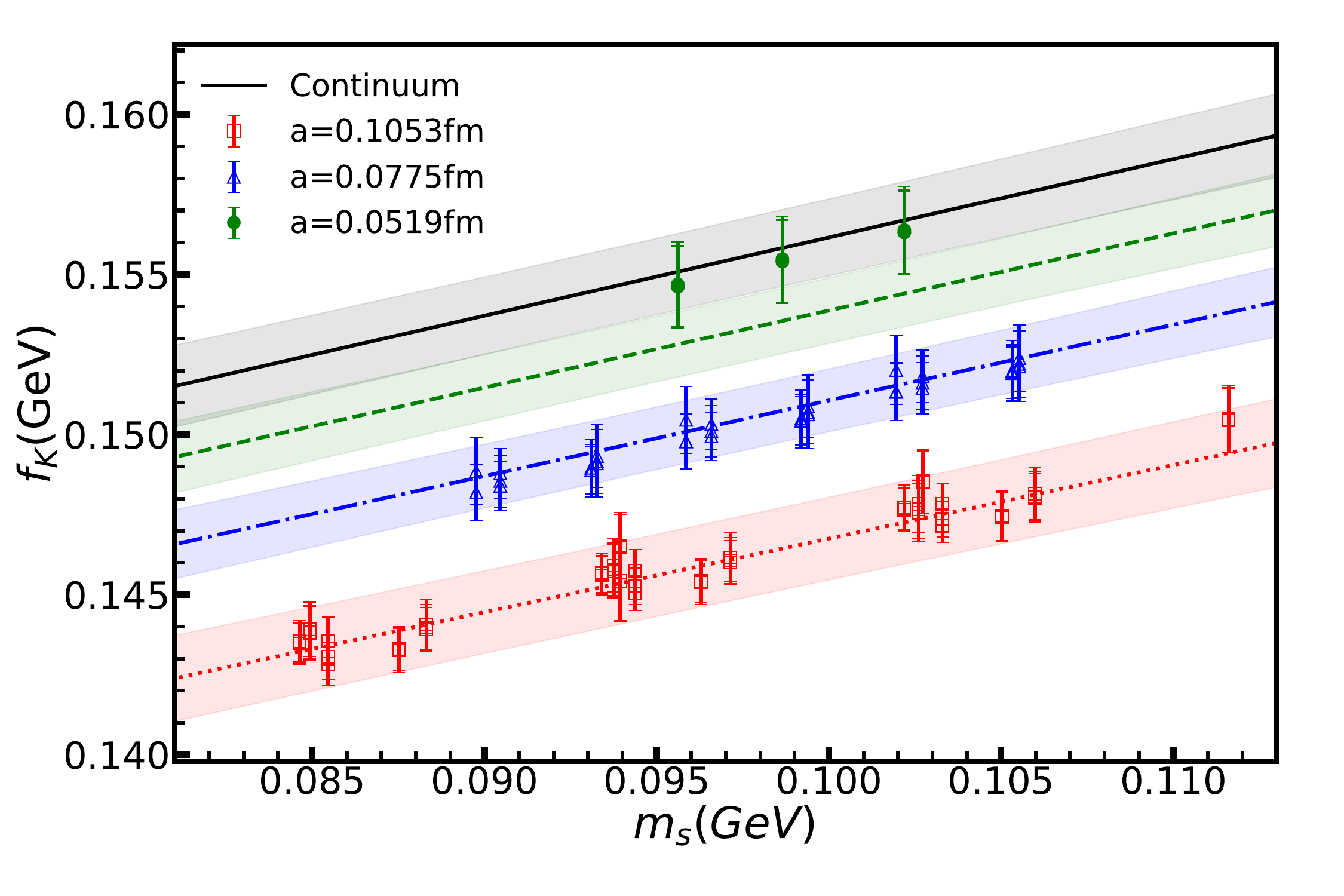}
    \caption{The corrected kaon mass $m^{\rm cr}_K$ and the decay constant $f^{\rm cr}_K$ with the physical light quark mass $m_l^{\rm phys}$, varies with the strange quark mass at three lattice spacing (colored data points and bands) and also continuum (gray band).}
    \label{fig:kaon_final}
\end{figure}

In Fig.~\ref{fig:kaon_final}, we show the corrected kaon mass $m^{\rm cr}_K$ and decay constant $f^{\rm cr}_K$ with the light quark mass $m_l$ corrected to its physical value $m_l^{\rm phys}$. The finite volume and partially quenched effects are also subtracted. We can found that $f_K$ also exhibits a strong lattice spacing dependence, similar to the $f_{\pi}$ case, while the Kaon mass is insensitive to the lattice spacing. 

As illustrated in Fig.~\ref{fig:pion_final} and Fig.~\ref{fig:kaon_final}, all the global fits of the pseudoscalar meson mass and decay constant provide reasonable $\chi^2$/d.o.f. More information on the global fit can be found in the supplemental materials~\cite{ref_sm}.

The physical quark masses $m_{u,d,s}$ and also corresponding $f_{\pi,K}$ using $m_l^{\rm phys}$ and intermediate RI/MOM scheme, are collected in Table~\ref{tab:final}. In addition, Table~\ref{tab:final} shows the global fit results using the $Z_{A,P}$ through the SMOM scheme for comparison. As we can see from the continuum extrapolation tests using a 317 MeV pion mass, the SMOM scheme yields quark masses that are 3-4\% lower and decay constants that are $\sim$2\% lower compared to the RI/MOM scheme. However, the ratio of the quark masses or decay constants remains unchanged within the uncertainty as the renormalization constants are cancelled.

Therefore, we consider the result using the RI/MOM scheme as the central value due to its smaller discretization error, and treat the difference between the results obtained using the two schemes as systematic uncertainties. Such a systematic uncertainty can also be considered as an estimate of the residual discretiation error, as the correct continuum limit should be independent of the intermediate renormalization scheme. All our determinations are consistent with the present lattice averages~\cite{FlavourLatticeAveragingGroupFLAG:2021npn} and/or PDG~\cite{ParticleDataGroup:2020ssz} within 1-2~ sigma.

\section{Summary}\label{sec:summary}

In this work, we determine the up, down, and strange quark masses, along with several low energy constants, using the 2+1 flavor full-QCD ensembles with tadpole-improved clover and Symanzik actions. The major results are summarized in Table~\ref{tab:final}.

Similar to one of the most precise works~\cite{BMW:2010skj} with the clover fermion, we have skipped the axial current improvement~\cite{DellaMorte:2005aqe} since the improvement coefficient itself strongly depends on the lattice spacing $a$ and bring the improvement term to be consistent with an ${\cal O}(a^2)$ correction. As evidence, both $f_{\pi}$ and $m_q$ 
show good consistency
with a simple linear $a^2$ lattice spacing dependence. Then assigning $c_A\sim 0.05 a/(0.105~\mathrm{fm})$ in the improved axial vector current $A^{\rm imp}_{\mu}=A_{\mu}+c_A a\partial_{\mu} P$~\cite{DellaMorte:2005aqe} can eliminate the discretization error in $f_{\pi,K}$, however, this error will be transferred to the quark mass. Thus simulations at more lattice spacings would be a more systematic solution to enhance the accuracy of our predictions in the continuum than the axial current improvement.

On the other hand, the additive chiral symmetry breaking makes the renormalization of the quark mass to be highly non-trivial. Our study suggests that the $Z_S$ and $Z_P$ obtained through the SMOM scheme are closer than those through the RI/MOM scheme, while the latter one can make the discretization error of both the $m^R_q=Z_A/Z_Pm^{\rm PC}_q$ and $g_S^R=Z_Sg_S$ to be smaller. Our final prediction of the quark masses are \red{$5.6(2.8)$\%} higher than the current 2+1 flavor lattice averages but consistent with the previous 2+1 and 2+1+1 flavor results using the RI/MOM scheme. At the same time, the RI/MOM scheme can also cause the Feynman-Hellman theorem $g_{S,\pi}\simeq \frac{m_{\pi}}{4m_q}$ to be violated by 7(3)\% after the linear ${\cal O}(a^2)$ continuum extrapolation. 

Using the SMOM scheme can eliminate the violation and bring the quark mass prediction closer to the current 2+1 flavor lattice average. However, the SMOM scheme introduces larger discretization errors for all the renormalized quantities we investigated and causes the decay constants $f_{\pi, K}$ to be 2-3\% smaller than the physical values after the linear ${\cal O}(a^2)$ continuum extrapolation. 

    The above observations indicate that renormalization is a significant issue that requires careful investigation, and conducting similar calculations using chiral fermions would be essential to gain a better understanding of these violations. At the same time, non-perturbative renormalization should remove all the ${\cal O}(\alpha_s)$ effects, but not all the cross terms like the residual ${\cal O}(a\alpha_s)$ effect of the clover action, which can cause the ${\cal O}(a^2)$ continuum extrapolation to fail. Thus, we consider the difference between the results obtained by the two schemes as systematic uncertainties in our final determination of the aforementioned quantities, which are larger than the statistical uncertainties in various cases. We anticipate that additional research utilizing ensembles with a greater number of lattice spacings can encompass both the ${\cal O}(a^2)$ and ${\cal O}(a\alpha_s)$ terms in the continuum extrapolation, resulting in a more dependable and uniform continuum limit.

It is worth mentioning that in Ref.~\cite{BMW:2010skj}, the trace-subtraction trick $\bar{S}=S-\frac{1}{4}\mathrm{Tr}[S]$ is applied {into renormalization procedure}, and the quark mass is renormalized at RI/MOM 2 GeV, followed by perturbative matching at a much higher scale. This approach is crucial in suppressing their truncation error to the sub-percent level. However, in our case, it appears to be inefficient due to significant non-perturbative effects observed at 2 GeV. We plan to conduct a more systematic investigation once the CLQCD ensembles at more lattice spacings are generated.

\section*{Acknowledgment}
We thank Ying Chen, Hengtong Ding, Xu Feng, Chuan Liu, Zhaofeng Liu, Qi-An Zhang, and the other CLQCD members for valuable comments and suggestions. The calculations were performed using the Chroma software suite~\cite{Edwards:2004sx} with QUDA~\cite{Clark:2009wm,Babich:2011np,Clark:2016rdz} through HIP programming model~\cite{Bi:2020wpt}. The numerical calculation were carried out on the ORISE Supercomputer, HPC Cluster of ITP-CAS, the Southern Nuclear Science Computing Center(SNSC), the Siyuan-1 cluster supported by the Center for High Performance Computing at Shanghai Jiao Tong University, and Advanced Computing East China Sub-center. This work is supported in part by NSFC grants No. 1229060, 1229062, 12293061, 12293065 and 12047503, the science and education integration young faculty project of University of Chinese Academy of Sciences, the Strategic Priority Research Program of Chinese Academy of Sciences, Grant No.\ XDB34030303 and YSBR-101, and also a NSFC-DFG joint grant under Grant No.\ 12061131006 and SCHA 458/22, Guangdong Major Project of Basic and Applied Basic Research No. 2020B0301030008. The numerical calculations in this paper have been done on the supercomputing system in the Dongjiang Yuan Intelligent Computing Center.

\bibliographystyle{apsrev4-1}
%\bibliography{reference.bib}

\clearpage

\section*{Appendix}

\subsection{Simulation details}\label{sec:simulation}

This section will be organized in the following: 
The dimensionless joint fit on the pseudo-scalar meson mass, its decay constant, and the corresponding partially conserved axial currennt (PCAC) quark mass will be discussed in Sec.~\ref{sec:dimless}; Based on the determination of an uniform lattice spacing at given $\hat{\beta}$ detailed in Sec.~\ref{sec:lat}, the mistuning effect of the tadpole improvement factors is not always negligible and will be addressed in Sec.~\ref{sec:tadpole}.

\subsubsection{Dimensionless joint fit}\label{sec:dimless}

We construct two kinds of the two-point functions for the meson states:
\begin{align}
C_{2,wp}^{\Gamma\Gamma'}(t;m_q)&=\sum_{\vec{x}}\langle\mathrm{Tr}[S^{\dagger}_C(\vec{x},y_0+t;y_0;m_q)\gamma_5\Gamma \nonumber\\
&\quad\quad\quad  S_C(\vec{x},y_0+t;y_0;m_q) \Gamma' \gamma_5]\rangle/L^3,\\
C_{2,ww}^{\Gamma\Gamma'}(t;m_q)&=\sum_{\vec{x},\vec{z}}\langle\mathrm{Tr}[S^{\dagger}_C(\vec{x},y_0+t;y_0;m_q)\gamma_5\Gamma \nonumber\\
&\quad\quad\quad S_C(\vec{z},y_0+t;y_0;m_q) \Gamma' \gamma_5]\rangle/L^3,
\end{align}
where $C_2$ is independent of the source time slice $y_0$ after taking expectation value, the Coulomb gauge fixed wall source propagators is defined as 
\bal
S_C(x;y_0;m_q)=\sum_{\vec{y}}  S(x;\vec{y},y_0;m_q,U_C),
\eal
$S(x;y;m_q,U)\equiv \psi(x;m_q,U)\bar{\psi}(y;m_q,U)$ is the quark propagator of the quark field $\psi$ with bare quark mass $m_q$ on a given gauge configuration $U$, and $U_C$ is the Coulomb gauge fixed configuration satisfying the gauge fixing condition $\mathrm{Im}[\sum_{i=1,2,3} (U_C(x)-U_C(x-a\hat{n}_i))]=0$.

    For the Clover fermion action, the PCAC quark mass $m_q$ is then defined through the pion correlation functions~\cite{JLQCD:2007xff}:
\bal
m_q^{\rm PC}&=\frac{m_{\rm PS}\sum_{\vec{x}}\langle A_4(\vec{x},t)P^\dag(
\vec{0},0) \rangle}{2\sum_{\vec{x}}\langle P(\vec{x},t)P^\dag(\vec{0},0) \rangle}|_{t\rightarrow \infty}\label{eq:mass_v1_sm}.
\eal
The renormalized quark mass is subsequently defined as $m^R_q=Z_A/Z_P m^{\rm PC}_q$. 

\begin{figure}[thb]
\includegraphics[width=0.5\textwidth]{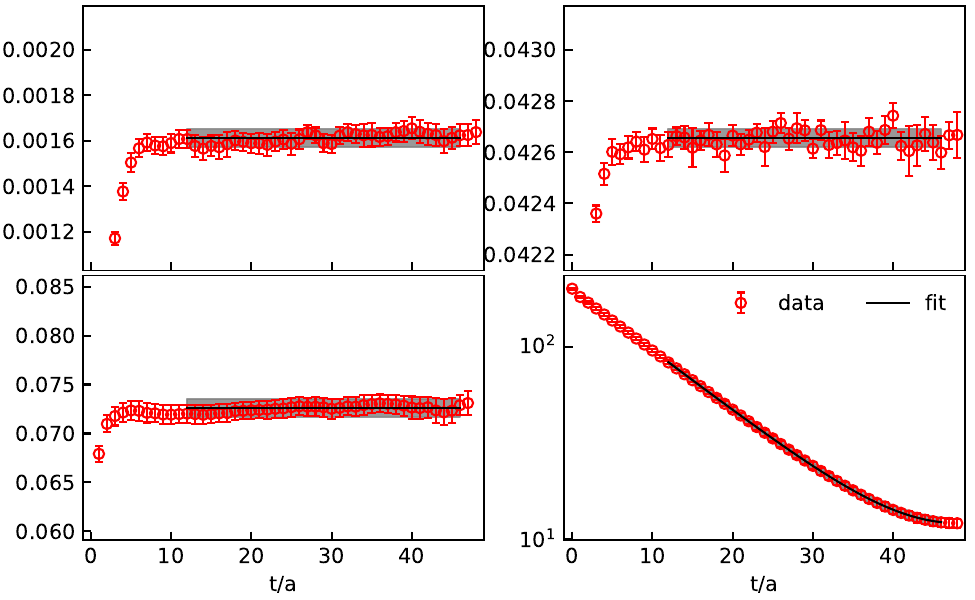}
     \caption{The ratios for $\tilde{m}_q$ (left top), $\tilde{f}_{\rm PS}$ (right top), $\tilde{m}_{\rm PS}$ (left bottom) defined in Eq.~(\ref{eq:fit_quark_mass}-\ref{eq:fit_pion_mass}), and the correlator $\tilde{C}_{2,wp}^{{\cal PP}}$ (right bottom) as functions of $\tilde{t}$, for the physical light quark mass at the coarsest lattice spacing. The joint fit results are shown on the plots as gray bands.}
\label{fig:joint_fit_l}
\end{figure}

\begin{figure}[thb]	\includegraphics[width=0.5\textwidth]{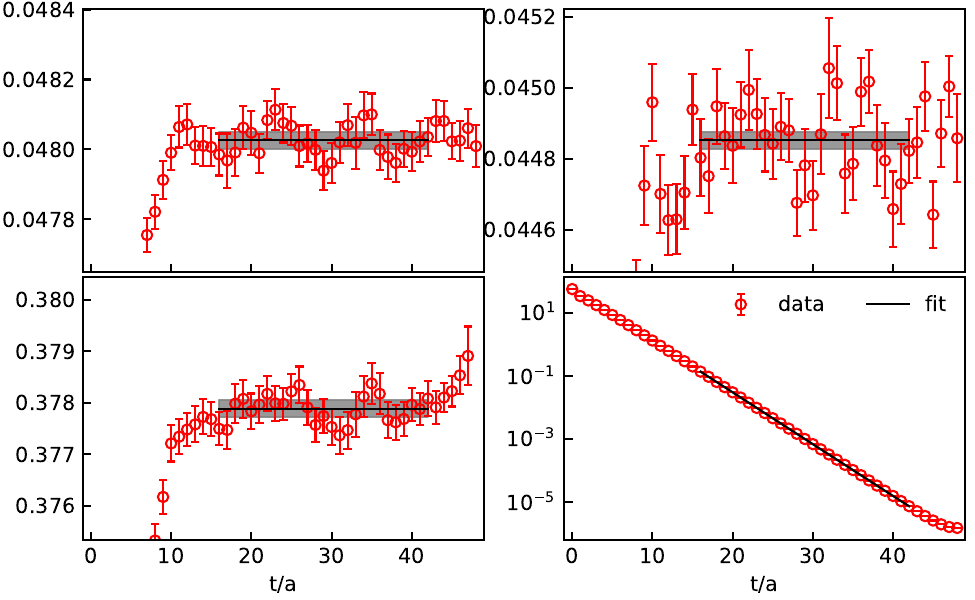}
    \caption{Similar to Fig.~\ref{fig:joint_fit_l} but for the strange quark mass and fewer sources ($n_{\rm src}=3$).} 
\label{fig:joint_fit_s}
\end{figure}

Through a joint fit ($\tilde{O}$ is the dimensionless value of any quantity $O$), 
\bal
&\frac{\tilde{C}_{2,wp}^{{\cal A}_4{\cal P}}(\tilde{t}-1)-\tilde{C}_{2,wp}^{{\cal A}_4{\cal P}}(\tilde{t}+1)}{4\tilde{C}_{2,wp}^{{\cal PP}}(\tilde{t})}|_{0\ll \tilde{t} \ll \tilde{T}} =\frac{\mathrm{Sinh}(\tilde{m}_{\rm PS})}{\tilde{m}_{\rm PS}}\tilde{m}_q^{\rm PC},\label{eq:fit_quark_mass}\\
&\sqrt{\frac{\tilde{C}_{2,wp}^{{\cal PP}}(\tilde{t})}{\tilde{C}_{2,ww}^{{\cal PP}}(\tilde{t})}}|_{0\ll \tilde{t} \ll \tilde{T}}=\frac{\tilde{m}_{\rm PS}^2}{2\tilde{m}_q^{\rm PC}\sqrt{Z_{wp}}}\tilde{f}_{\rm PS},\label{eq:fit_fpi}\\
&\tilde{C}_{2,wp}^{{\cal PP}}(\tilde{t})|_{0\ll \tilde{t} \ll \tilde{T}}=\frac{Z_{wp}}{2\tilde{m}_{\rm PS}}(e^{-\tilde{m}_{\rm PS} \tilde{t}}+e^{-\tilde{m}_{\rm PS}(\tilde{T}-\tilde{t})}), \label{eq:fit_2pt}
\eal
the PCAC quark mass $\tilde{m}_q^{\rm PC}$, pseudo-scalar mass 
\bal
\tilde{m}_{\rm PS}=\mathrm{Cosh}^{-1}\frac{\tilde{C}_{2,wp}^{{\cal PP}}(\tilde{t}-1)+\tilde{C}_{2,wp}^{{\cal PP}}(\tilde{t}+1)}{2\tilde{C}_{2,wp}^{{\cal PP}}(\tilde{t})}|_{0\ll \tilde{t} \ll \tilde{T}},\label{eq:fit_pion_mass}
\eal
and decay constant $\tilde{f}_{\rm PS}$ are extracted as fit parameters, alongside an additional unphysical fit parameter $Z_{wp}$ for the Coulomb gauge-fixed wall source.

In Fig.~\ref{fig:joint_fit_l}, we shown the joint fit result of the unitary light quark on the physical point ensemble C48P14. To suppress the statistical uncertainty, we repeated the calculation on $n_{\rm src}$=48 of 96 times slides on $n_{\rm cfg}$=203 configurations. The values of $n_{\rm src}$ and $n_{\rm cfg}$ of the other ensembles can be found in Table.~\ref{tab:lattice}. The ratios (red data points) defined in Eq.~(\ref{eq:fit_quark_mass}) for $\tilde{m}_q$ are consistent with a constant in the region of $\tilde{t}>10$, as shown in the top left panel. The situation is similar for $\tilde{f}_{\rm PS}$ (top right panel) and also $\tilde{m}_{\rm PS}$ (bottom left panel). Since the statistics on C48P14 are limited, we performed a joint uncorrelated fit in the range $\tilde{t}\in [10,40]$ for Eq.~(\ref{eq:fit_quark_mass}-\ref{eq:fit_2pt}) and used bootstrap resampling to estimate the uncertainty, represented by the gray band in the figure. We can see that the fit agrees very well with the data points, and the uncertainties of the fitted bands are comparable to the original data. Thus, it is unlikely that the uncertainty has been underestimated by the uncorrelated fit.

The strange quark mass case on the ensemble C48P14 is shown in Fig.~\ref{fig:joint_fit_s}. Since the statistical uncertainty decreases with with a heavier quark mass, we only repeated the calculation on 3 time slides for each configuration, resulting in more noticeable fluctuations compared to the light quark case.

\begin{figure}[thb]
\includegraphics[width=0.5\textwidth]{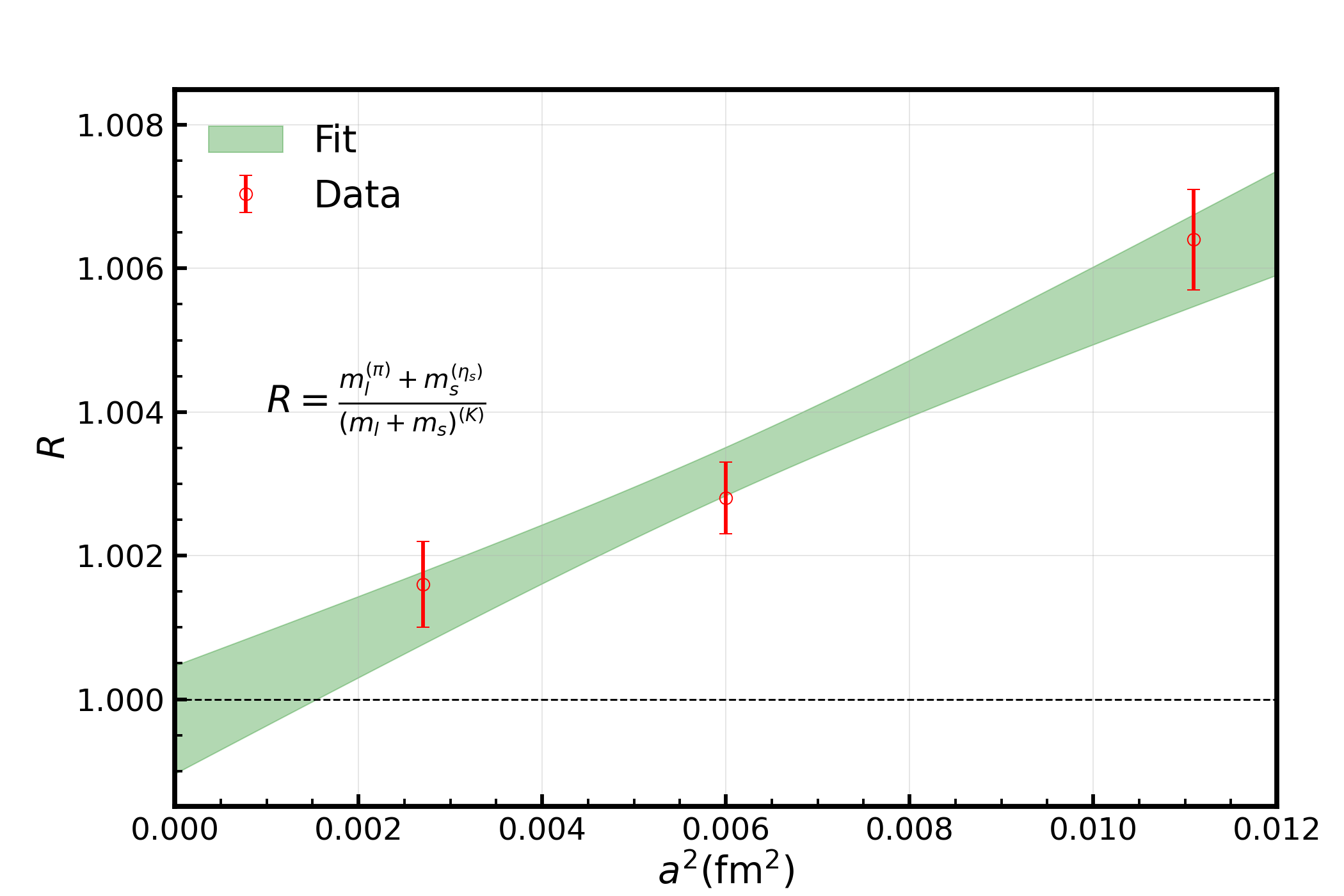}
\caption{The ratio of the PCAC quark mass determined from the quarkoinum (pion and $\eta_s$), and also kaon. Two definitions are consistent upto ${\cal O}(a^2)$ correction. 
}
\label{fig:quark_mass_def}
\end{figure}

Through the above fit, we can extract the light and strange quark masses through the pion and $\eta_s$ correlators, respectively:
\bal
\partial_{\mu}(\bar{l}\gamma_5\gamma_{\mu}l)=2m^{\rm PC}_l \bar{l}\gamma_5l,\ \partial_{\mu}(\bar{s}\gamma_5\gamma_{\mu}s)=2m^{\rm PC}_s \bar{s}\gamma_5l.
\eal
Alternatively, the sum of the light and strange quark masses can also be extracted from the kaon correlator,
\bal
\partial_{\mu}(\bar{s}\gamma_5\gamma_{\mu}l)=(m^{\rm PC}_l+m^{\rm PC}_s) \bar{s}\gamma_5l.
\eal
Thus we should verify two extraction provide consistent result, at least in the continuum. 
Fig.~\ref{fig:quark_mass_def} shows the ratio of these two determinations of $m_l+m_s$ at three lattice spacings, with $m_{\pi}\sim 300$ MeV and $m_{\eta_s}\sim 700$ MeV for better signals.
Two determinations deviate from each other by approximately 0.6\% at the coarsest lattice spacing but are consistent within the statistical uncertainty of 0.1\% after a linear $a^2$ extrapolation to the continuum limit. 

{At three lattice spacings with $m_{\pi}\sim 300$ MeV, we fit the dimensionless PCAC quark mass $\tilde{m}_q^{\rm PC}=m_q^{\rm PC}a$ with the following form:
\bal
\tilde{m}^{\rm PC}=k_m (\tilde{m}^{b}-\tilde{m}_{\rm crti}),
\eal
where $\tilde{m}_q^b=m_q^{\rm b}a$ is the original input quark mass parameters, and} $\tilde{m}_{\rm crti}$ corresponds to the critical pion mass that makes the pion mass and $\tilde{m}^{\rm PC}$ vanish.
The parameter $k_m=1+{\cal O}(a^2,\alpha_s,\blue{a\alpha_s})$ approaches $1/Z_A$ determined by non-perturbative RI/MOM renormalization (due to the relation $Z_mZ_P=1$) in the continuum limit, while it is affected by the ${\cal O}(a^2)$ discretization error and ${\cal O}(\alpha_s)$ loop effects at finite lattice spacing.

\begin{table}[t]
	\caption{\label{tab:quark_mass_def} The bare coupling $\alpha_s^{\rm b}$, critical quark mass $\tilde{m}_{\rm crti}$ and slope $k_m$ at three lattice spacing $a$ and $m_{\pi}\sim$ 300 MeV.}	
		\begin{tabular}{l | ccc}
		  $a$ (fm)  & 0.105 & 0.077 & 0.052 \\   	
         $\alpha_s^{\rm b}(a)$  & 0.2397 & 0.2234 & 0.2035 \\   	
		 \hline 
         $\tilde{m}_{\rm crti}$ & 
         -0.28560(4)
         & -0.23545(3) 
         & -0.18885(1)\\	  
		$k_m$  & \ 0.881 (1) & \ 0.953 (1) & \ 1.009(1)\\	
		\end{tabular}  
 \end{table}

Based on the numerical results listed in Table~\ref{tab:quark_mass_def}, we observe that $\tilde{m}_{\rm crti}$ remains negative even after a naive ${\cal O}(a)+{\cal O}(a^2)$ extrapolation to the continuum, with a value of -0.0865. Therefore, it is crucial to include the ${\cal O}\big((\alpha^b_s)^2\big)$ term, where $\alpha_s^b=\frac{g_0^2}{4\pi}=\frac{10}{4\pi\hat{\beta}u_0^4}$, in the continuum extrapolation to ensure that $\tilde{m}_{\rm crti}\ _{\overrightarrow{a\rightarrow0}} 0$, as predicted by lattice perturbative theory.

On the other hand, the discrepancy between $k_m$ and 1 diminishes as the lattice spacing becomes smaller. We will discuss this further in the following section on renormalization.

\subsubsection{lattice spacing determination}\label{sec:lat}

\begin{figure}[thb]
\includegraphics[width=0.45\textwidth]{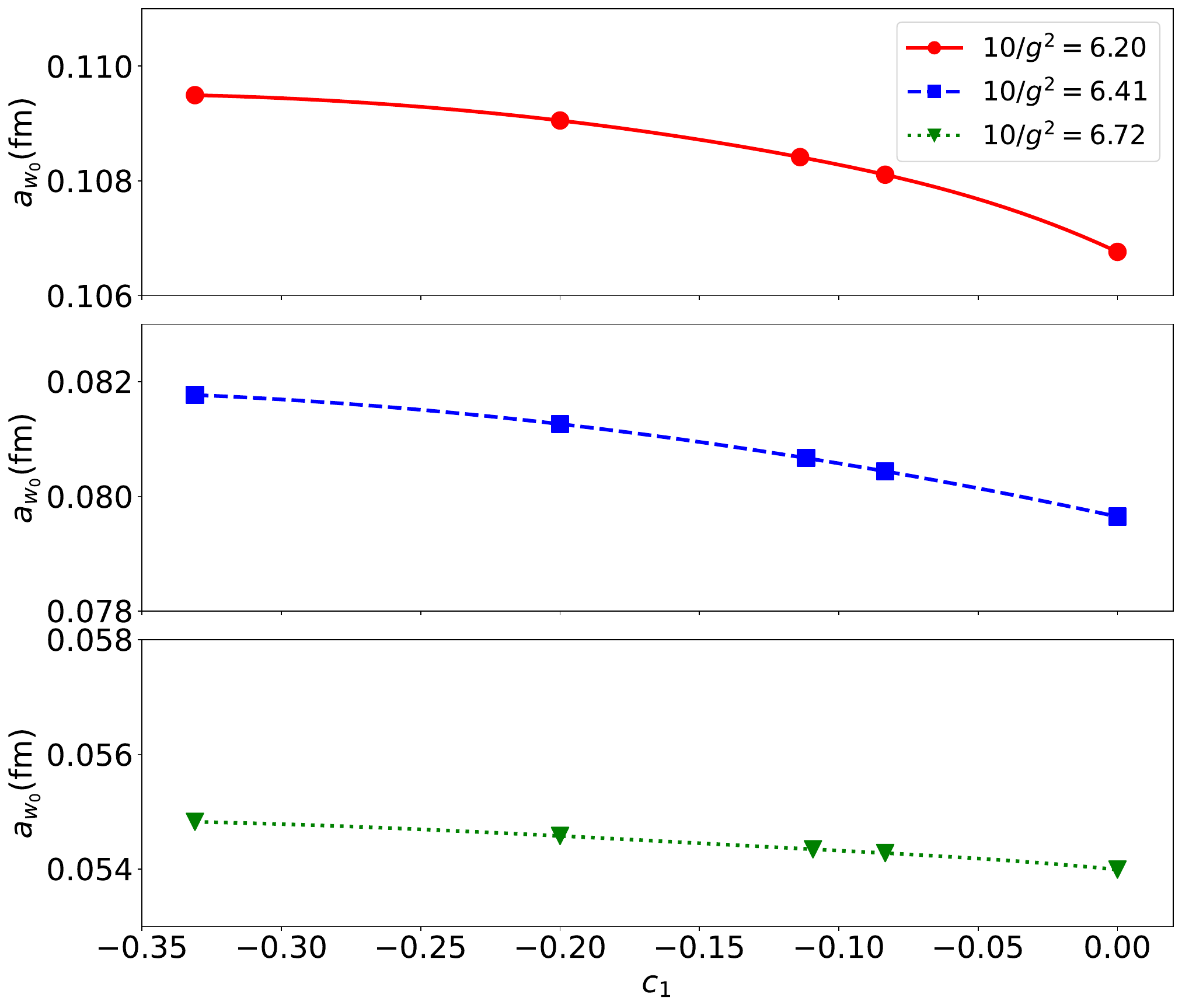}
\caption{The lattice spacings determined by $w_0$ with different gauge action improvement factor $c_1$.}
\label{fig:lattice_spacing_in_diff_flow}
\end{figure}

Fig.~\ref{fig:lattice_spacing_in_diff_flow} illustrates the lattice spacing determined by the gradient flow~\cite{Luscher:2010iy} with $w_0$~\cite{BMW:2012hcm} at three bare couplings $\hat{\beta}$ and $m_{\pi}\sim 300$ MeV, using different gauge action improvement coefficient $c_1$ in the flow: $c_1=-0.331$ (Iwasaki action), $-0.2$ and $-1/(12u_0^2)$ for interpolation, $-1/12$ (Symanzik action), and 0 (Wilson action). The $c_1$ dependence becomes weaker at smaller lattice spacings,  indicating that it is a discretization effect. As the tadpole improvement factor approaches 1 with increasing gradient flow $t$, implementing tadpole improvement for the action used by the flow only affects the small flow time region and is unnecessary. Thus, we consistently use the standard $c_1=-1/12$ in the gradient flow to match the gauge action employed in HMC and mitigate the discretization error.

\begin{table}[h]
\centering
\caption{$a_{w_0}$ on different ensembles and the fitted values through the functional form in Eq.~(\ref{eq:a_global_fit}). }
\begin{tabular}{l|l|l||c|c}
 & $a_{w_0}$ & $a^{\rm fit}_{w_0}$ & \multicolumn{2}{c}{fit parameters}  \\
\hline
C24P34 & $0.11198 (17)$ & $0.11202 (16)$ & $a(6.20)$ & $0.10530 (18)$  \\
C24P29 & $0.10811 (15)$ & $0.10834 (09)$ & $a(6.41)$ & $0.07746 (18)$\\
C32P29 & $0.10858 (11)$ & $0.10836 (06)$ & $a(6.72)$ & $0.05187 (26)$\\
C32P23 & $0.10637 (13)$ & $0.10637 (06)$ & $c_{l}$ & $0.507 (18)$ \\
C48P23 & $0.10631 (06)$ & $0.10635 (05)$ & $c_{s}$ & $0.110 (23)$ \\
C48P14 & $0.10583 (07)$ & $0.10582 (06)$ & $c_{L}$ & $0.001 (09)$ \\
F32P30 & $0.08044 (10)$ & $0.08044 (08)$ & $c_{u_0}$ & $-408 (125)$ \\
F48P30 & $0.08017 (05)$ & $0.08017 (05)$ & $c_{v_0}$ &\ \ $379 (110)$ \\
F32P21 & $0.07758 (16)$ & $0.07757 (15)$ \\
F48P21 & $0.07810 (07)$ & $0.07809 (06)$ \\
H48P32 & $0.05430 (11)$ & $0.05430 (11)$ \\
\end{tabular}
\label{tab:lat_a_fit}
\end{table}

Using the FLAG average value of the gradient flow scale, $w_0=0.1736(9)$ fm~\cite{FlavourLatticeAveragingGroupFLAG:2021npn}, the lattice spacing $a_{w_0}$ for each ensemble is determined using the Symanzik flow and summarized in Table~\ref{tab:lat_a_fit}. It is evident that $a_{w_0}$ primarily depends on the gauge coupling $\hat{\beta}$, while the quark mass also exhibits significant effects, as indicated by the precise $a_{w_0}$ results with a statistical uncertainty of approximately 0.1\%. Empirically, $a_{w_0}$ can be described by the following parameterization, yielding a $\chi^2$/d.o.f. of 2.2 (with a corresponding p-value of 0.09, still larger than the standard lower bound of 0.05):
\begin{align}
&a_{w_0}(\hat{\beta},\tilde{m}_{\pi}, \tilde{m}_{\eta_s},\delta u_0, \delta v_0)=a(\hat{\beta})\big[1+c_l(\frac{\tilde{m}_{\pi}^2}{a(\hat{\beta})^2}-m_{\pi,{\rm phys}}^2)\nonumber\\
&\quad\quad\quad + c_s(\frac{\tilde{m}_{\eta_s}^2}{a(\hat{\beta})^2}-m_{\eta_s,{\rm phys}}^2)+c_Le^{-\tilde{m}_{\pi}\tilde{L}}\nonumber\\
&\quad\quad\quad + c_{u_0}(u_0-u_0^I)+c_{v_0}(v_0-v_0^I)\big]\;,\label{eq:a_global_fit}
\end{align}
Here, $m_{\pi,{\rm phys}}=134.98$ MeV~\cite{ParticleDataGroup:2020ssz} represents the physical pion mass without QED correction, and $m_{\eta_s,{\rm phys}}=689.63(18)$ MeV~\cite{Borsanyi:2020mff} corresponds to the pseudoscalar meson mass of the strange quark with only connected insertions. The tadpole improvement factors used in the actions are denoted as $u^I_0$ and $v^I_0$, while $u_0$ and $v_0$ are tadpole improvement factors obtained from the generated configurations. The lattice spacing $a(\beta)$ with physical quark masses at each $\beta$, along with the other fitting parameters $c_{l,s,L,u_0,v_0}$ are also provided in Table~\ref{tab:lat_a_fit}.

\begin{table*}[t]
	\centering
        \label{tab:tadpole}
	\caption{ The tadpole improvement factors $u^{\rm I}_0$ and $v^{\rm I}_{0}$ used in the gauge and fermion actions, $u_0$ and $v_{0}$ measured from the realistic configurations generated using $u^{\rm I}_0$ and $v^{\rm I}_{0}$, and their averages $\bar{u}_0$ and $\bar{v}_{0}$.}
	 \resizebox{2.1\columnwidth}{!}{
		\begin{tabular}{l | llllll | llll | l}
		    & C24P34 & C24P29 &  C32P29 & C32P23 & C48P23 & C48P14 & F32P30  & F48P30  & F32P21  & F48P21 & H48P32 \\  
\hline      
$u^{\rm I}_0$ & 0.855453 &  0.855453 & 0.855453 & 0.855520 & 0.855520 & 0.855548 
& 0.863437 & 0.863473 & 0.863488 & 0.863499    & 0.873378  \\
  		$u_0$&0.855255(7)&0.855439(2)&0.855429(2)&0.855528(2)&0.855523(1)&0.855530(2)
&0.863460(1)&0.863459(1)&0.863519(2)&0.863515(1)&0.873372(1)\\
		$\bar{u}_0$
&0.855354(4)&0.855446(1)&0.855441(1)&0.855524(1)&0.855522(1)&0.855539(1)
&0.863449(1) &0.863466(1)&0.863504(1) &0.863507(1)&0.873375(1)
\\
		\hline
		$v^{\rm I}_{0}$ & 0.951479 &0.951479 &0.951479 &0.951545&0.951545 &0.951570 
  & 0.956942 & 0.956984 & 0.957017 & 0.957006  & 0.963137   \\
		$v_0$&0.951275(6)&0.951461(2)&0.951452(2)&0.951550(2)&0.951547(1)&0.951554(2)
&0.956968(1)&0.956967(1)&0.957024(1)&0.957019(1)&0.963134(1)\\
		$\bar{v}_0$&0.951377(3)&0.951470(1)& 0.951466(1)&0.951548(1)&0.951546(1)&0.951562(1)
&0.956955(1)&0.956976(1)&0.957021(1)&0.957013(1)&0.963136(1)
\\
		\end{tabular}
	}
\end{table*}

We observe that the volume dependence is consistent with zero, but there is still a non-vanishing mismatch effect ($u_0^I\neq u_0$ and $v_0^I\neq v_0$) in the tadpole improvement factors. It is also interesting to note that the dependence of $a_{w_0}$ on the strange quark mass is weaker compared to that of the light quark mass by a factor of approximately 4, instead of 2 from two light flavors.

\subsubsection{Mismatch effect of the tadpole improvement factors}\label{sec:tadpole}

As shown in Table~\ref{tab:lat_a_fit}, the mismatch effect of the tadpole improvement factors can have a non-zero impact on the determination of $a_{w_0}$ through the gradient flow. Both $u_0$ and $v_0$ represent vacuum expectation values and cannot be accurately determined before generating gauge configurations using the realistic Hybrid Monte Carlo (HMC) production. Therefore, for each ensemble, we initiate the HMC with an initial guess for $u_0$ and $v_0$, measure their values on each trajectory until they stabilize, and subsequently restart the HMC production with the updated values of $u_0$ and $v_0$. After several iterations, the input values $u^{\rm I}_0$ and $v^{\rm I}_0$ become consistent with $u_0$ and $v_0$, as measured from the configurations, at a level of approximately 0.002\%, resulting in an effect of around 0.6\% on the lattice spacing. The only exception is the test ensemble C24P34, which exhibits a larger deviation in both $u_0$ and $v_0$; however, their impact on $a{w_0}$ is mainly canceled out due to the opposite signs of $c_{u_0}$ and $c_{v_0}$.

\begin{figure}[thb]
 \includegraphics[width=0.5\textwidth]{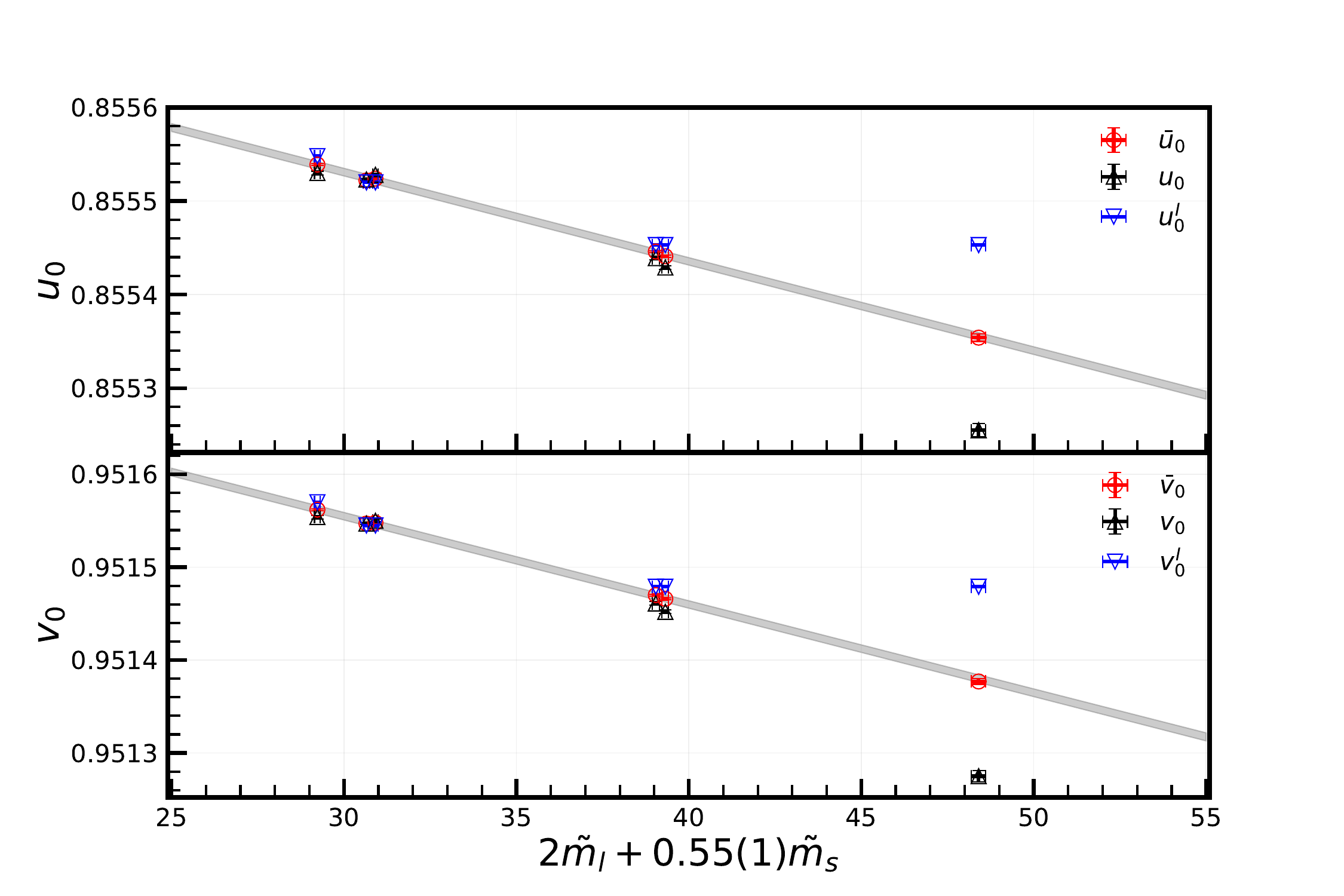}
\caption{The combined quark mass $2\tilde{m}_l^{\rm PC}+0.55(1)\tilde{m}^{\rm PC}_s$ dependence of $\bar{u}_0=\frac{u_0+u_0^I}{2}$ (upper panel) and $\bar{v}_0=\frac{v_0+v_0^I}{2}$ (lower panel), at $10/g^2$=6.2. 
}
\label{fig:TI_factors}
\end{figure}

\begin{table}[h]
\centering
\caption{The joint fit results with the same $\tilde{m}_l^b$ and either $c_{\rm sw}=1/(v^I_0)^3$ or $1/\bar{v}_0^3$, on the test ensemble C24P34 which has the largest mismatch effect. }
\begin{tabular}{c|c|c|c|c}
 & $\tilde{m}_{\rm PS}$ & $\tilde{m}_{l}^{\rm PC}$ & $\tilde{f}_{\rm PS}$ & $Z_{wp}$ \\
\hline
$1/(v^I_0)^3$=1.1609 & 0.1832(12) & 0.01191(13) & 0.0768(10) & 10.50(30) \\
$1/(\bar{v}_0)^3$=1.1613 & 0.1822(12) & 0.01178(13) & 0.0768(11) & 10.50(31) \\
\hline
difference & 0.0011(01) & 0.00013(01) & 0.0000(00) &\ \  0.00(01) \\
\end{tabular}
\label{tab:mismatch}
\end{table}

After averaging $u_0$ and $u^{\rm I}_0$ to obtain an estimate of the self-consistent tadpole improvement factor $\bar{u}_0$, we observe that the volume dependence is at the level of 0.001\% based on the spatial sizes utilized. On the other hand, the quark mass dependence appears to be more pronounced.

In the upper panel of Fig.\ref{fig:TI_factors}, the value of $\bar{u}_0$ exhibits a linear behavior with respect to the combined quark mass $2\tilde{m}_l^{\rm PC}+0.55(1)\tilde{m}^{\rm PC}_s\propto 0.5 \tilde{m}_{\pi}^2+0.14\tilde{m}_{\eta_s}^2$. This suggests that the dependence on the strange quark mass exhibits a similar suppression as observed in $a_{w_0}$, as shown in Table~\ref{tab:lat_a_fit}. In the lower panel, we observe that $v_0$ exhibits the same quark mass dependence within the uncertainty.

However, such a mismatch will not significantly alter the physical observables. To illustrate this, let's consider the C24P34 ensemble with the largest mismatch effect. We calculate the pion two-point function $C^{PP}_{2,wp}(t;\tilde{m}_l^b,c_{\rm sw})$ using the same $\tilde{m}_l^b=-0.2770$, but with either $c_{\rm sw}=1/(v^I_0)^3$ or $1/\bar{v}_0^3$. The joint fit parameters defined in Eq.~(\ref{eq:fit_quark_mass}-\ref{eq:fit_2pt}) are listed in Table\ref{tab:mismatch}. We find that the matrix elements $\tilde{f}_{\rm PS}$ and $Z{wp}$ remain unchanged within 0.05\% uncertainties. However, $\tilde{m}_{\rm PS}$ experiences a 0.6\% change, resulting in a 1.1\% shift in the quark mass. This can be understood by noting that increasing $c_{\rm sw}$ leads to an increase in $\tilde{m}_{\rm crti}$, thereby making the multiplicative renormalizable quark mass $\tilde{m}^{\rm PC}_l$ larger for the same $\tilde{m}_l^b$. Consequently, the mismatch effect falls within the statistical uncertainty of C24P34 and is an order of magnitude smaller in other ensembles. Hence, we can safely disregard this mismatch effect in the subsequent discussions.

\subsection{Renormalization}\label{sec:renorm}

Unlike the hadron spectrum, the determination of hadron matrix elements on the lattice using discretized actions requires additional renormalization. The RCs defined under the $\MSbar$ scheme, can only be obtained through regularization-independent (RI) schemes such as RI/MOM~\cite{Martinelli:1994ty} or SMOM~\cite{Aoki:2007xm,Sturm:2009kb}. These RCs should be independent of intermediate schemes.

For the overlap fermion action, which possesses explicit chiral symmetry, the relations $Z_V=Z_A$ and $Z_P=Z_S=1/Z_m$ are guaranteed. The consistency of using RI/MOM or SMOM schemes has been verified within systematic uncertainties~\cite{He:2022lse}. However, in the case of the clover fermion action, which exhibits additive chiral symmetry breaking, additional considerations and discussions regarding its impact on renormalization are necessary in this section.

\begin{table}[h]
\caption{Summary of uncertainties of RCs in percentage on
the C32P29 ensemble through the intermediate RI/MOM scheme.}
\label{tab:err_summary_rc}
\resizebox{1.0\columnwidth}{!}{
\begin{tabular}{c|cccc}
    \hline
    \hline
    Source &
    $Z^{\rm \bar{MS}}_{q}/Z_{V}$  &
    $Z^{\rm \bar{MS}}_{S}/Z_{V}$  &
    $Z^{\rm \bar{MS}}_{P}/Z_{V}$  &
    $Z^{\rm \bar{MS}}_{T}/Z_{V}$  \\
    \hline
    $\rm{Statistical\ error}$ & $0.13\%$ & $0.37\%$ & $0.92\%$ & $0.06\%$ \\
   
    $\rm{Lattice\ spacing}$ & $<0.01\%$ & $<0.01\%$ & $<0.01\%$ & $<0.01\%$ \\    
    
    $\rm{Finite\ volume\ effect}$ & $0.03\%$ & $0.24\%$ & $0.62\%$ & $0.02\%$ \\    
    
    $\rm{Fit\ range\ of}\ a^2\mu^2 $ & $0.11\%$ & $0.47\%$ & $1.21\%$ & $0.31\%$ \\    
    \hline
    $\Lambda_{QCD}^{\overline{\rm{MS}}}$ & $<0.01\%$ & $1.64\%$ & $1.65\%$ & $<0.01\%$ \\   
    
    $\rm{Truncation\ in\ matching} $ & $0.24\%$ & $3.45\%$ & $3.45\%$ & $0.24\%$ \\ 
        
    $\rm{Perturbative\ running} $ & $0.02\%$ & $0.07\%$ & $0.07\%$ & $0.02\%$ \\    
    \hline \hline
    $\rm{Total} $ & $0.32\%$ & $3.74\%$ & $4.26\%$ & $0.31\%$ \\

\end{tabular}
}
\end{table}

Following a similar strategy employed by Ref.~\cite{He:2022lse}, the values of $Z_{S,P,T}$ incorporate two sources of uncertainty. The first one encompasses ensemble-independent statistical and systematic uncertainties, including lattice spacing, finite volume effects, and $a^2\mu^2$ fit range. The second source of uncertainty arises from perturbative matching, including uncertainties associated with $\Lambda_{QCD}$, truncation in the perturbative matching, and perturbative scale running. These uncertainties are fully correlated across different ensembles. As shown in Table~\ref{tab:err_summary_rc} for the C32P29 ensemble, the truncation error is the largest source of  uncertainty for $Z_{S,P}$.

In Ref.~\cite{He:2022lse}, the truncation error resulting from the 3-loop perturbative matching between the RI/MOM and $\MSbar$ schemes is estimated by introducing a fake 4-loop correction. As an illustration, considering the scalar/pseudoscalar case with the largest truncation error, the 3-loop matching and the fake 4-loop one are expressed as follows:
\bal
C_{S, n_f=3, {\rm 3-loop}}^{\MSbar,\MOM}&=1+0.4244\alpha_s+1.007\alpha_s^2 +2.722\alpha_s^3,\\
C_{S, n_f=3, {\rm fake 4-loop}}^{\MSbar,\MOM}&=1+0.4244\alpha_s+1.007\alpha_s^2 +2.722\alpha_s^3\nonumber\\
&\quad \quad +7.358\alpha_s^4.
\eal
The coefficient of fake $\alpha_s^4$ term $2.722^2/1.007=7.358$ is the same as that provided by the Pad$\acute{\rm e}$ approximation,
\bal
C_{S, n_f=3, {\rm Pade}_3}^{\MSbar,\MOM}&=\frac{1-2.279\alpha_s-0.1402\alpha^2_s}{1-2.703\alpha_s}\nonumber\\
&=1+0.4244\alpha_s+1.007\alpha_s^2 +2.722\alpha_s^3\nonumber\\
&\quad \quad +7.358\alpha_s^4+\mathcal{O}(\alpha_s^5).
\eal

Based on the precise 4-loop calculation, the precise 4-loop matching is~\cite{Gracey:2022vjc},
\bal
C_{S, n_f=3, {\rm 4-loop}}^{\MSbar,\MOM}&=1+0.4244\alpha_s+1.007\alpha_s^2 +2.722\alpha_s^3\nonumber\\
&\quad \quad +8.263\alpha_s^4,
\eal
which is larger than $C_{S, n_f=3, {\rm fake 4-loop}}^{\MSbar,\MOM}$ but smaller than $C_{S, n_f=3, {\rm Pade}_3}^{\MSbar,\MOM}$. Considering that the existence of the 5-loop correction, we anticipate that the Pad$\acute{\rm e}$ approximation will provide a more accurate estimate for $C_{S, n_f=3}$.

Thus we shall use the 4-loop Pad$\acute{\rm e}$ approximation,
\bal
C_{S, n_f=3, {\rm Pade}_4}^{\MSbar,\MOM}&=\frac{1-2.611\alpha_s-0.2813\alpha^2_s-0.3349\alpha^3_s}{1-3.036\alpha_s}\nonumber\\
&=1+0.4244\alpha_s+1.007\alpha_s^2 +2.722\alpha_s^3\nonumber\\
&\quad \quad +8.263\alpha_s^4+25.084\alpha_s^5 +\mathcal{O}(\alpha_s^6),
\eal
as the central value of the matching coefficent, and taking the difference between $C_{S, n_f=3, {\rm Pade}_4}$ and $C_{S, n_f=3, {\rm Pade}_3}$ as systematic uncertainty of the truncation on the perturbative matching.

In this section, we will commence with the vector current normalization and explore different choices of quark field renormalization in Section~\ref{sec:zv_zq}. Section~\ref{sec:zv_zq} will then present the investigation of the chiral symmetry breaking effect between $Z_V$ and $Z_A$. Subsequently, in Section~\ref{sec:zp_zs}, we will examine the analogous investigation of $Z_P$ and $Z_S$, with additional discussions on $Z_m$ in Section~\ref{sec:zp_zm}. Finally, for completeness, the case of the tensor current will be discussed in Section~\ref{sec:zt}.

\subsubsection{Vector normalization and $Z_q$}\label{sec:zv_zq} 

\begin{figure}[thb]
 \includegraphics[width=0.5\textwidth]{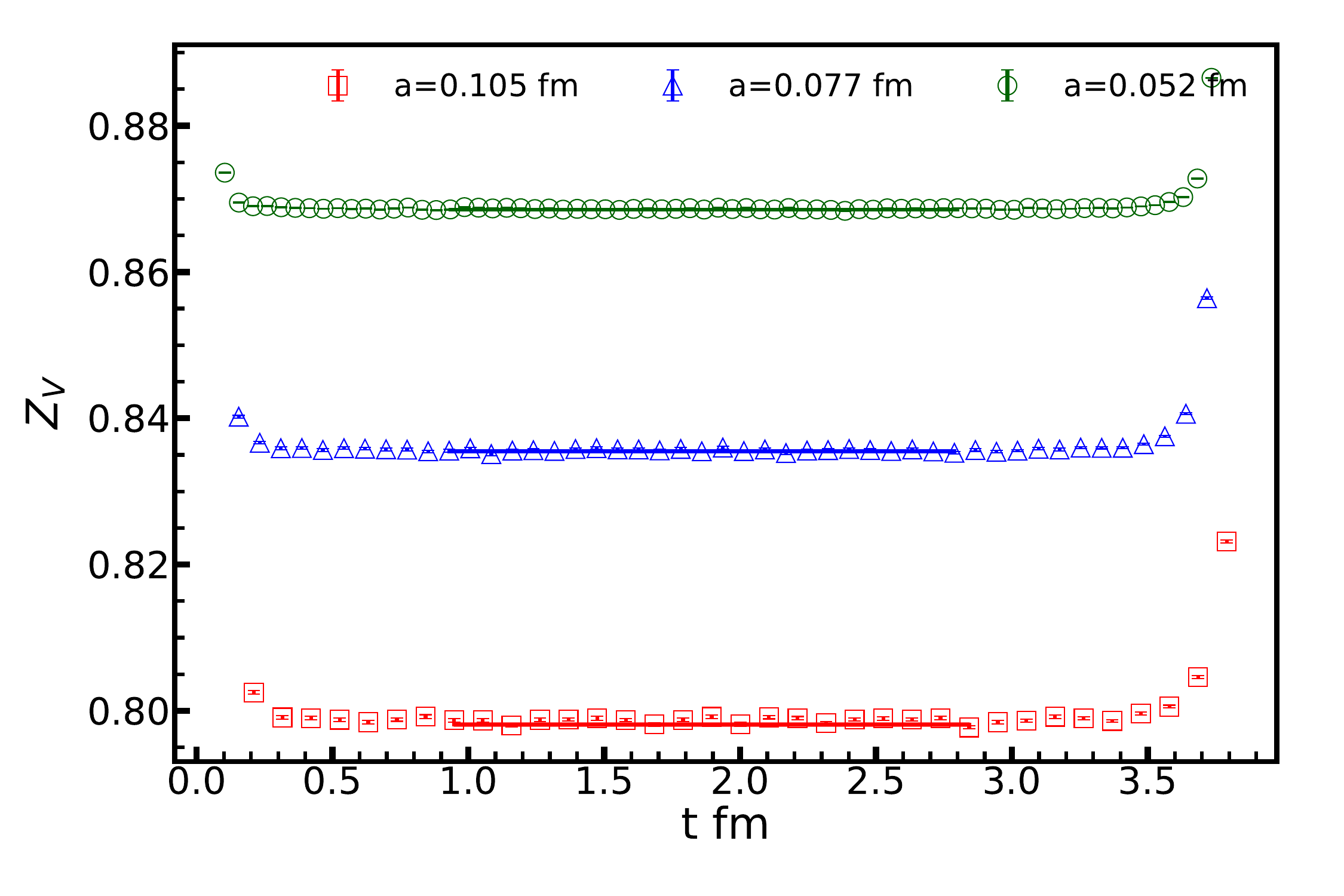}
\caption{$Z_V$ from the vector current conservation of pion correlator at three lattice spacing. 
{
    The plateau values are summarized in Table \ref{tab:renorm} as the $Z_V$ of the ensembles C24P29, F32P30 and H48P32.}
}
\label{fig:ZV}
\end{figure}

Unlike its continuum counterpart, the local vector current under lattice regularization is subject to both the ${\cal O}(a^2)$ discretization error and ${\cal O}(\alpha_s)$ loop corrections~\cite{Zhang:2020rsx}, and then requires additional normalization. The normalization constant $Z_V$ can be determined from the vector current conservation condition,
\bal
Z_V\frac{\langle H|V_4|H\rangle}{\langle H|H\rangle}=1,
\eal
where $V_{\nu}=\bar{\psi}\gamma_{\nu}\psi$ and $H$ represents an arbitrary hadronic state. Thus extracting $Z_V$ from the pion correlator in the rest frame will be the cheapest choice
~\cite{Zhang:2020rsx},
\bal
Z_V=&\sum_{\vec{x}}\langle \mathrm{Tr}[S^{\dagger}_C(\vec{x},\tilde{T}/2;0;m_q)
S_C(\vec{x},\tilde{T}/2;0;m_q) ]\rangle/\nonumber\\ 
&\big\{2\sum_{\vec{x},\vec{z}}\langle \mathrm{Tr}[S^{\dagger}_C(\vec{x},\tilde{T}/2;0;m_q)S(\vec{x},\tilde{T}/2;\vec{z},t;m_q)\nonumber\\
&\quad \gamma_tS_C(\vec{z},t;y_0;m_q) ]\rangle\big\}|_{\tilde{T}\gg t\gg 0},
\eal
where the additional propagator in the denominator can be obtained using the sequential source technique. The ratio at three lattice spacing and their constant fits in the range of $0\ll t\ll \tilde{T}/2$, are shown in Fig.~\ref{fig:ZV}. As demonstrated in\cite{Zhang:2020rsx}, $Z_V$ is affected by $\alpha_s$ corrections and cannot be accurately extrapolated to 1 using a simple naive $a^{2n}$ continuum extrapolation.

\begin{figure}[thb]
 \includegraphics[width=0.5\textwidth]{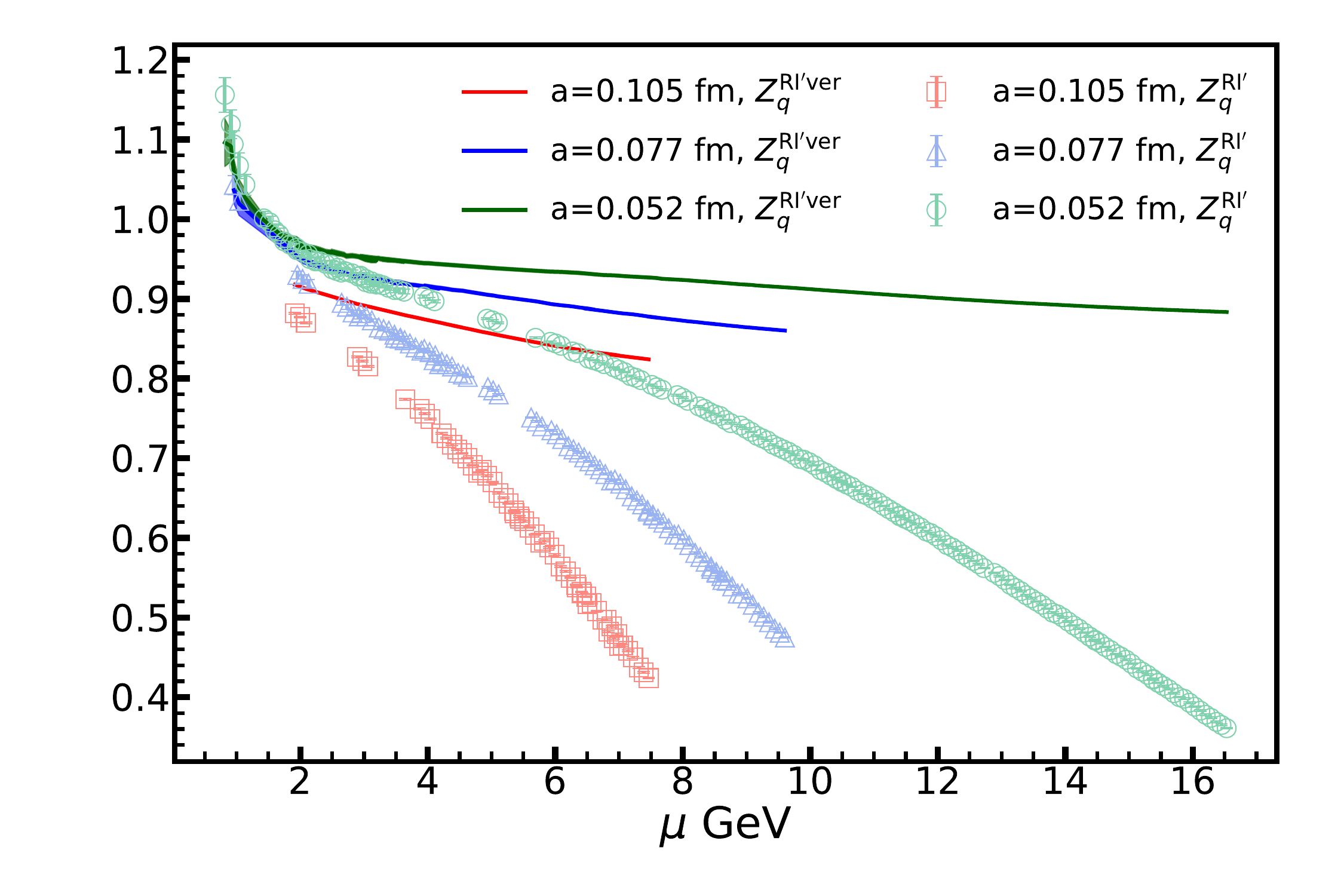}
\caption{Comparison of $Z_q^{\rm RI'}$ (data points) and $Z_q^{\rm RI', ver}$  (bands)  at three lattice spacing with different scale $\mu$. The discrepancy observed in the figure is primarily attributed to the discretization error.}
\label{fig:Zq}
\end{figure}

The quark field RC $Z_q^{\rm RI'}$ in the regularization independent (RI) scheme can be accessed through either its definition
\bal
Z_q^{\rm RI'}(\mu)&=\mathop{\textrm{lim}}\limits_{m\rightarrow0}\frac{-i}{12p^2}\textrm{Tr}\Big[S^{-1}(p)\slashed{p}\Big]_{p^2=\mu^2},
\eal
with $S(p)=\sum_xe^{-ip\cdot x}\langle \psi(x)\bar{\psi}(0)\rangle$, or the vertex correction of the vector current which is equivalent in the continuum,
\bal
Z_q^{\rm RI', ver}(\mu)&=\mathop{\text{lim}}\limits_{m\rightarrow0} \frac{Z_V}{36}\text{Tr}[\Lambda^\mu_{V}(p,p)(\gamma_\mu-\frac{\slashed{p}p_\mu}{p^2})]_{p^2=\mu^2},
\eal
where 
\bal
\Lambda_\mathcal{O}(p_1,p_2)&=S^{-1}(p_1)G_\mathcal{O}(p_1,p_2)S^{-1}(p_2),\nonumber\\
G_\mathcal{O}(p_1,p_2)&=\sum_{x,y}e^{-i(p_1\cdot x-p_2\cdot y)}\langle
\psi(x)\mathcal{O}(0)\bar{\psi}(y)\rangle.
\eal
These two definitions are equivalent in the continuum, but they are subject to different discretization errors. Fig.\ref{fig:Zq} illustrates the comparison between $Z_q^{\rm RI'}$ and $Z_q^{\rm RI', ver}$ at three lattice spacings. The two definitions become closer either at smaller $\mu$ or smaller $a$, with $Z_q^{\rm RI', ver}$ exhibiting a significantly smaller $a^2p^2$ error. This situation is reminiscent of the overlap fermions case~\cite{He:2022lse}, where the $a^2\mu^2$ errors have opposite signs.

Given that $Z_q^{\rm RI', ver}$ involves an off-diagonal projection that introduces additional discretization errors, it is more convenient to define the quark field renormalization through the standard RI definition~\cite{Martinelli:1994ty},
\bal
Z_q^{\omega}(\mu^2)&=\mathop{\text{lim}}\limits_{m\rightarrow0} \frac{Z_V}{48}\text{Tr}[\Lambda^\mu_{V}(p_1,p_2)\gamma_\mu]_{p_1^2=p_2^2=\mu^2,(p_1-p_2)^2=\omega \mu^2},
\eal
The discrepancy between $Z_q^{\omega}$ with different $\omega$ and $Z_q^{\rm RI'}$ should be eliminated through their respective perturbative matchings, which can be calculated in the continuum. For the SMOM condition with $\omega=1$, the convergence of the perturbative matching is poorer compared to that of $Z_q$ using the q-projection definition~\cite{Sturm:2009kb}, but the discretization error will be smaller.

\subsubsection{Chiral symmetry breaking between $Z_A$ and $Z_V$}\label{sec:zv_za}

\begin{figure}[thb]
 \includegraphics[width=0.45\textwidth]{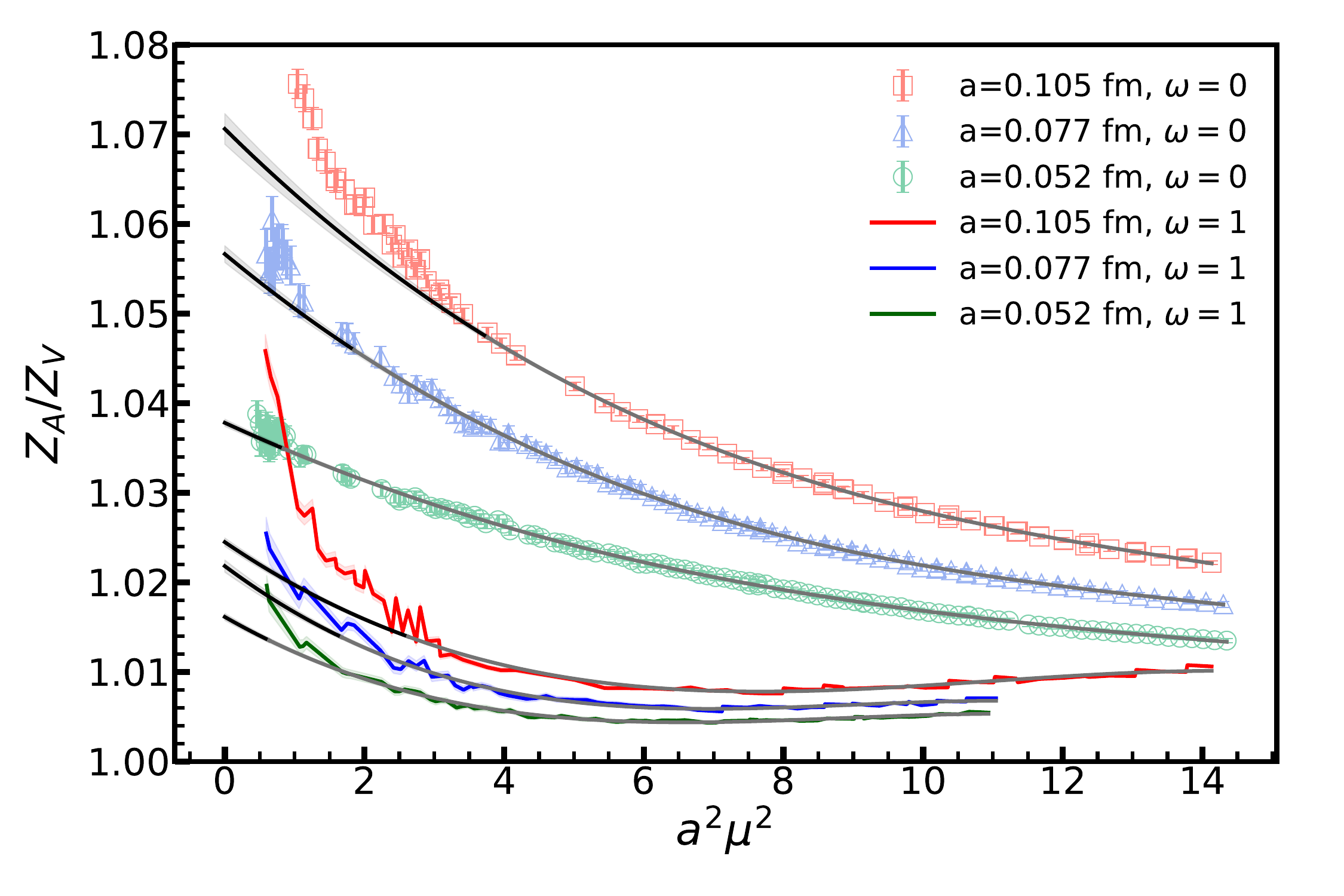}
\caption{The ratio $Z_A/Z_V$ through the RI/MOM scheme (data points) or SMOM scheme (bands) at three lattice spacing as functions of $a^2p^2$.}
\label{fig:ZA_ZV}
\end{figure}

For the clover fermion, the ratio
\bal
\frac{Z_A}{Z_V}(\mu^2,\omega)
&=\frac{\text{Tr}[\Lambda^{\mu, \MSbar}_{A}(p_1,p_2)\gamma_5\gamma_{\mu}]}{\text{Tr}[\Lambda^{\mu, \MSbar}_{V}(p_1,p_2)\gamma_{\mu}]}|_{p_1^2=p_2^2=\mu^2,(p_1-p_2)^2=\omega \mu^2}
\eal
can deviate from unity due to the additive chiral symmetry breaking present in the action. As depicted in Fig.~\ref{fig:ZA_ZV} for the RI/MOM scheme (data points, $\omega=0$) and SMOM scheme (bands, $\omega=1$) at three different lattice spacings with $m_{\pi}\sim$ 300 MeV, we observe that the breaking diminishes as $a^2\mu^2$ increases, indicating that the chiral symmetry breaking in the mass term becomes less important. However, the ratio does not approach unity in the continuum limit without any $\alpha_s$ corrections. It is worth noting that the breaking using the SMOM scheme is much smaller than that with the RI/MOM scheme, while increases rapidly at small $a^2\mu^2$.

\begin{figure}[thb]
 \includegraphics[width=0.45\textwidth]{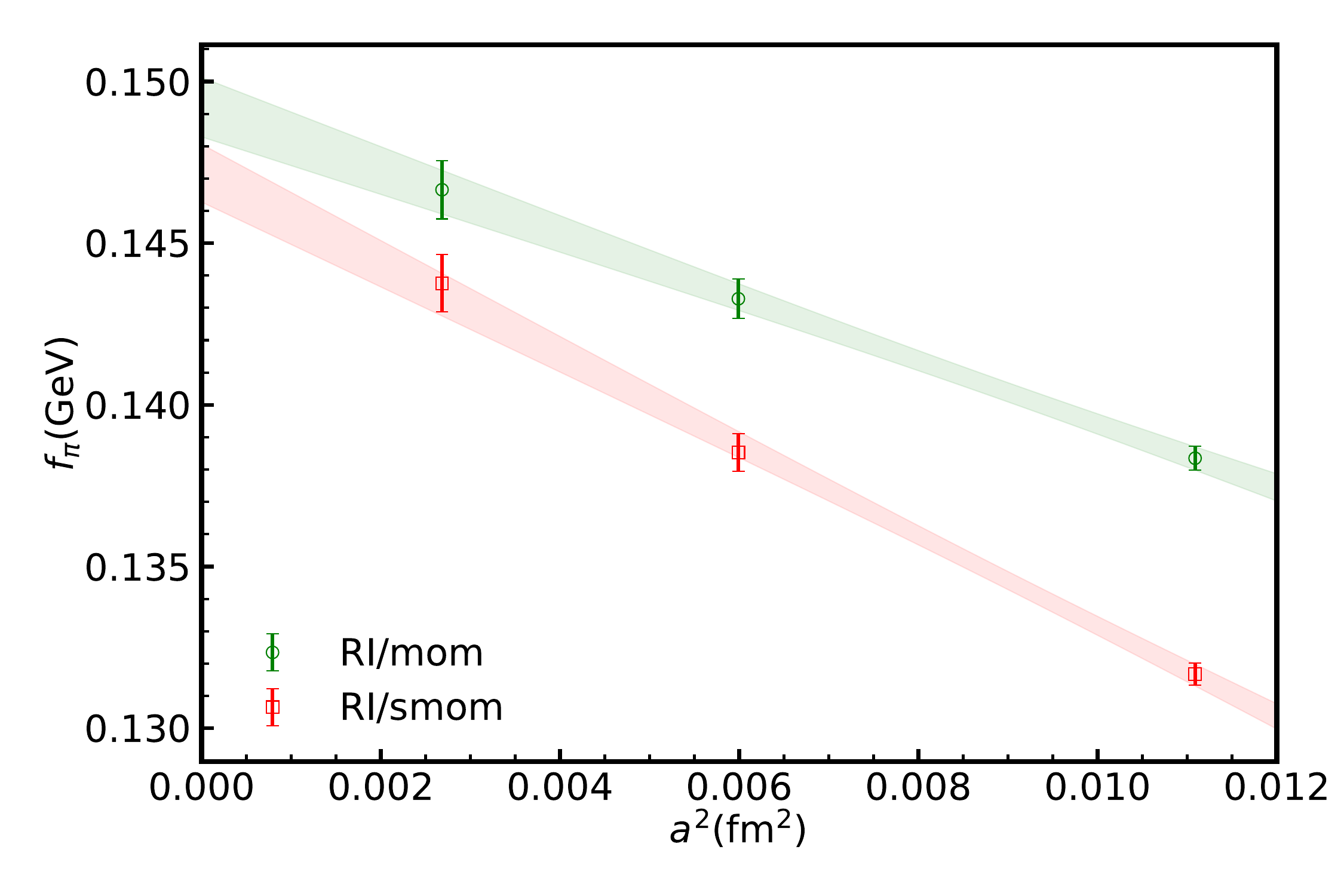}
\caption{Renormalized pion decay constant $f_{\pi}$ with $m_{\pi}=317$ MeV at three lattice spacing, using $Z_A=Z_V\frac{Z_A}{Z_V}$ through either RI/MOM or SMOM scheme. The extrapolated values deviate by 1.3(8)\%. }
\label{fig:fpi_lat}
\end{figure}

If we fit the $Z_A/Z_V$ data with a polynomial form in the range of $9\mathrm{GeV}^2\le p^2\le 15/a^2$ and extrapolate it to $a^2p^2=0$, the obtained result will be sensitive to the choice of $\omega$. Therefore, we  extrapolate the $f_{\pi}$ at two coarser lattice spacings to the unitary pion mass at the finest lattice spacing, which is $m_{\pi}=317$ MeV. The renormalized $f^R_{\pi}=Z_A f_{\pi}$ is shown in Fig.~\ref{fig:fpi_lat}, using both the extrapolated $Z_A/Z_V$ obtained through RI/MOM and SMOM. The results suggest that the scheme sensitivity is approximately 1\% after a linear $a^2$ continuum extrapolation, with the discretization error through RI/MOM being 25\% smaller.

\begin{figure}[thb]
 \includegraphics[width=0.45\textwidth]{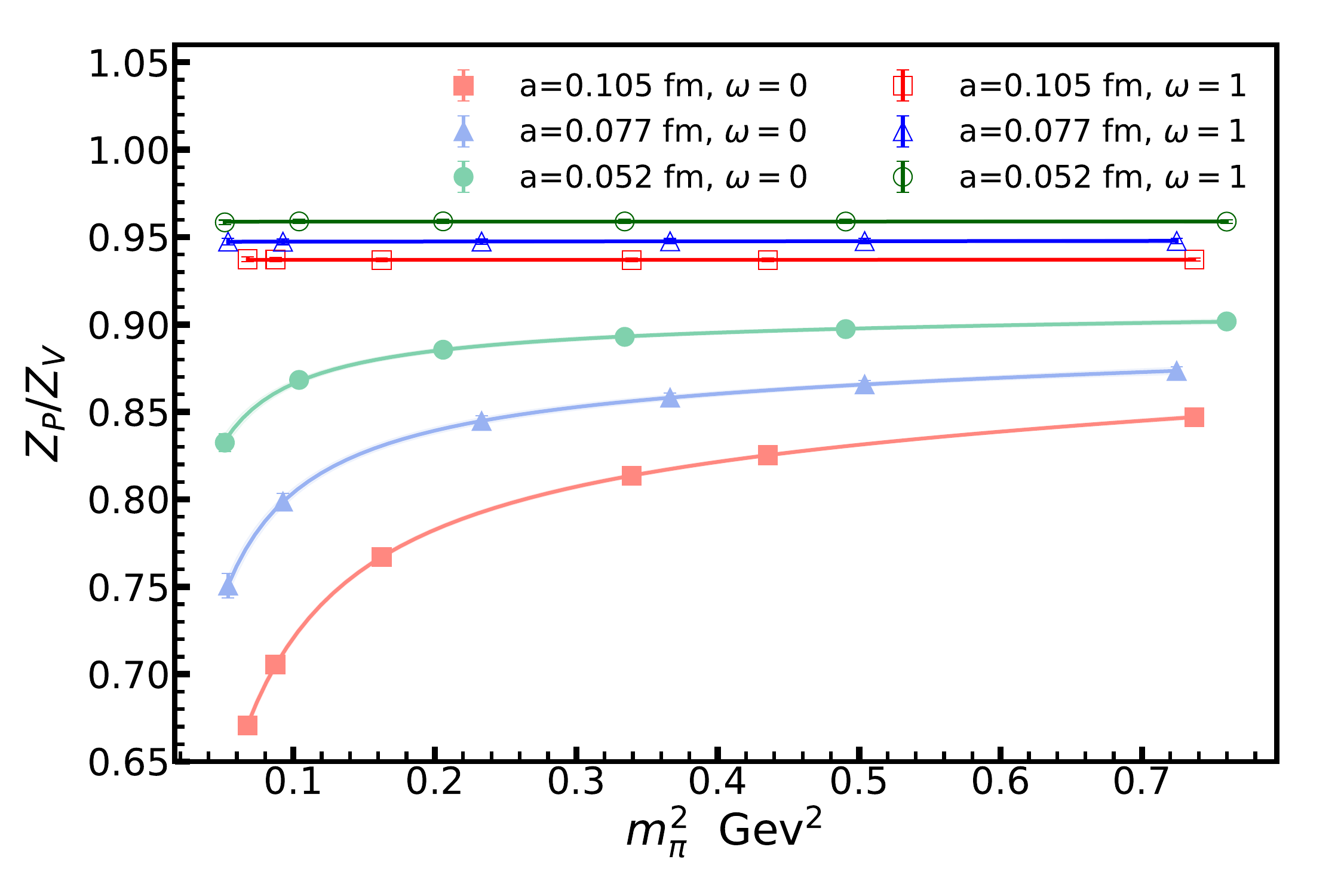}
\caption{Valence pion mass square $m_{\pi,vv}^2$ dependence of $\hat{Z}^{\omega=0}_P$ (RI/MOM, lighter points) and $\hat{Z}^{\omega=1}_P$ (SMOM, darker points) with $a^2\mu^2=4$, at three lattice spacing and $m_{\pi,ss}\sim 300$ MeV}
\label{fig:ZP_dep_mass}
\end{figure}

\subsubsection{Pesudoscalar current renormalization $Z_P$ and induced quark mass}\label{sec:zp_zm}

The renormalization of $m_q^{\rm PC}$ requires both $Z_A$ and $Z_P$. By defining $\hat{Z}^{\omega}_P$ under either the RI/MOM ($\omega=0$) or SMOM ($\omega=1$) scheme,
\bal
\hat{Z}^{\omega}_P(\mu)&=\frac{Z^{\omega}_q}{\frac{1}{12}\text{Tr}[\Lambda_{P}(p_1,p_2)\gamma_5]}|_{p_1^2=p_2^2=\mu^2, (p_1-p_2)^2=\omega\mu^2}\nonumber\\
&=Z_V\frac{\frac{1}{48}\text{Tr}[\Lambda^\mu_{V}(p_1,p_2)\gamma_\mu]}{\frac{1}{12}\text{Tr}[\Lambda_{P}(p_1,p_2)\gamma_5]}|_{p_1^2=p_2^2=\mu^2, (p_1-p_2)^2=\omega\mu^2},
\eal
$Z_P$ in the chiral limit can be extracted from the following parameterization,
\bal
\hat{Z}^{\omega}_P(\mu;m_q^{\rm PC})=\left[(\frac{A_P^{\omega}(\mu)}{m_q^{\rm PC}}+(Z^{\omega}_P(\mu))^{-1}+C_P^{\omega}(\mu) m_q^{\rm PC}\right]^{-1},
\eal
where $m_q^{\rm PC}$ can be replaced by $m_{\pi}^2$ based on the GMOR relation. As shown in Fig.\ref{fig:ZP_dep_mass}, $A_P^{\omega=1}$ is negligible in the fit, but $A_P^{\omega=0}$ will be non-zero due to the mass pole of the Goldstone meson and is related to the dynamical quark mass under the Landau gauge\cite{Chang:2021vvx}.

\begin{figure}[thb]
 \includegraphics[width=0.45\textwidth]{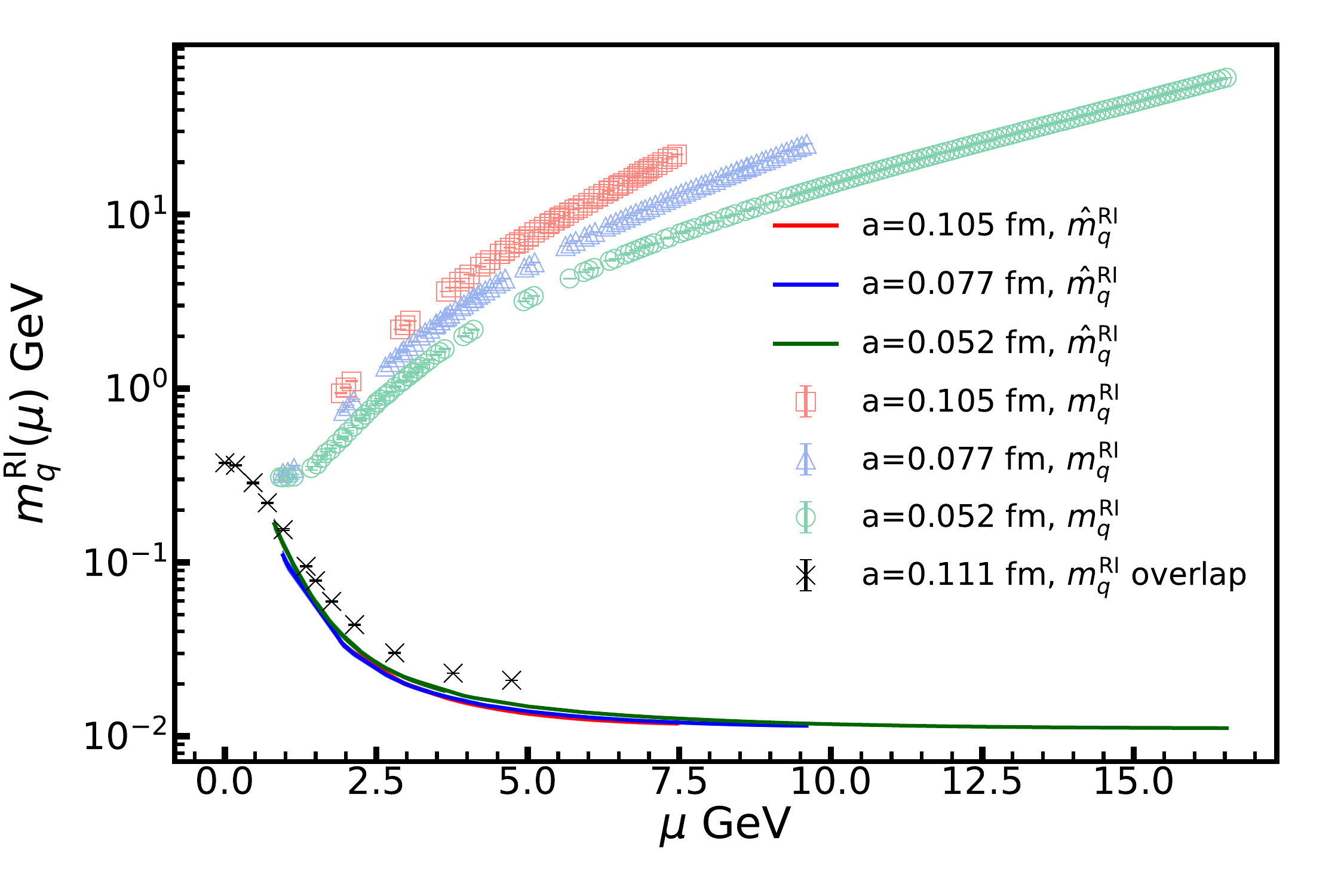}
\caption{Values of $m^{\rm RI}_q$ (data points) and $\hat{m}^{\rm RI}_q$ (bands) at three lattice spacing with $m_{\pi}\sim 300$ MeV at different RI/MOM scale $\mu$. $\hat{m}^{\rm RI}_q$ (black crosses) using the overlap fermion are also shown on the figure for comparison.}
\label{fig:Zm_ZP}
\end{figure}

If one define the RI quark mass at given scale as~\cite{Chang:2021vvx},
\bal
m^{\rm RI}_q=\frac{\frac{1}{12}\text{Tr}[S^{-1}(p)]|_{p^2=\mu^2}}{Z^{\rm RI}_q(\mu)},
\eal
then for the overlap fermion action with exact chiral symmetry, we can further define $Z^{\rm RI}_m = m^{\rm RI}_q/m_q^b$ and have the relation $Z_m\hat{Z}_p=1$ holding for arbitrary $m_q$ and $\mu$. Thus, the following definition of the RI quark mass:
\bal
\hat{m}^{\rm RI}_q=\frac{Z_Am_q^{\rm PC}}{\hat{Z}^{\rm MOM}_P(\mu)}, 
\eal
ensures that $m^{\rm RI}_q=\hat{m}^{\rm RI}_q$ given the relation $Z_A m_q^{\rm PC} = m_q^b$ for the overlap fermion. However, the case of clover fermions can be quite different due to its additive chiral symmetry breaking.

In Fig.\ref{fig:Zm_ZP}, we present both $m^{\rm RI}_q$ and $\hat{m}^{\rm RI}q$ at three lattice spacings with $m_{\pi}\sim 300$ MeV, and compare them with the results obtained from the overlap fermion at $a=0.111$ fm. We observe that $\hat{m}^{\rm RI}_q$ seems to be insensitive to the lattice spacing and  is close to the overlap result $\hat{m}^{\rm RI}_q=m^{\rm RI}_q$\cite{Chang:2021vvx}; but $m^{\rm RI}_q$ exhibits a significant $a^2\mu^2$ error and some unknown UV effects, resulting in large values at large $\mu$.

\begin{figure}[thb]
 \includegraphics[width=0.45\textwidth]{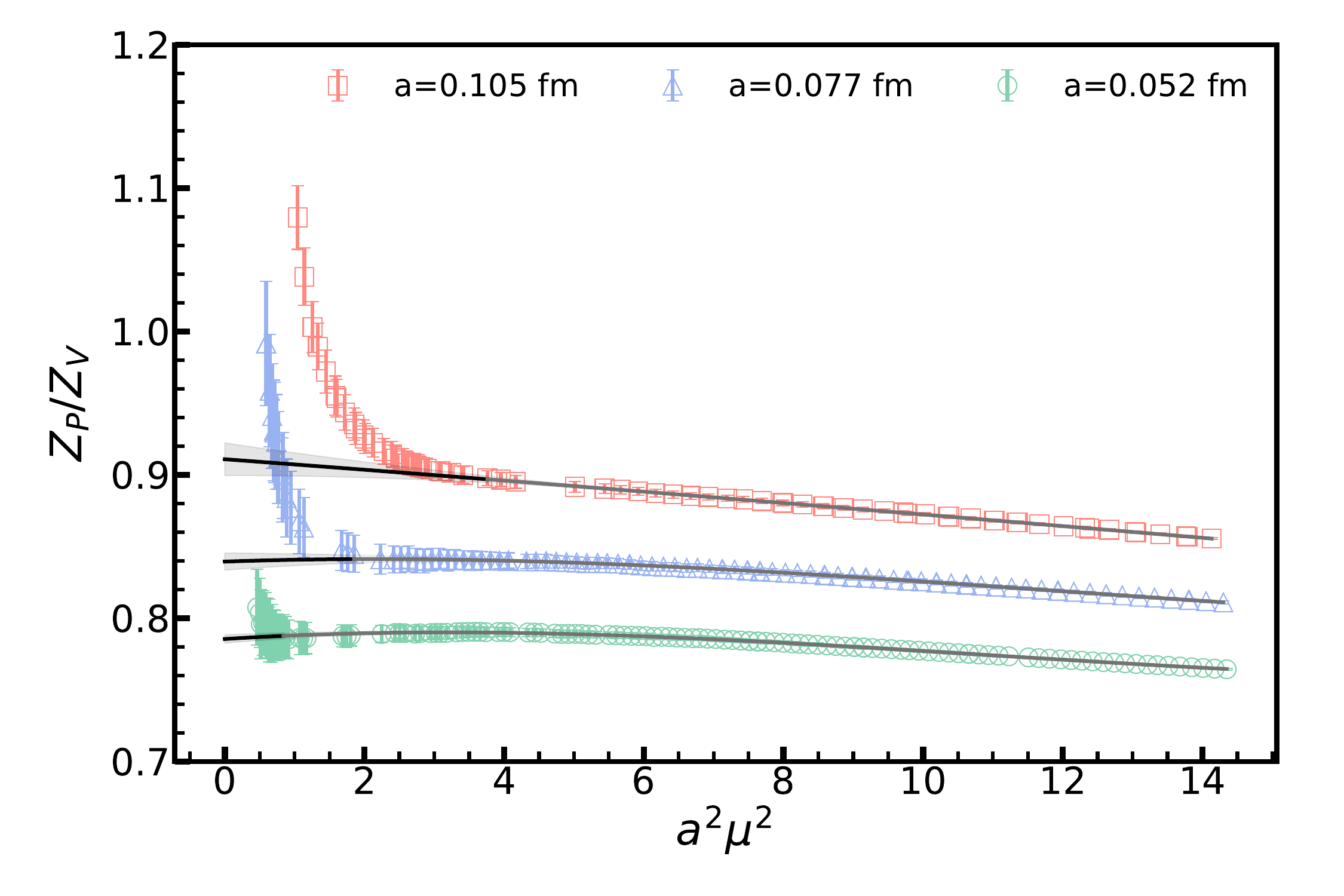}
 \includegraphics[width=0.45\textwidth]{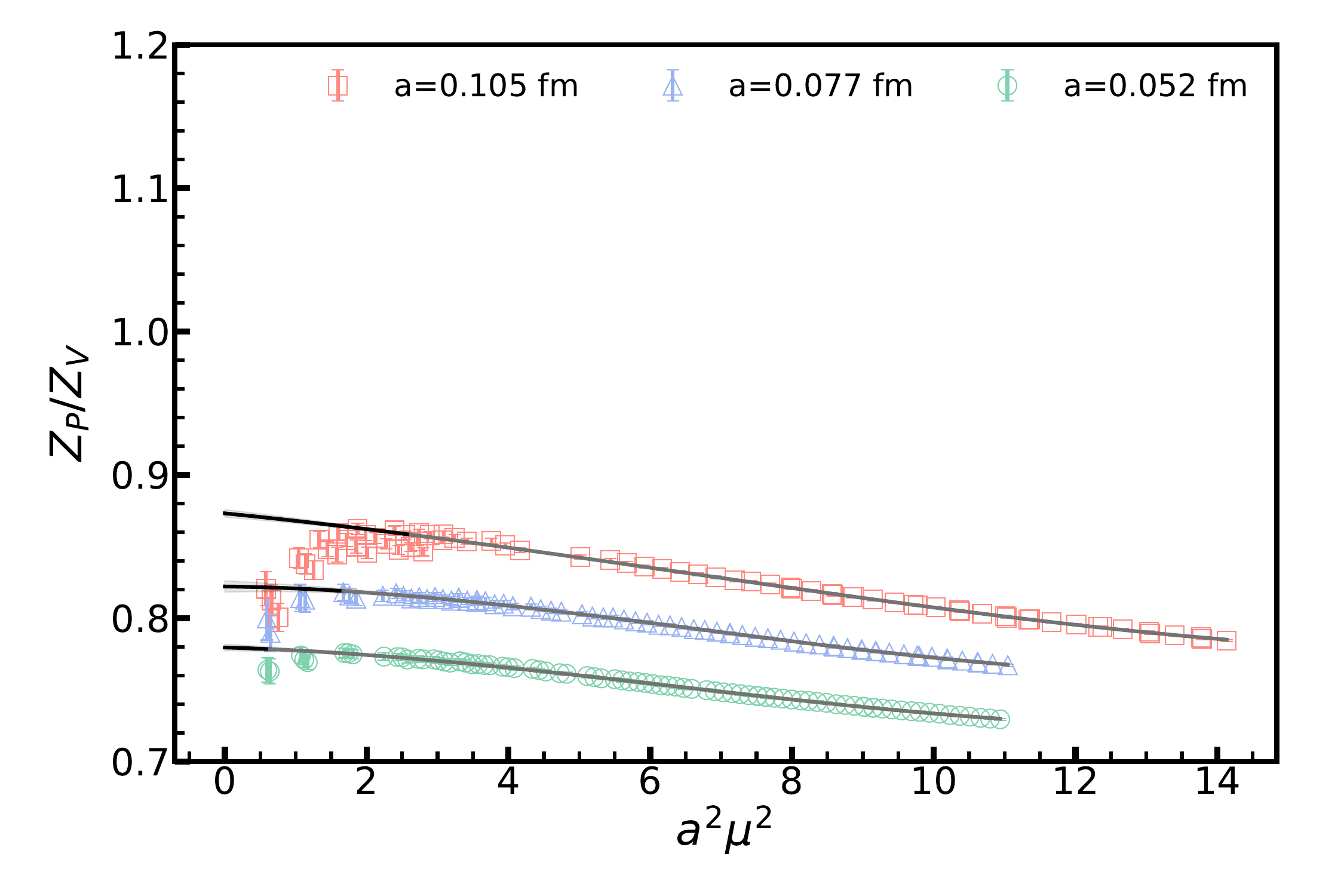}
\caption{$a^2\mu^2$ dependence of $Z^{\MSbar\rm{(2\ GeV)}}_P$ through the RI/MOM ($\omega=0$, upper panel) and SMOM ($\omega=1$, lower panel) schemes, at three lattice spacing and $m_{\pi,ss}\sim 300$ MeV.}
\label{fig:ZP_dep_mu}
\end{figure}

\begin{figure}[thb]
 \includegraphics[width=0.45\textwidth]{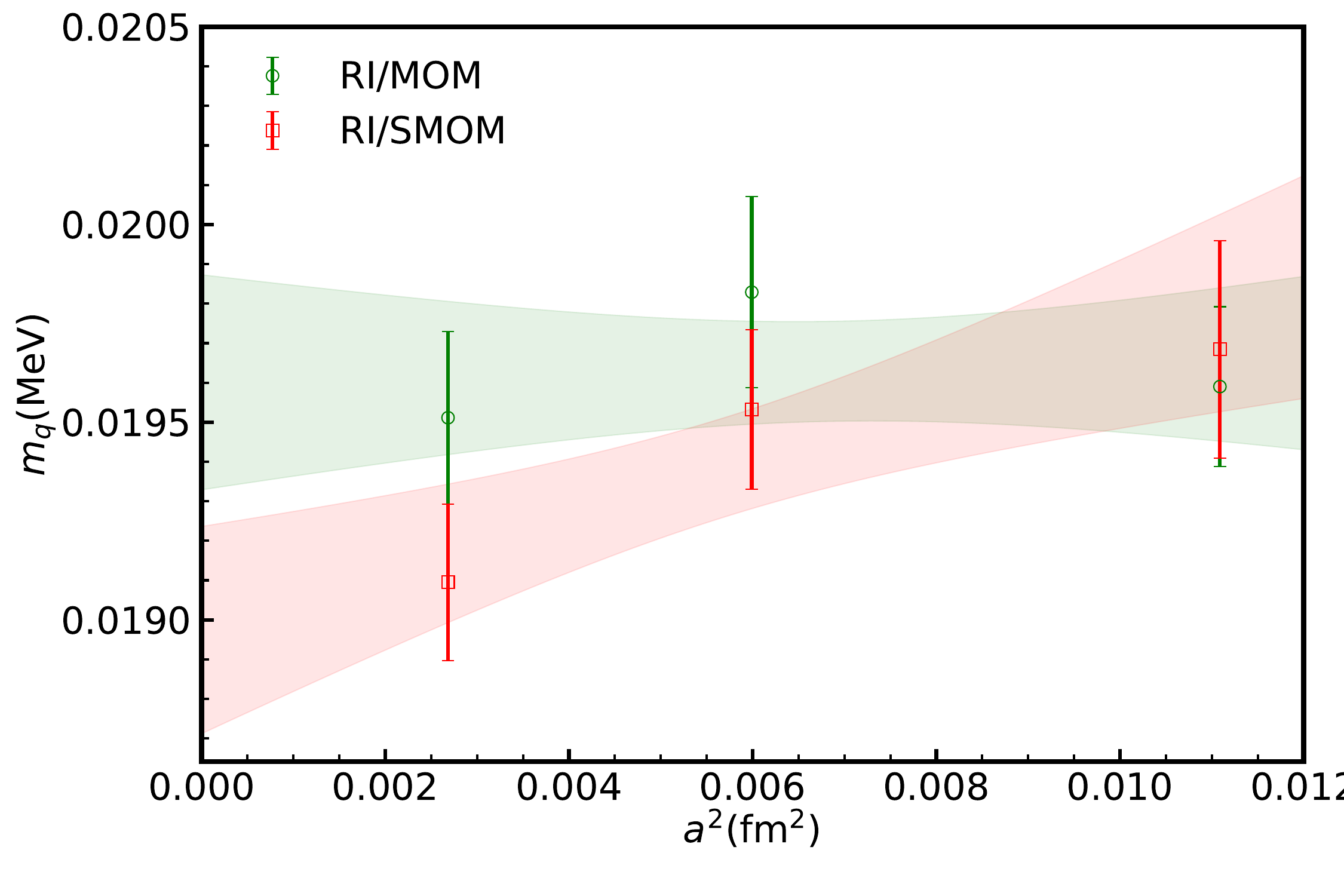}
\caption{Renormalized quark mass with $m_{\pi}$=317 MeV at three lattice spacing, using $Z^{\MSbar\rm{(2\ GeV)}}_P$ through either RI/MOM or SMOM scheme. The extrapolated values deviate by 3.1(1.5)\%.}
\label{fig:mass_lat}
\end{figure}

After $Z^{\omega}_P$ is extracted, it can be further converted to the $\MSbar$ scheme at given scale likes $\mu_0$= 2~GeV through the perturbative matching, and shall be independent of both $\mu$ and $\omega$. In Fig.~\ref{fig:ZP_dep_mu}, $Z^{\MSbar\rm{(2\ GeV)}}_P$ obtained through the RI/MOM scheme at the coarsest lattice spacing $a$=0.105 fm can be consistent with that using the SMOM scheme, with the 3-4\% systematic uncertainties from the perturabative matching. The situation improves at smaller lattice spacing, and systematic uncertainties are also smaller there. In Fig.~\ref{fig:mass_lat}, we show the renormalized quark mass at $\MSbar\rm{(2\ GeV)}$ with $m_{\pi}$=317 MeV at three lattice spacing, using $Z^{\MSbar\rm{(2\ GeV)}}_P$ through either RI/MOM or SMOM scheme. We can see that the lattice spacing dependence using the RI/MOM scheme is consistent with zero, while SMOM shows a non-vanishing dependence and make the continuum extrapolated value to be 3.1(1.5)\% lower.

\begin{figure}[thb]
 \includegraphics[width=0.45\textwidth]{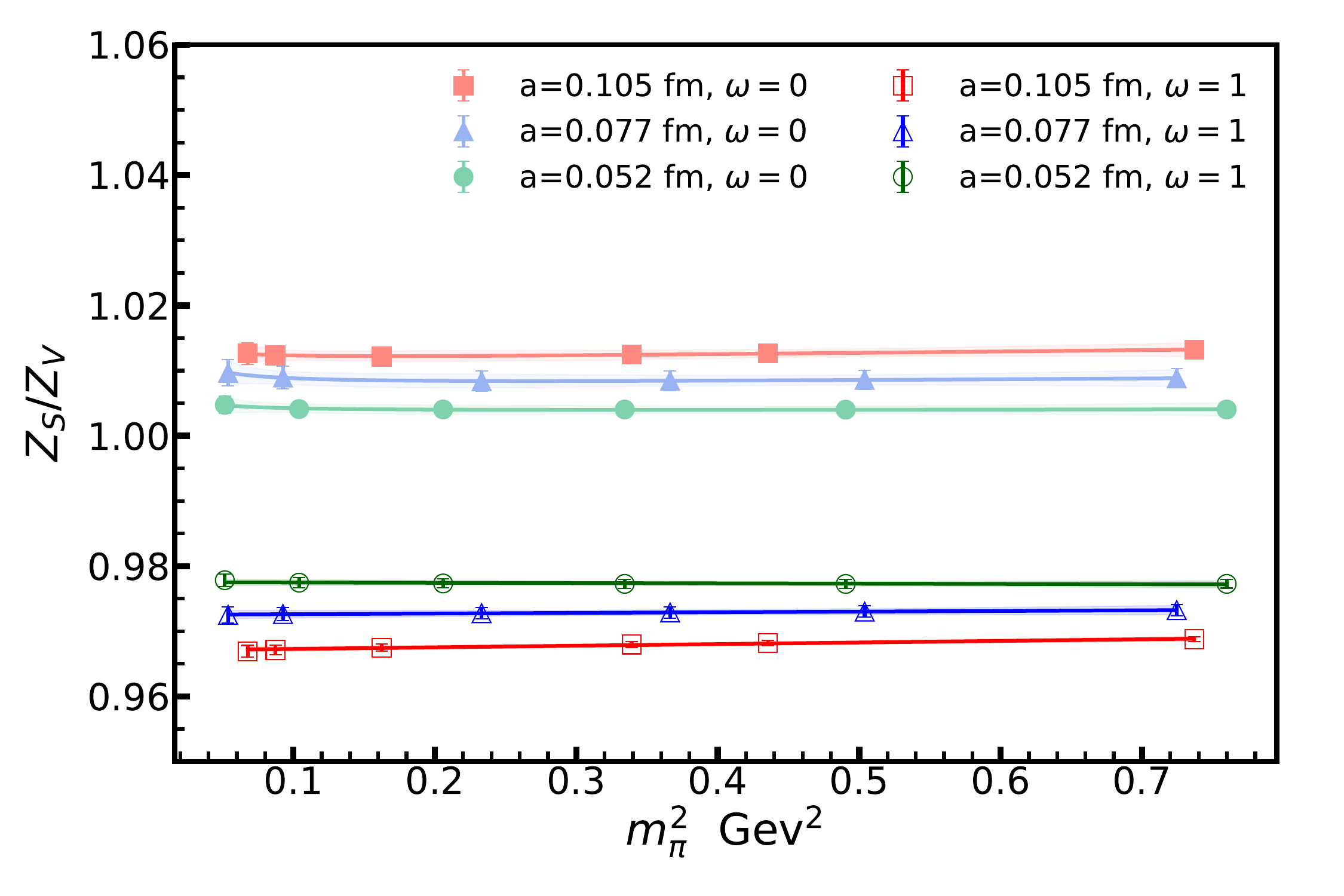}
\caption{Valence pion mass square $m_{\pi,vv}^2$ dependence of $\hat{Z}^{\omega=0}_S$ (RI/MOM, lighter  points) and $\hat{Z}^{\omega=1}_P$ (SMOM, darker points) with $a^2\mu^2=4$, at three lattice spacing and $m_{\pi,ss}\sim 300$ MeV. }
\label{fig:ZS_dep_mass}
\end{figure}

\subsubsection{Chiral symmetry breaking between $Z_P$ and $Z_S$}\label{sec:zp_zs}

The scalar current RC can be defined similarly with a slightly different parameterization,
\bal
&\hat{Z}^{\omega}_S(\mu)=Z_V\frac{\frac{1}{48}\text{Tr}[\Lambda^\mu_{V}(p_1,p_2)\gamma_\mu]}{\frac{1}{12}\text{Tr}[\Lambda_{S}(p_1,p_2)]}|_{p_1^2=p_2^2=\mu^2, (p_1-p_2)^2=\omega\mu^2},\nonumber\\
&\hat{Z}^{\omega}_S(\mu;m_q^{\rm PC})=\frac{A_S^{\omega}(\mu)}{m_q^{\rm PC}}+(Z^{\omega}_S(\mu))+C_S^{\omega}(\mu) m_q^{\rm PC},
\eal
where $S=\bar{\psi}\psi$, and the $A_S^{\omega}$ term can be dropped for $\omega=1$. However, as shown in Fig.~\ref{fig:ZS_dep_mass}, the effect of the $A_S^{\omega}$ term is consistently negligible regardless of the value of $\omega$.

\begin{figure}[thb]
 \includegraphics[width=0.45\textwidth]{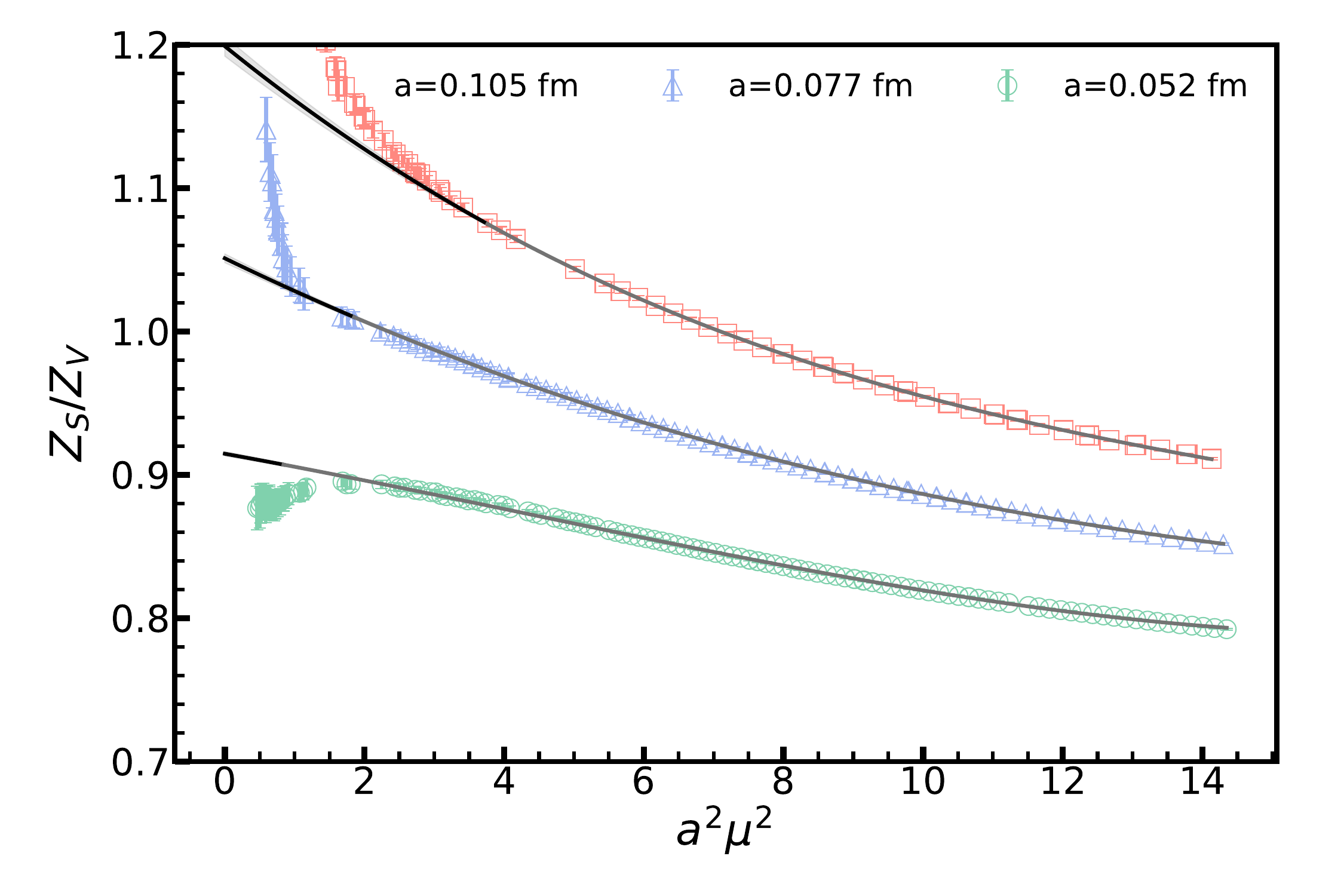}
 \includegraphics[width=0.45\textwidth]{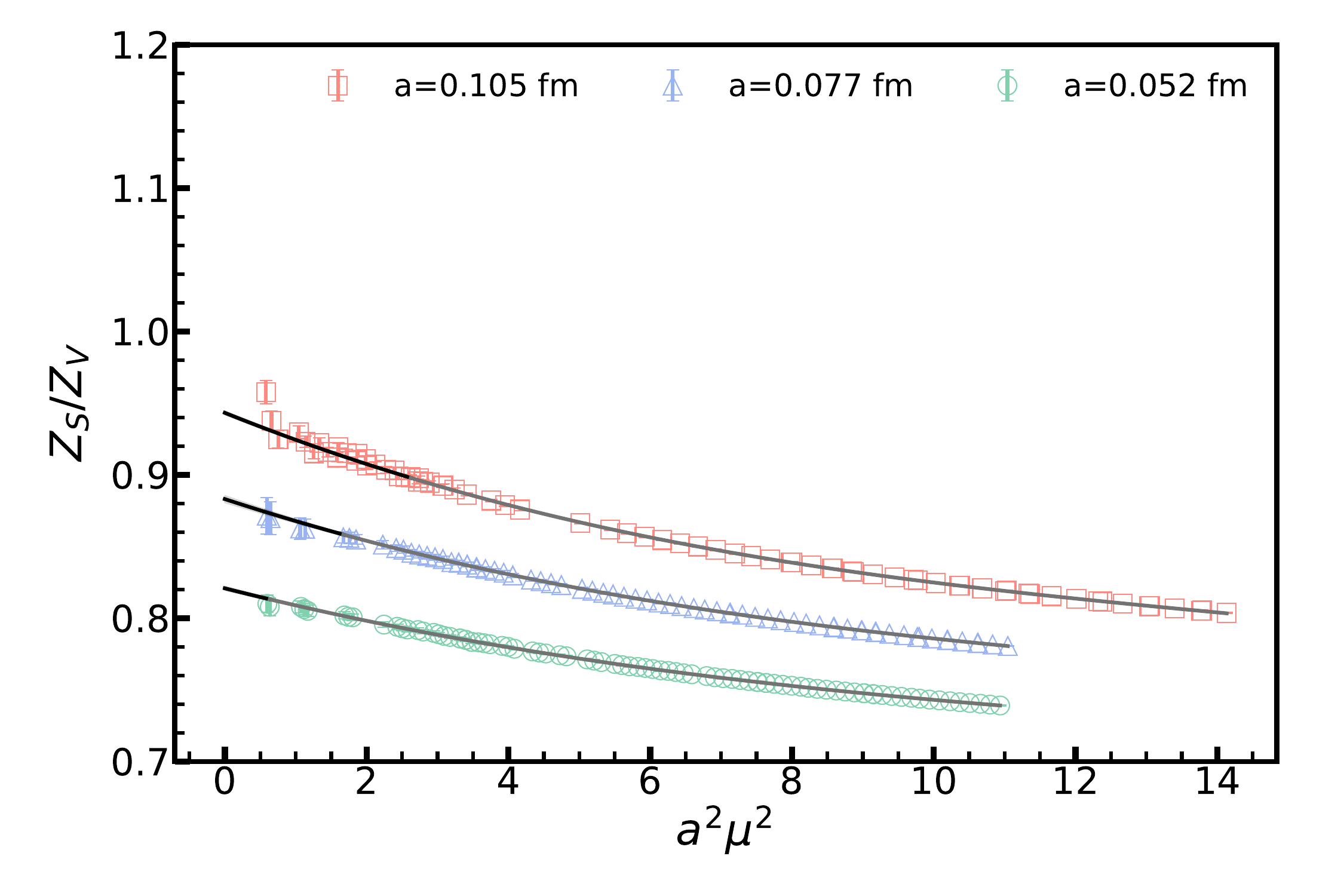}
\caption{$a^2\mu^2$ dependence of $Z^{\MSbar\rm{(2\ GeV)}}_S$ through the RI/MOM ($\omega=0$, upper panel) and SMOM ($\omega=1$, lower panel) schemes, at three lattice spacing and $m_{\pi,ss}\sim 300$ MeV. }
\label{fig:ZS_dep_mu}
\end{figure}

After converting to the $\MSbar$ scheme at a given scale using the same matching procedure as in the pseudoscalar case, we obtain $Z^{\MSbar\rm{(2\ GeV)}}_S$ \blue{as shown in Fig.~\ref{fig:ZS_dep_mu},} through either the RI/MOM scheme (upper panel) or the SMOM scheme (lower panel) with the corresponding $a^2\mu^2$ errors. We observe that the scheme dependence of $Z_S$ is much stronger than that of $Z_P$, but the difference diminishes as the lattice spacing decreases.

\begin{figure}[thb]
 \includegraphics[width=0.45\textwidth]{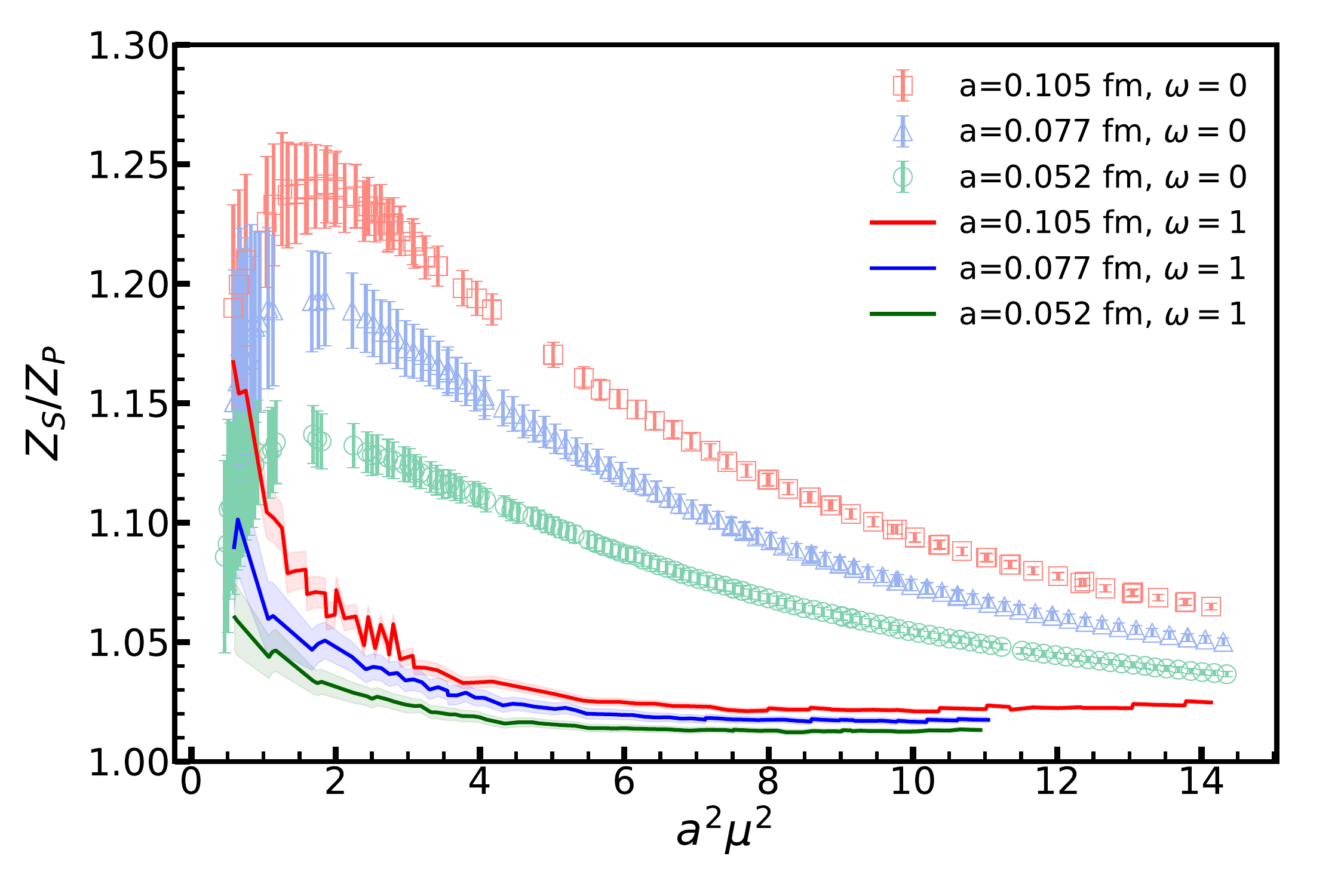}
\caption{The ratio $Z_S/Z_P$ through the RI/MOM scheme at three lattice spacing as functions of $a^2p^2$.}
\label{fig:ZS_ZP}
\end{figure}

Another comparison we can make is the ratio $Z_S/Z_P$ obtained through either the RI/MOM or SMOM scheme. As depicted in Fig.~\ref{fig:ZS_ZP}, the chiral symmetry breaking effect is significantly smaller when using the SMOM scheme (bands) compared to the RI/MOM scheme (data points). Moreover, both schemes exhibit further suppression at smaller lattice spacings. Similar to the $Z_A/Z_V$ case, the inclusion of the ${\cal O}(\alpha_s)$ term is necessary to restore chiral symmetry in the continuum.

\subsubsection{$Z_T$ and brief summary on renormalization}\label{sec:zt}

\begin{figure}[thb]
 \includegraphics[width=0.45\textwidth]{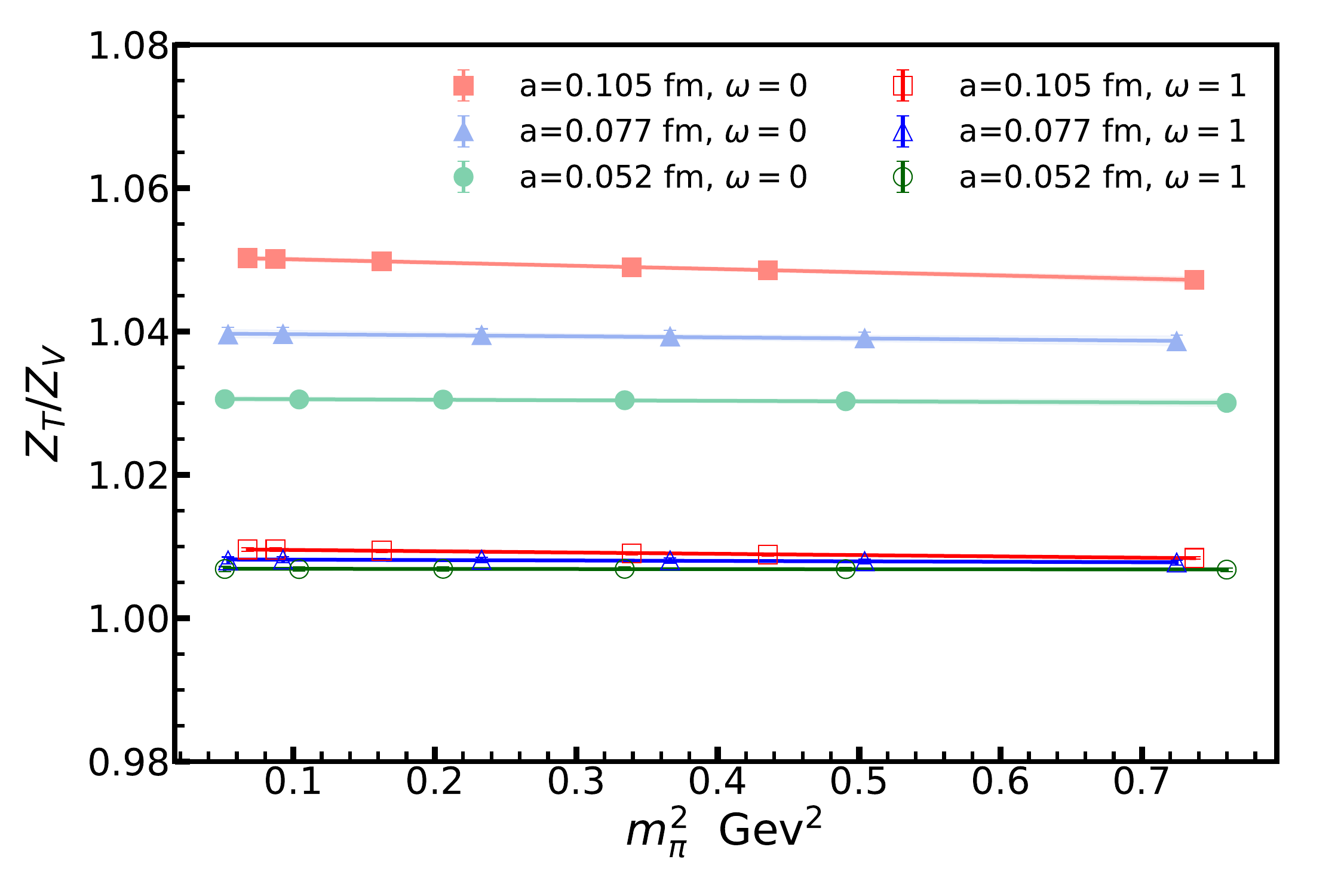}
\caption{Valence pion mass square $m_{\pi,vv}^2$ dependence of $\hat{Z}^{\omega=0}_S$ (RI/MOM, lighter  points) and $\hat{Z}^{\omega=1}_T$ (SMOM, darker points) with $a^2\mu^2=4$, at three lattice spacing and $m_{\pi,ss}\sim 300$ MeV.  }
\label{fig:ZT_dep_mass}
\end{figure}

\begin{figure}[thb]
 \includegraphics[width=0.45\textwidth]{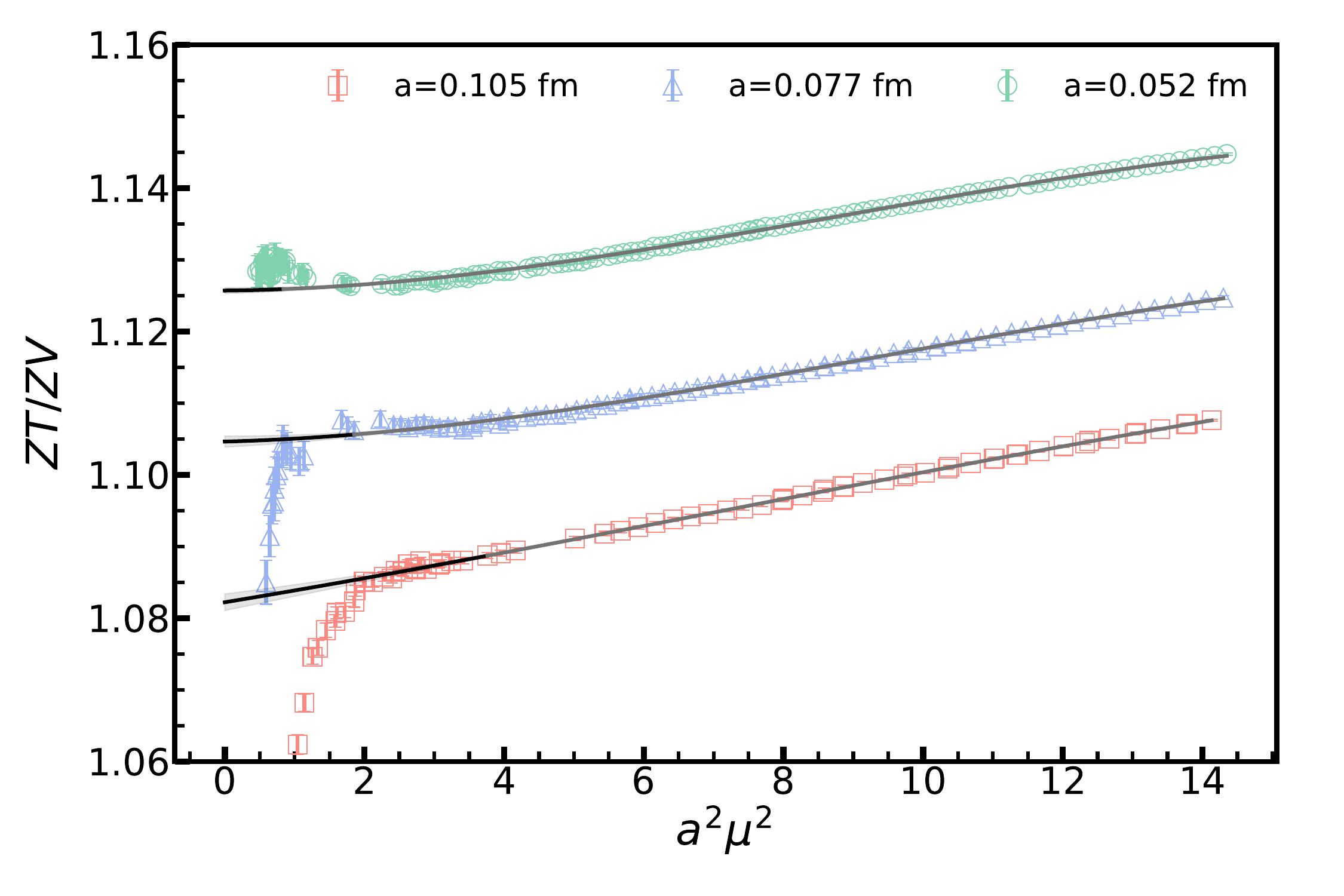}
 \includegraphics[width=0.45\textwidth]{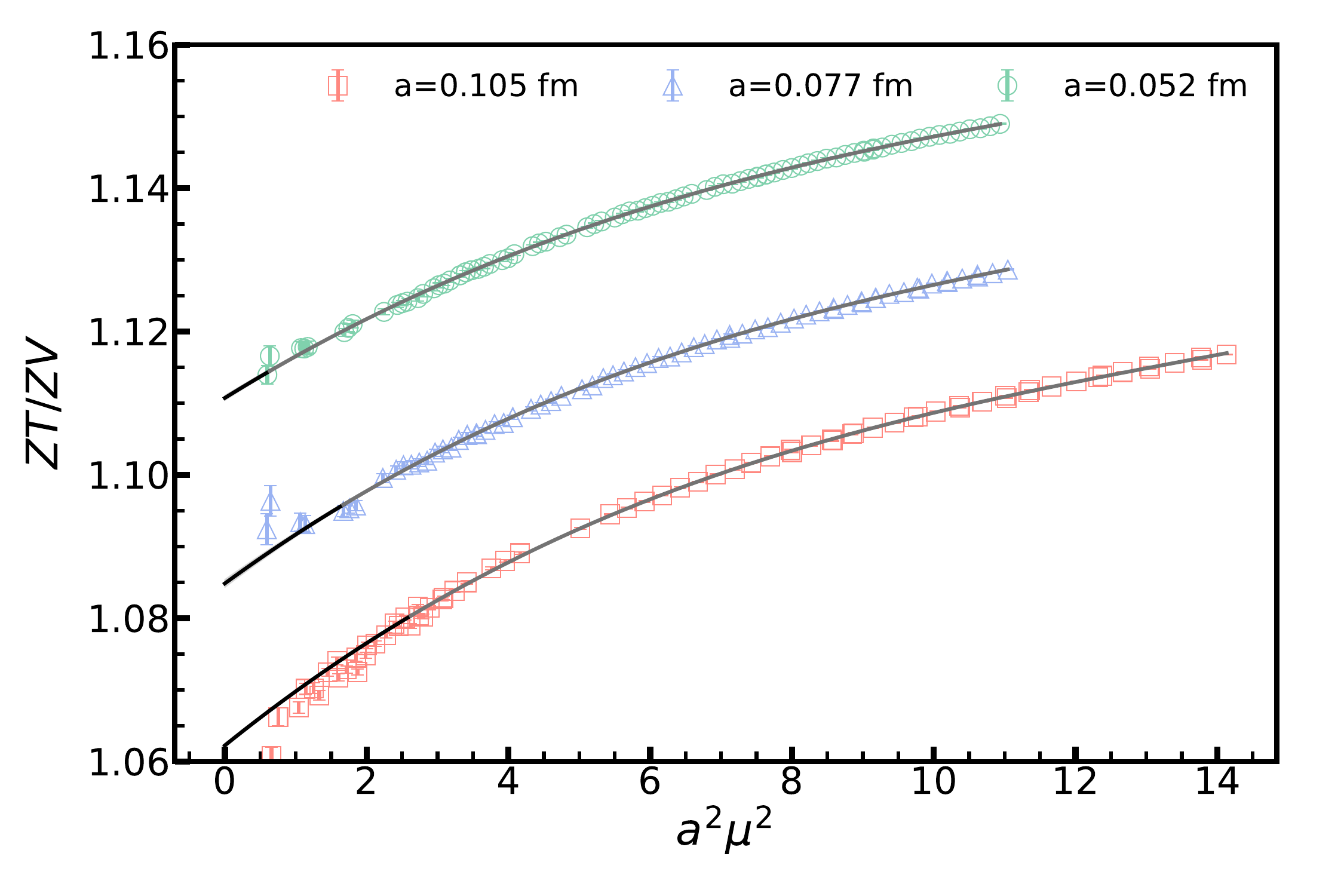}
\caption{$a^2\mu^2$ dependence of $Z^{\MSbar\rm{(2\ GeV)}}_T$ through the RI/MOM ($\omega=0$,  upper panel) and SMOM ($\omega=1$, lower panel) schemes, at three lattice spacing and $m_{\pi,ss}\sim 300$ MeV.}
\label{fig:ZT_dep_mu}
\end{figure}

For completeness, we also show the tensor current renormalization using both the RI/MOM and SMOM schemes,
\bal
&\hat{Z}^{\omega}_T(\mu)=Z_V\frac{\frac{1}{48}\text{Tr}[\Lambda^\mu_{V}(p_1,p_2)\gamma_\mu]}{\frac{1}{72}\text{Tr}[\Lambda_{T_{\mu\nu}}(p_1,p_2)\sigma_{\mu\nu}]}|_{p_1^2=p_2^2=\mu^2, (p_1-p_2)^2=\omega\mu^2},
\eal
where $T_{\mu\nu}=\bar{\psi}\sigma_{\mu\nu}\psi$. Then $Z^{\MSbar\rm{(2\ GeV)}}_T$ can be obtained through the corresponding perturbative matching after performing a linear $m_{\pi}^2$ extrapolation to the continuum, as shown in Fig.~\ref{fig:ZT_dep_mass}. The systematic uncertainty arising from the SMOM matching is estimated to be 2-3\%, which implies that $Z^{\MSbar\rm{(2\ GeV)}}_T$ is considered independent of the intermediate scheme within this uncertainty.

\begin{table}[h]
\centering
\caption{Normalization and renormalization constant at $\MSbar\rm{(2\ GeV)}$ at three lattice spacing with $m_{\pi,ss}\sim$ 300 MeV, using RI/MOM ($\omega=0$) or SMOM ($\omega=1$) as intermediate scheme.) In the RI/MOM case, the first line represents the result fit without $1/p^{2}$ term, while the second line represents the fit with that term.}\label{tab:mom_vs_smom}
\resizebox{1.0\columnwidth}{!}{
\begin{tabular}{l|l|llll}
 a(fm) & $\omega$ & $Z_A$ & $Z_S$ & $Z_P$ & $Z_T$ \\
\hline
\multirow{3}*{0.1053(2)} & 0 & 0.8547(13)& 0.957(08)(34)& 0.727(09)(26)& 0.864(01)(05)\\
&  & 0.8388(39)& 0.897(20)(44)& 0.672(36)(31)& 0.867(04)(04)\\
\cline{2-6}
& 1& 0.8177(04)& 0.754(02)(09)& 0.698(05)(08)& 0.848(01)(27)\\
\hline
\multirow{3}*{0.0775(2)} & 0 & 0.8821(08)&  0.879(05)(18)& 0.701(05)(15)& 0.923(01)(03)\\
& & 0.8792(25)& 0.892(12)(11)& 0.689(22)(09)& 0.919(05)(02)\\
\cline{2-6}
& 1 & 0.8533(06)& 0.738(01)(07)& 0.689(02)(06)& 0.906(01)(22)\\
\hline
\multirow{3}*{0.0519(3)} & 0 & 0.9011(04)& 0.796(09)(10)& 0.683(05)(08)& 0.978(01)(02)\\
& & 0.8998(07)& 0.822(08)(04)& 0.683(08)(03)& 0.973(01)(01)\\
\cline{2-6}
& 1 & 0.8838(02)& 0.720(01)(05)& 0.684(01)(05)& 0.961(00)(19) \\
\end{tabular}
}
\end{table}

\begin{table*}
	\centering \caption{ Normalization and renormalization constants at $\MSbar\rm{(2\ GeV)}$ on all the ensembles. The RCs include two uncertainties where the former one is the ensemble independent statistical and systematic uncertainties, and the latter one is the systematic uncertainties from the perturbative matching which are fully correlated on different ensembles.}\label{tab:renorm}
	%\begin{ruledtabular}
 \resizebox{2.0\columnwidth}{!}{
		\begin{tabular}{l | llllll | llll | l}
		    & C24P34 & C24P29 &  C32P29 & C32P23 & C48P23 & C48P14 & F32P30  & F48P30  & F32P21  & F48P21 & H48P32 \\  
		\hline 		
         $Z_V$ &0.79676(32) & 0.79814(23) & 0.79810(13) & 0.79957(13) & 0.79954(05) & 0.79957(06) & 0.83548(12) & 0.83511(04) &  0.83579(09) & 0.83567(05) & 0.86855(04)\\
		 $Z_A$ &0.85698(89) & 0.85442(85) & 0.8547(13) & 0.85593(87) & 0.85674(96) & 0.85605(94) & 0.88161(64) & 0.88213(77) & 0.88069(60) & 0.88063(98) & 0.90113(36)\\
		 $Z_S$ &0.960(05)(33) & 0.953(06)(31) & 0.957(08)(34) & 0.957(06)(33) & 0.966(10)(34) & 0.963(04)(34) & 0.871(09)(22) & 0.879(05)(18) & 0.871(06)(21) & 0.878(04)(17) & 0.796(09)(10)\\
		 $Z_P$ &0.732(11)(26) &  0.733(13)(24) & 0.727(09)(26) & 0.733(08)(26) & 0.733(09)(27) & 0.727(11)(26) & 0.699(07)(17) & 0.701(05)(15) & 0.704(05)(17) & 0.701(08)(15) & 0.683(05)(08)\\	
		 $Z_T$ &0.865(02)(05) & 0.864(01)(05) & 0.864(01)(05)  & 0.865(01)(05) & 0.865(01)(05) & 0.866(01)(05) & 0.922(01)(04) & 0.923(01)(03) & 0.920(01)(04) & 0.922(01)(03) & 0.978(01)(02)\\ 

		\end{tabular}
  }
	%\end{ruledtabular}
\end{table*}

Based on the values of $Z_{A,S,P,T}$ obtained in this work and collected in Table~\ref{tab:mom_vs_smom}, it can be observed that the dependence on the intermediate scheme (RI/MOM or SMOM) is reduced as the lattice spacing decreases. However, it is important to note that the scheme dependence is not completely eliminated even after performing the continuum extrapolation. Considering the lattice spacing dependence of the renormalized $f_{\pi}$, $m_q$, and $g_S$, it is recommended to use the RI/MOM scheme to suppress the discretization error.

\red{In Table~\ref{tab:mom_vs_smom}, we also show the renormalization constants using the RI/MOM scheme but with an additional $1/p^2$ term~\cite{Hasan:2019noy}. This term can have an obvious impact at the coarsest lattice spacing but is consistent with zero at the finest lattice spacing. This is understandable as the fitting range at finer lattice spacing is much larger, reducing the possible influence of a $1/p^2$ pole in the inferred region. Thus, it can be considered a discretization effect that will be eliminated in the continuum extrapolation.}

As a summary of this section, Table~\ref{tab:renorm} summarizes the normalization and renormalization constants for all the ensembles used in this work, obtained through the intermediate RI/MOM scheme. The first uncertainty of RCs is ensemble-independent, while the second one is fully correlated on different ensembles and can be suppressed after the continuum extrapolation.

\subsection{Global fit}\label{sec:global}

In order to process this continuum extrapolation systematically, we calculate the quark propagators with unitary light quark mass and also 2 partially quenched quark masses with the constraint $m_{\pi}L>3.5$, on each of the 11 ensembles. Then we use the following NLO partially quenched $\chi$PT form~\cite{Sharpe:1997by} to describe the pion masses and decay constants with different valence and sea quark masses, in addition to extra parameters $c_{m/f,a/l}$ for the finite lattice spacing/volume corrections,

\begin{align}
    m^2_{\pi,{\rm vv}} &= \Lambda_{\chi}^2  2  y_{\rm v} 
    \left\{
        1+\frac{2}{N_f} 
        \left[
            (2 y_{\rm v}-y_{\rm s})\mathrm{ln} (2y_{\rm v})+(y_{\rm v}-y_{\rm s})
        \right]
    \right. \nonumber
    \\
    &
    \left.
        ~~ + 2y_{\rm v} (2\alpha_8-\alpha_5) + 2y_{\rm s} N_f (2\alpha_6-\alpha_4)
    \right\}            \notag
    \\
        &\quad \times\left[ 1+c^{\pi}_{L}e^{-m_{\pi}L}+c^{\pi}_{s}(m_{\eta_s}^2-m_{\eta_s,{\rm phys}}^2) 
    \right]\nonumber\\
    &\quad \times(1+c^{\pi}_{a}a^2),\\
    F_{\pi,{\rm vv}}&= 
    F \left(
        1-\frac{N_f}{2}(y_{\rm v}+y_{\rm s})\mathrm{ln}(y_{\rm v}+y_{\rm s})+y_{\rm v}\alpha_5  + y_{\rm s}N_f\alpha_4
        \right)
        \nonumber
    \\
    &\quad \times 
    \left[ 
        1+d^{\pi}_{L}e^{-m_{\pi}L}+d^{\pi}_{s}(m_{\eta_s}^2-m_{\eta_s,{\rm phys}}^2)
    \right]         \nonumber\\
    &\quad \times(1+d^{\pi}_{a}a^2)
\end{align}
where $N_f=2$ for two light flavors and $\Lambda_{\chi}=4\pi F$ is the intrinsic scale of $\chi$PT with $F$ being the pion decay constant in the chiral limit.
The dimensionless expansion parameters $y_{\rm v/s}=\frac{\Sigma m^{\rm v/s}_l}{F^2 \Lambda_{\chi}^2}$ involve the chiral condensate $\Sigma$ and quark mass $m_{v/s}$ for the valence and sea quark masses, respectively. Additionally, $\alpha_i$  represents the NLO low energy constants of $\chi$PT at the intrinsic scale $\Lambda_{\chi}$.  These constants can be converted to the usual scale-dependent partially quenched $\chi$PT NLO coefficients as $L_i(\mu)=\frac{1}{128 \pi^2}(\alpha_i+c_i\mathrm{ln}\frac{\mu}{\Lambda_{\chi}})$ with $c_{4,5,6,8}=\{-\frac{1}{2},-\frac{N_f}{2},-\frac{2+N_f^2}{N_f^2},\frac{N_f^2-4}{4N_f}\}$, respectively~\cite{Gasser:1984gg,Boyle:2015exm}. We also introduce additional corrections terms with the coefficients $c_{m/f,a/L/s}$ to account for the discretization, finite volume and strange quark mismatch effects, with $m_{\eta_s}=689.63(18)$ MeV from Ref.~\cite{Borsanyi:2020mff}.

Since the statistics on each ensemble are different, we perform 4000 bootstrap re-samplings on each ensemble and conduct the correlated global fit based on these bootstrap samples. In such a strategy, the correlation between different ensembles vanishes within the statistical uncertainty of the resampling and is suppressed by the number of bootstrap samples. The lattice spacing and renormalization constants are sampled for each bootstrap sample using a Gaussian distribution with their uncertainties as the width of the distribution. These values are then applied to the dimensionless quantities extracted from the joint fit defined in Eq.~(\ref{eq:fit_quark_mass}-\ref{eq:fit_pion_mass}).

  \begin{table}[ht!]                   
    \caption{Global fits with and without non-perturabtive uncertainty $\delta_{\rm np}Z_P$ and perturabtive uncertainty $\delta_{\rm p}Z_P$.}  
    %\begin{ruledtabular}
    \resizebox{1.0\columnwidth}{!}{

    \begin{tabular}{c| c c c c}              
    \hline
    \hline
    $\delta Z_P$ included & N/A & $\delta_{\rm np}$ & $\delta_{\rm p}$ & $\delta_{\rm np}$+$\delta_{\rm p}$ \\
    \hline 
                   %& xx& xx& xx & 1.5 \\ 
    $\chi^2$/d.o.f.     & \multicolumn{4}{c}{1.3} \\
    \hline
    $F({\rm GeV})$                      & 0.08659(69)& 0.08659(72)& 0.08659(71) & 0.08660(74) \\ 
    $\Sigma^{1/3}({\rm GeV})$           & 0.2684(14) & 0.2685(30)& 0.2685(14)  & 0.2686(36) \\ 
    \hline
    $\alpha_4$                          & 0.342(93) &  0.343(99)&  0.340(99)    & 0.34(10) \\ 
    $\alpha_5$                          & -0.38(18)  & -0.39(19)&-0.38(18)     & -0.38(20) \\ 
    $\alpha_6$                          & 0.056(41)  & 0.058(84)& 0.054(47)     & 0.054(86) \\ 
    $\alpha_8$                          & 0.482(79)  & 0.48(18)& 0.483(81)     & 0.48(18) \\ 
    \hline
    $c^{\pi}_{a}({\rm fm}^{-2})$        &  2.12(67) &  2.1(1.8) & 2.1(3.7)  &  2.0(4.1) \\ 
    $c^{\pi}_{L}$                       & 0.64(20)  &  0.65(53)& 0.64(20)  &  0.65(51)\\ 
    $c^{\pi}_{s}({\rm GeV}^2)$          &0.058(32)   & 0.06(14) &0.101(32)  &  0.07(14)\\
    $d^{\pi}_{a}({\rm fm}^{-2})$        & -5.56(45)  & -5.55(47)& -5.57(55) &  -5.57(57)     \\ 
    $d^{\pi}_{L}$                       & -0.73(14)  & -0.73(15)&-0.73(14) &  -0.73(15)\\ 
    $d^{\pi}_{s}(\rm {GeV}^{2})$        &0.197(28)   & 0.197(31)&0.197(29)  &  0.198(32)\\
    \hline  
    $m_{l,{\rm phys}}( {\rm MeV})$      & 3.600(31)  & 3.60(11)  & 3.60(03)      &  3.60(11)\\ 
    $f_{\pi,{\rm phys}}({\rm MeV})$     & 130.73(89) & 130.73(90)& 130.73(90)    & 130.74(92)\\ 
    \hline
    \hline
    \end{tabular}  
    
    }
    %\end{ruledtabular}
    \label{tab:pion_fit_test}
\end{table}

The lattice spacing uncertainty is sampled with an uniform seed at a given lattice spacing, since this uncertainty is fully corrected on all the ensembles at this lattice spacing. Similarly, 
The perturbative matching uncertainty of $Z_P$ is sampled with a uniform seed on all the ensembles, but with respective rescale factors on each ensemble. On the other hand, the uncertainty of lattice spacing and the non-perturbative uncertainty of $Z_P$ are sampled independently on each ensemble. In order to show the impact of two $Z_P$ uncertainties in the global fit, we preform the following four cases of fit, and show the results of $\Sigma$, $F$, $\alpha_{4,5,6,8}$ and also $c_{m/f,a/L/s}$ in Table~\ref{tab:pion_fit_test}:

(1) Fitting with the statistical uncertainty $\delta\tilde{O}$ from the dimensionless observable $\tilde{O}$ with $O=m_{\pi},\ m_q^{\rm PC}$ and $f_{\pi}$, and also lattice spacing uncertainty $\delta a$;

(2) Fitting with $\delta\tilde{O}$, $\delta a$ and non-perturbative uncertainty $\delta_{\rm np}Z_P$;

(3) Fitting with $\delta\tilde{O}$, $\delta a$ and perturbative uncertainty $\delta_{\rm p}Z_P$;

(4) Fitting with $\delta\tilde{O}$, $\delta a$, $\delta_{\rm np}Z_P$ and also $\delta_{\rm p}Z_P$.

All four cases provide reasonable $\chi^2$/d.o.f. values, and similar values of $F$ which are irrelevant to $Z_P$. However, the uncertainty of $\Sigma$ is highly sensitive to $\delta Z_P$, as expected. By imposing the conditions $y_{\rm s}=y_{\rm v}$, $M_{\pi,{\rm vv}}=M_{\pi, {\rm phys}}=134.98~\mathrm{MeV}$, $a\rightarrow 0$, and $L\rightarrow \infty$, we can extract the light quark mass $m_{l}^{\MSbar\rm{(2\ GeV)}}$ in the continuum and infinite volume limits from the global fit. The $m_l$ obtained from different cases of fits are also listed in Table~\ref{tab:pion_fit_test}.

As shown in the table, the extrapolated quark mass $m_l={\color{black}3.60(3)}$ obtained in Case (1) is consistent with the value 3.64(8)(11)~MeV at $a=0.105$ fm, as we argued before. Additionally, it has a smaller uncertainty due to the constraints from the other ensembles and the exclusion of $\delta Z_P$. However, the uncertainty is enlarged to 2.7\% in Case (4) where both the non-perturbative and perturbative uncertainties of $Z_P$ are included, while the central value remains almost unchanged. Treating $\delta_{\rm p}Z_P$ as correlated across all the ensembles significantly suppresses its impact from 3\% (at $a=0.105$ fm) to 0.5\% after the continuum extrapolation, as shown in Case (3); However, the Case (2) suggests that the $\delta_{\rm np}Z_P$ is enlarged from 1\% (at $a=0.105$ fm) to 2.5\% simultaneously since it is independent for different ensembles.

It is worth mentioning that the relative uncertainty of $\Sigma$ and $m_l$ in Case (2-4) is almost the same and will be cancelled when we consider the renormalization independent combination $\Sigma m_l$.

Currently, 90\% of the uncertainty of $m_l={\color{black}3.60(11)}$ comes from the independent statistical uncertainty of $Z_P$ using the RI/MOM scheme on different ensembles, which can be suppressed if we adopt a more aggressive treatment on the renormalization constants. 
However, the current 3\% uncertainty of $m_l$ is similar to the difference between the continuum extrapolation of $m_l$ at $m_{\pi}=317$ MeV using either the RI/MOM or SMOM scheme. Therefore, we reserve this opinion for future study, where more lattice spacings can be utilized to gain a better understanding of the chiral symmetry breaking effect in the renormalization constants.

With the $m_l$ extracted above, we can also determine the physical $f_{\pi}$ in the continuum and infinite volume limits to be {130.7(9)} MeV, which is consistent with the experimental value of 130.4(2) MeV~\cite{ParticleDataGroup:2020ssz}. Based on $F$ extracted above, we predict $F_{\pi}/F=\sqrt{1/2}f_{\pi}/F={\color{black}1.0675(19)}$. 

As shown in Table~\ref{tab:pion_fit_test} and inspired by the previous sections, the discretization error in quark mass is consistent with zero, while this error in $f_{\pi}$ is much more significant. On the other hand, the finite volume effect appears to be compariable with the discretization effect, with the correction on the F32P21 ensemble with the smallest $m_{\pi}L$ being about \red{4.6(1.1)}\%. 

It is worth mentioning that the dependence of the pion mass on the strange quark mass is almost consistent with zero, as expected based on the direct calculation of the strange content in the pseudo-scalar meson and the Feynman-Hellman theorem~\cite{Yang:2014xsa}. However, the dependence of the pion decay constant on the strange quark mass is more significant, which is partially attributed to the smaller uncertainty. Currently, the strange quark mass on the C24P34, C48P14, and H48P32 ensembles is higher than the physical value, while on all the other ensembles it is lower. Therefore, this dependence may arise from other systematic effects, and we cannot exclude this possibility with the present ensembles. Further studies with tuned quark masses would be helpful in verifying this.

To illustrate the lattice spacing dependence and the unitary quark mass dependence, we subtracted the partially quenching effect using bootstrap samples of the fit parameters from the original data points $m^{\rm data}_{\pi,{\rm vv}}$ and $f^{\rm data}_{\pi,{\rm vv}}$. We define the corrected $m^{\rm uni}_{\pi}$ and $f^{\rm uni}_{\pi}$ as follows,

\bal
    (m^{\rm uni}_{\pi})^2 &= \frac{(m^{\rm data}_{\pi,{\rm vv}})^2}{1+c_{m,L}e^{-m_{\pi}L}+c_{m,s}(m_{\eta_s}^2-m_{\eta_s,{\rm phys}}^2)} \nonumber 
    \\
    & \quad\quad  -\Lambda_{\chi}^2y_{\rm v}(y_{\rm s}-y_{\rm v})\big\{-\frac{2}{N_f}[\mathrm{ln} (2y_{\rm v})+1]\nonumber 
    \\
    & \quad\quad \quad +2N_f(2\alpha_6-\alpha_4)\big\}(1+c_{m,a}a^2),\nonumber
    \\
    f^{\rm uni}_{\pi}&=\frac{f^{\rm data}_{\pi,{\rm vv}}}{1+c_{f,l}e^{-m_{\pi}L}+c_{f,s}(m_{\eta_s}^2-m_{\eta_s,{\rm phys}}^2)}\nonumber\\&\quad \quad -F(y_{\rm s}-y_{\rm v})\{-\frac{N_f}{2} \mathrm{ln}(y_{\rm v}+y_{\rm s})+N_f\alpha_4\}
    \nonumber\\&\quad \quad\quad (1+c_{f,a}a^2).
\eal

The $m_{\pi}$ and $f_{\pi}$ with unitary valence and sea quark masses $y=y_{\rm v}=y_{\rm s}$ have another widely used parameterization,
\begin{align}
    m_\pi^2&=\Lambda_{\chi}^2  2 y
    \left[ 
        1+y
        \left(
            \ln\frac{2y\Lambda^2_{\chi}}{m^2_{\pi, {\rm phys}}}-{\ell}_3
        \right)
        +\mathcal{O}(y^2)
    \right],\\
    F_\pi &= F 
        \left[
            1-2y 
            \left(
                \ln\frac{2y\Lambda^2_{\chi}}{m^2_{\pi, {\rm phys}}}-\ell_4
            \right)
            +\mathcal{O}(y^2)
        \right].
\end{align}
where $\ell_{3,4}$ is related to $\alpha_{4,5,6,8}$ by
\bal\label{eq:NLO_LEC}
\ell_3=&\ln\frac{\Lambda^2_{\chi}}{m_{\pi,{\rm phys}}^2}-2[(2\alpha_8-\alpha_5)+2(2\alpha_6-\alpha_4)],\nonumber\\
\ell_4=&\ln\frac{\Lambda^2_{\chi}}{m_{\pi,{\rm phys}}^2}+\frac{1}{2}(\alpha_5+2\alpha_4).
\eal
Our determination of $\ell_{3,4}$ are also collected in Table~\ref{tab:final}\red{, consistent with the current FLAG average but have smaller uncertainties}.

In this work, we use the $m_{K^{\pm}}$ and $m_{K^0}$ with the constraint $m^{\rm phys}_u+m^{\rm phys}_d=2m^{\rm phys}_l$, to determine the up, down and strange quark masses $m_{u,d,s}$. The partially quenched Kaon masses and decay constants on all the ensembles are fitted with the following form proposed in a recent work~\cite{ExtendedTwistedMass:2021gbo},
\bal
    \label{eq:kaon_form}
    &m_K^2(m^{\rm v}_l,m^{\rm s}_l,m^{\rm v}_s,m^{\rm s}_s,a) 
     =(b_s^{\rm v} m^{\rm v}_s+b_s^{\rm s} m^{\rm s}_s+b_l^{\rm v} m^{\rm v}_l+b_l^{\rm s} m^{\rm s}_l) \notag \\ 
    &~~~~~~ \times\left[ 1+c^K_l m^{\rm v}_l+c^K_m a^2 + c_{L}^K \exp{(-m_{{\pi}} L)} \right],\\
    &f_K(m^{\rm v}_l,m^{\rm s}_l,m^{\rm v}_s,m^{\rm s}_s,a)\nonumber
    =(d_f+d_s^{\rm v} m^{\rm v}_s+d_s^{\rm s} m^{\rm s}_s+d_l^{\rm v} m^{\rm v}_l+d_l^{\rm s} m^{\rm s}_l) \notag\\
    &~~~~~~ \times\left[1+d^K_a a^2 +d_{L}^K \exp{(-m_{{\pi}} L)}   \right].
\eal 

\begin{table}[ht!]          
\caption{Global fit of the Kaon mass and decay constant, with both $\delta_{\rm np}Z_P$ and $\delta_{\rm p}Z_P$. }  
    \resizebox{1.0\columnwidth}{!}{
    \begin{tabular}{c| c c c c c cc }   
        \hline
        \hline
        $\chi^2$/d.o.f. & $b_l^v({\rm GeV})$ & $b_l^s({\rm GeV})$ & $b_s^v({\rm GeV})$  & $b_s^s({\rm GeV})$ & $c_l^K({\rm GeV^{-1}})$ & $c_a^K({\rm fm^{-2}})$ & $c_{L}^k$ \\
        0.94            &2.36(94)            & 0.23(25)           & 2.30(12)            & 0.11(12)           & 0.07(13)                & 1.2(3.3)               & 0.24(35) \\
        \hline
        $\chi^2$/d.o.f. & $d_f({\rm GeV})$   & $d_l^v$            & $d_l^s$             & $d_s^v$            & $d_s^s$                 & $d_a^K({\rm fm^{-2}})$ & $d_{L}^k$ \\
        0.98            & 0.1291(32)         & 0.161(51)          & 0.528(44)           & 0.573(49)          & 0.085(30)               & -5.43(87)              & -0.386(94) \\
        \hline
        \hline
    \end{tabular}  
    }
\label{tab:Kaon fit table1}
\end{table}

Based on the QED correction $\Delta_{\rm QED} m_K$ obtained in previous literature~\cite{Giusti:2017dmp},
\bal
\Delta_{\rm QED} m_{K^{\pm}}-\Delta_{\rm QED} m_{K^{0}}=2.07(15)\ {\rm MeV}, \nonumber\\
\Delta_{\rm QED} m^2_{K^{0}}=0.174(24)\times 10^{-3}\ {\rm  GeV}^2,
\eal
we can obtain $\Delta_{\rm QED} m_{K^{0}}=0.17(2)$ MeV and $\Delta_{\rm QED} m_{K^{\pm}}=2.24(15)$ MeV, respectively. Thus, the pure QCD Kaon mass with physical quark mass will be
\bal
m_{K^{0}, {\rm QCD}}=m_{K^{0}}^{\rm phys}-\Delta_{\rm QED} m_{K^{0}}=497.44(02)\ {\rm MeV},\nonumber\\
m_{K^{\pm}, {\rm QCD}}=m_{K^{\pm}}^{\rm phys}-\Delta_{\rm QED} m_{K^{\pm}}=491.44(15)\ {\rm MeV},\nonumber\\
\eal
and the physical quark masses $m^{\rm phys}_u$, $m^{\rm phys}_d$ and $m^{\rm phys}_s$ can be determined using the following three conditions
\bal
m_K(m^{\rm phys}_d,m^{\rm phys}_l,m^{\rm phys}_s,m^{\rm phys}_s,0)&=m_{K^0, {\rm QCD}}, \nonumber\\
m_K(m^{\rm phys}_u,m^{\rm phys}_l,m^{\rm phys}_s,m^{\rm phys}_s,0)&=m_{K^{\pm}, {\rm QCD}},\nonumber\\
m^{\rm phys}_u+m^{\rm phys}_d&=2m^{\rm phys}_l.
\eal
Note that we ignored the isospin symmetry breaking effect in the sea quark masses, since it should be ${\cal O}((m_d-m_u)^2)$
 and negligible given the current uncertainties.

%\begin{figure}[thb]
%    \centering
%    \includegraphics[width=0.45\textwidth]{figures/kaon_vs_lattice_fit_a2_volbeh_Pion.pdf}
%    \includegraphics[width=0.45\textwidth]{figures/fka_vs_lattice_fit_a2_volbeh_Pion.pdf}
%    \caption{The corrected kaon mass $m^{\rm cr}_K$ and the decay constant $f^{\rm cr}_K$ with the physical light quark mass $m_l^{\rm phys}$, varies with the strange quark mass at three lattice spacing (colored data points and bands) and also continuum (gray band).}
%    \label{fig:kaon_final_sm}
%\end{figure}

{The corrected kaon mass $m^{\rm cr}_K$ and the decay constant $f^{\rm cr}_K$ shown in Fig.~\ref{fig:kaon_final} are defined as
\bal
(m^{\rm cr}_K)^2&=(m^{\rm data}_K)^2-m^2_K(m^{\rm v}_l,m^{\rm s}_l,m^{\rm v}_s,m^{\rm s}_s,a,L)\nonumber\\
&\quad +m^2_K(m^{\rm phys}_l,m^{\rm phys}_l,m^{\rm v}_s,m^{\rm v}_s,a,L\rightarrow \infty),\\
f^{\rm cr}_K&=f^{\rm data}_K-f_K(m^{\rm v}_l,m^{\rm s}_l,m^{\rm v}_s,m^{\rm s}_s,a,L)\nonumber\\
&\quad +f_K(m^{\rm phys}_l,m^{\rm phys}_l,m^{\rm v}_s,m^{\rm v}_s,a,L\rightarrow \infty),
\eal
with the light quark mass $m_l$ corrected to its physical value $m_l^{\rm phys}$ using the bootstrap samples of the fitted $m_K$.}

\red{As a consistency check, we also calculate $m_{\eta_s}$ on each ensemble and perform a similar global fit with the following functional form,
\begin{align}
m_{\eta_s}^2 = &[h_l^{\rm s} m_l^{\rm s} + h_s^{\rm v} m_s^{\rm v} +h_s^{\rm s} m_s^{\rm s}+h_{l^2}(m_l^{\rm s})^2]\nonumber\\
&\quad \quad (1+h_{a}a^2+h_{L}e^{m_{\eta_s}L}),
\end{align}
The $\chi^2$/d.o.f. of the fit is 1.4 an all the coefficients except $h_s^l$ are consistent with zero within the uncertainty, but dropping more terms will enlarge the $\chi^2$/d.o.f. significantly and then we keep all the coefficients. 
Using the physical light and strange quark masses, we predict $m_{\eta_s}=687.4(2.2)$ MeV in the infinite volume and continuum limit, which is consistent with the BMWc value of 689.63(18) MeV~\cite{Borsanyi:2020mff} we used in the fits for lattice spacing and $f_{\pi}$. Additively, we extract the dependence of $m_{\eta_s}^2$ on the strange quark mass, $\frac{\partial m_{\eta_s}^2}{\partial m_s}=4.9(1)$ GeV, which is slightly lower than the value of $m_{\pi}^2/m_q$ shown in Fig.~\ref{fig:pion_final}.}

The physical quark masses $m_{u,d,s}$ and also corresponding $f_{\pi,K}$ using $m_l^{\rm phys}$ and intermediate RI/MOM scheme, are collected in Table~\ref{tab:final}. \red{In addition, Table~\ref{tab:final} shows the global fit results using the $z_{A,P}$ through the SMOM scheme for comparison. As we can see from the continuum extrapolation tests using a 317 MeV pion mass, the SMOM scheme yields quark masses that are 3-4\% lower and decay constants that are $\sim$2\% lower compared to the RI/MOM scheme. However, the ratio of the quark masses or decay constants remains unchanged within the uncertainty as the renormalization constants are cancelled.}

\red{Therefore, we consider the result using the RI/MOM scheme as the central value due to its smaller discretization error, and treat the difference between the results obtained using the two schemes as systematic uncertainties. Such a systematic uncertainty can also be considered as an estimate of the residual discretiation error, as the correct continuum limit should be independent of the intermediate renormalization scheme. With this systematic uncertainty, all our determinations are consistent with the present lattice averages~\cite{FlavourLatticeAveragingGroupFLAG:2021npn} and/or PDG~\cite{ParticleDataGroup:2020ssz} within 1-2~$\sigma$, and the low energy constants $\Sigma$ and $\ell_{3,4}$ have smaller uncertainties.}


\begin{thebibliography}{55}%
\makeatletter
\providecommand \@ifxundefined [1]{%
 \@ifx{#1\undefined}
}%
\providecommand \@ifnum [1]{%
 \ifnum #1\expandafter \@firstoftwo
 \else \expandafter \@secondoftwo
 \fi
}%
\providecommand \@ifx [1]{%
 \ifx #1\expandafter \@firstoftwo
 \else \expandafter \@secondoftwo
 \fi
}%
\providecommand \natexlab [1]{#1}%
\providecommand \enquote  [1]{``#1''}%
\providecommand \bibnamefont  [1]{#1}%
\providecommand \bibfnamefont [1]{#1}%
\providecommand \citenamefont [1]{#1}%
\providecommand \href@noop [0]{\@secondoftwo}%
\providecommand \href [0]{\begingroup \@sanitize@url \@href}%
\providecommand \@href[1]{\@@startlink{#1}\@@href}%
\providecommand \@@href[1]{\endgroup#1\@@endlink}%
\providecommand \@sanitize@url [0]{\catcode `\\12\catcode `\$12\catcode
  `\&12\catcode `\#12\catcode `\^12\catcode `\_12\catcode `\%12\relax}%
\providecommand \@@startlink[1]{}%
\providecommand \@@endlink[0]{}%
\providecommand \url  [0]{\begingroup\@sanitize@url \@url }%
\providecommand \@url [1]{\endgroup\@href {#1}{\urlprefix }}%
\providecommand \urlprefix  [0]{URL }%
\providecommand \Eprint [0]{\href }%
\providecommand \doibase [0]{http://dx.doi.org/}%
\providecommand \selectlanguage [0]{\@gobble}%
\providecommand \bibinfo  [0]{\@secondoftwo}%
\providecommand \bibfield  [0]{\@secondoftwo}%
\providecommand \translation [1]{[#1]}%
\providecommand \BibitemOpen [0]{}%
\providecommand \bibitemStop [0]{}%
\providecommand \bibitemNoStop [0]{.\EOS\space}%
\providecommand \EOS [0]{\spacefactor3000\relax}%
\providecommand \BibitemShut  [1]{\csname bibitem#1\endcsname}%
\let\auto@bib@innerbib\@empty
%</preamble>
\bibitem [{\citenamefont {Aubin}\ \emph
  {et~al.}(2004{\natexlab{a}})\citenamefont {Aubin}, \citenamefont {Bernard},
  \citenamefont {DeTar}, \citenamefont {Osborn}, \citenamefont {Gottlieb},
  \citenamefont {Gregory}, \citenamefont {Toussaint}, \citenamefont {Heller},
  \citenamefont {Hetrick},\ and\ \citenamefont {Sugar}}]{MILC:2004qnl}%
  \BibitemOpen
  \bibfield  {author} {\bibinfo {author} {\bibfnamefont {C.}~\bibnamefont
  {Aubin}}, \bibinfo {author} {\bibfnamefont {C.}~\bibnamefont {Bernard}},
  \bibinfo {author} {\bibfnamefont {C.~E.}\ \bibnamefont {DeTar}}, \bibinfo
  {author} {\bibfnamefont {J.}~\bibnamefont {Osborn}}, \bibinfo {author}
  {\bibfnamefont {S.}~\bibnamefont {Gottlieb}}, \bibinfo {author}
  {\bibfnamefont {E.~B.}\ \bibnamefont {Gregory}}, \bibinfo {author}
  {\bibfnamefont {D.}~\bibnamefont {Toussaint}}, \bibinfo {author}
  {\bibfnamefont {U.~M.}\ \bibnamefont {Heller}}, \bibinfo {author}
  {\bibfnamefont {J.~E.}\ \bibnamefont {Hetrick}}, \ and\ \bibinfo {author}
  {\bibfnamefont {R.}~\bibnamefont {Sugar}} (\bibinfo {collaboration} {MILC}),\
  }\href {\doibase 10.1103/PhysRevD.70.114501} {\bibfield  {journal} {\bibinfo
  {journal} {Phys. Rev. D}\ }\textbf {\bibinfo {volume} {70}},\ \bibinfo
  {pages} {114501} (\bibinfo {year} {2004}{\natexlab{a}})},\ \Eprint
  {http://arxiv.org/abs/hep-lat/0407028} {arXiv:hep-lat/0407028} \BibitemShut
  {NoStop}%
\bibitem [{\citenamefont {Aubin}\ \emph
  {et~al.}(2004{\natexlab{b}})\citenamefont {Aubin} \emph
  {et~al.}}]{HPQCD:2004hdp}%
  \BibitemOpen
  \bibfield  {author} {\bibinfo {author} {\bibfnamefont {C.}~\bibnamefont
  {Aubin}} \emph {et~al.} (\bibinfo {collaboration} {HPQCD, MILC, UKQCD}),\
  }\href {\doibase 10.1103/PhysRevD.70.031504} {\bibfield  {journal} {\bibinfo
  {journal} {Phys. Rev. D}\ }\textbf {\bibinfo {volume} {70}},\ \bibinfo
  {pages} {031504} (\bibinfo {year} {2004}{\natexlab{b}})},\ \Eprint
  {http://arxiv.org/abs/hep-lat/0405022} {arXiv:hep-lat/0405022} \BibitemShut
  {NoStop}%
\bibitem [{\citenamefont {Mason}\ \emph {et~al.}(2006)\citenamefont {Mason},
  \citenamefont {Trottier}, \citenamefont {Horgan}, \citenamefont {Davies},\
  and\ \citenamefont {Lepage}}]{Mason:2005bj}%
  \BibitemOpen
  \bibfield  {author} {\bibinfo {author} {\bibfnamefont {Q.}~\bibnamefont
  {Mason}}, \bibinfo {author} {\bibfnamefont {H.~D.}\ \bibnamefont {Trottier}},
  \bibinfo {author} {\bibfnamefont {R.}~\bibnamefont {Horgan}}, \bibinfo
  {author} {\bibfnamefont {C.~T.~H.}\ \bibnamefont {Davies}}, \ and\ \bibinfo
  {author} {\bibfnamefont {G.~P.}\ \bibnamefont {Lepage}} (\bibinfo
  {collaboration} {HPQCD}),\ }\href {\doibase 10.1103/PhysRevD.73.114501}
  {\bibfield  {journal} {\bibinfo  {journal} {Phys. Rev. D}\ }\textbf {\bibinfo
  {volume} {73}},\ \bibinfo {pages} {114501} (\bibinfo {year} {2006})},\
  \Eprint {http://arxiv.org/abs/hep-ph/0511160} {arXiv:hep-ph/0511160}
  \BibitemShut {NoStop}%
\bibitem [{\citenamefont {Bazavov}\ \emph {et~al.}(2009)\citenamefont {Bazavov}
  \emph {et~al.}}]{MILC:2009ltw}%
  \BibitemOpen
  \bibfield  {author} {\bibinfo {author} {\bibfnamefont {A.}~\bibnamefont
  {Bazavov}} \emph {et~al.} (\bibinfo {collaboration} {MILC}),\ }\href
  {\doibase 10.22323/1.086.0007} {\bibfield  {journal} {\bibinfo  {journal}
  {PoS}\ }\textbf {\bibinfo {volume} {CD09}},\ \bibinfo {pages} {007} (\bibinfo
  {year} {2009})},\ \Eprint {http://arxiv.org/abs/0910.2966} {arXiv:0910.2966
  [hep-ph]} \BibitemShut {NoStop}%
\bibitem [{\citenamefont {Davies}\ \emph {et~al.}(2010)\citenamefont {Davies},
  \citenamefont {McNeile}, \citenamefont {Wong}, \citenamefont {Follana},
  \citenamefont {Horgan}, \citenamefont {Hornbostel}, \citenamefont {Lepage},
  \citenamefont {Shigemitsu},\ and\ \citenamefont {Trottier}}]{Davies:2009ih}%
  \BibitemOpen
  \bibfield  {author} {\bibinfo {author} {\bibfnamefont {C.~T.~H.}\
  \bibnamefont {Davies}}, \bibinfo {author} {\bibfnamefont {C.}~\bibnamefont
  {McNeile}}, \bibinfo {author} {\bibfnamefont {K.~Y.}\ \bibnamefont {Wong}},
  \bibinfo {author} {\bibfnamefont {E.}~\bibnamefont {Follana}}, \bibinfo
  {author} {\bibfnamefont {R.}~\bibnamefont {Horgan}}, \bibinfo {author}
  {\bibfnamefont {K.}~\bibnamefont {Hornbostel}}, \bibinfo {author}
  {\bibfnamefont {G.~P.}\ \bibnamefont {Lepage}}, \bibinfo {author}
  {\bibfnamefont {J.}~\bibnamefont {Shigemitsu}}, \ and\ \bibinfo {author}
  {\bibfnamefont {H.}~\bibnamefont {Trottier}},\ }\href {\doibase
  10.1103/PhysRevLett.104.132003} {\bibfield  {journal} {\bibinfo  {journal}
  {Phys. Rev. Lett.}\ }\textbf {\bibinfo {volume} {104}},\ \bibinfo {pages}
  {132003} (\bibinfo {year} {2010})},\ \Eprint {http://arxiv.org/abs/0910.3102}
  {arXiv:0910.3102 [hep-ph]} \BibitemShut {NoStop}%
\bibitem [{\citenamefont {Bazavov}\ \emph {et~al.}(2010)\citenamefont {Bazavov}
  \emph {et~al.}}]{MILC:2009mpl}%
  \BibitemOpen
  \bibfield  {author} {\bibinfo {author} {\bibfnamefont {A.}~\bibnamefont
  {Bazavov}} \emph {et~al.} (\bibinfo {collaboration} {MILC}),\ }\href
  {\doibase 10.1103/RevModPhys.82.1349} {\bibfield  {journal} {\bibinfo
  {journal} {Rev. Mod. Phys.}\ }\textbf {\bibinfo {volume} {82}},\ \bibinfo
  {pages} {1349} (\bibinfo {year} {2010})},\ \Eprint
  {http://arxiv.org/abs/0903.3598} {arXiv:0903.3598 [hep-lat]} \BibitemShut
  {NoStop}%
\bibitem [{\citenamefont {Laiho}\ and\ \citenamefont {Van~de
  Water}(2011)}]{Laiho:2011np}%
  \BibitemOpen
  \bibfield  {author} {\bibinfo {author} {\bibfnamefont {J.}~\bibnamefont
  {Laiho}}\ and\ \bibinfo {author} {\bibfnamefont {R.~S.}\ \bibnamefont {Van~de
  Water}},\ }\href {\doibase 10.22323/1.139.0293} {\bibfield  {journal}
  {\bibinfo  {journal} {PoS}\ }\textbf {\bibinfo {volume} {LATTICE2011}},\
  \bibinfo {pages} {293} (\bibinfo {year} {2011})},\ \Eprint
  {http://arxiv.org/abs/1112.4861} {arXiv:1112.4861 [hep-lat]} \BibitemShut
  {NoStop}%
\bibitem [{\citenamefont {McNeile}\ \emph {et~al.}(2010)\citenamefont
  {McNeile}, \citenamefont {Davies}, \citenamefont {Follana}, \citenamefont
  {Hornbostel},\ and\ \citenamefont {Lepage}}]{McNeile:2010ji}%
  \BibitemOpen
  \bibfield  {author} {\bibinfo {author} {\bibfnamefont {C.}~\bibnamefont
  {McNeile}}, \bibinfo {author} {\bibfnamefont {C.~T.~H.}\ \bibnamefont
  {Davies}}, \bibinfo {author} {\bibfnamefont {E.}~\bibnamefont {Follana}},
  \bibinfo {author} {\bibfnamefont {K.}~\bibnamefont {Hornbostel}}, \ and\
  \bibinfo {author} {\bibfnamefont {G.~P.}\ \bibnamefont {Lepage}},\ }\href
  {\doibase 10.1103/PhysRevD.82.034512} {\bibfield  {journal} {\bibinfo
  {journal} {Phys. Rev. D}\ }\textbf {\bibinfo {volume} {82}},\ \bibinfo
  {pages} {034512} (\bibinfo {year} {2010})},\ \Eprint
  {http://arxiv.org/abs/1004.4285} {arXiv:1004.4285 [hep-lat]} \BibitemShut
  {NoStop}%
\bibitem [{\citenamefont {Basak}\ \emph {et~al.}(2016)\citenamefont {Basak}
  \emph {et~al.}}]{MILC:2015vfd}%
  \BibitemOpen
  \bibfield  {author} {\bibinfo {author} {\bibfnamefont {S.}~\bibnamefont
  {Basak}} \emph {et~al.} (\bibinfo {collaboration} {MILC}),\ }\href {\doibase
  10.22323/1.251.0259} {\bibfield  {journal} {\bibinfo  {journal} {PoS}\
  }\textbf {\bibinfo {volume} {LATTICE2015}},\ \bibinfo {pages} {259} (\bibinfo
  {year} {2016})},\ \Eprint {http://arxiv.org/abs/1606.01228} {arXiv:1606.01228
  [hep-lat]} \BibitemShut {NoStop}%
\bibitem [{\citenamefont {Maezawa}\ and\ \citenamefont
  {Petreczky}(2016)}]{Maezawa:2016vgv}%
  \BibitemOpen
  \bibfield  {author} {\bibinfo {author} {\bibfnamefont {Y.}~\bibnamefont
  {Maezawa}}\ and\ \bibinfo {author} {\bibfnamefont {P.}~\bibnamefont
  {Petreczky}},\ }\href {\doibase 10.1103/PhysRevD.94.034507} {\bibfield
  {journal} {\bibinfo  {journal} {Phys. Rev. D}\ }\textbf {\bibinfo {volume}
  {94}},\ \bibinfo {pages} {034507} (\bibinfo {year} {2016})},\ \Eprint
  {http://arxiv.org/abs/1606.08798} {arXiv:1606.08798 [hep-lat]} \BibitemShut
  {NoStop}%
\bibitem [{\citenamefont {Bazavov}\ \emph
  {et~al.}(2018{\natexlab{a}})\citenamefont {Bazavov} \emph
  {et~al.}}]{Bazavov:2017lyh}%
  \BibitemOpen
  \bibfield  {author} {\bibinfo {author} {\bibfnamefont {A.}~\bibnamefont
  {Bazavov}} \emph {et~al.},\ }\href {\doibase 10.1103/PhysRevD.98.074512}
  {\bibfield  {journal} {\bibinfo  {journal} {Phys. Rev. D}\ }\textbf {\bibinfo
  {volume} {98}},\ \bibinfo {pages} {074512} (\bibinfo {year}
  {2018}{\natexlab{a}})},\ \Eprint {http://arxiv.org/abs/1712.09262}
  {arXiv:1712.09262 [hep-lat]} \BibitemShut {NoStop}%
\bibitem [{\citenamefont {Bazavov}\ \emph
  {et~al.}(2018{\natexlab{b}})\citenamefont {Bazavov} \emph
  {et~al.}}]{FermilabLattice:2018est}%
  \BibitemOpen
  \bibfield  {author} {\bibinfo {author} {\bibfnamefont {A.}~\bibnamefont
  {Bazavov}} \emph {et~al.} (\bibinfo {collaboration} {Fermilab Lattice, MILC,
  TUMQCD}),\ }\href {\doibase 10.1103/PhysRevD.98.054517} {\bibfield  {journal}
  {\bibinfo  {journal} {Phys. Rev. D}\ }\textbf {\bibinfo {volume} {98}},\
  \bibinfo {pages} {054517} (\bibinfo {year} {2018}{\natexlab{b}})},\ \Eprint
  {http://arxiv.org/abs/1802.04248} {arXiv:1802.04248 [hep-lat]} \BibitemShut
  {NoStop}%
\bibitem [{\citenamefont {Basak}\ \emph {et~al.}(2019)\citenamefont {Basak}
  \emph {et~al.}}]{MILC:2018ddw}%
  \BibitemOpen
  \bibfield  {author} {\bibinfo {author} {\bibfnamefont {S.}~\bibnamefont
  {Basak}} \emph {et~al.} (\bibinfo {collaboration} {MILC}),\ }\href {\doibase
  10.1103/PhysRevD.99.034503} {\bibfield  {journal} {\bibinfo  {journal} {Phys.
  Rev. D}\ }\textbf {\bibinfo {volume} {99}},\ \bibinfo {pages} {034503}
  (\bibinfo {year} {2019})},\ \Eprint {http://arxiv.org/abs/1807.05556}
  {arXiv:1807.05556 [hep-lat]} \BibitemShut {NoStop}%
\bibitem [{\citenamefont {Allton}\ \emph {et~al.}(2008)\citenamefont {Allton}
  \emph {et~al.}}]{RBC-UKQCD:2008mhs}%
  \BibitemOpen
  \bibfield  {author} {\bibinfo {author} {\bibfnamefont {C.}~\bibnamefont
  {Allton}} \emph {et~al.} (\bibinfo {collaboration} {RBC-UKQCD}),\ }\href
  {\doibase 10.1103/PhysRevD.78.114509} {\bibfield  {journal} {\bibinfo
  {journal} {Phys. Rev. D}\ }\textbf {\bibinfo {volume} {78}},\ \bibinfo
  {pages} {114509} (\bibinfo {year} {2008})},\ \Eprint
  {http://arxiv.org/abs/0804.0473} {arXiv:0804.0473 [hep-lat]} \BibitemShut
  {NoStop}%
\bibitem [{\citenamefont {Aoki}\ \emph {et~al.}(2011)\citenamefont {Aoki} \emph
  {et~al.}}]{RBC:2010qam}%
  \BibitemOpen
  \bibfield  {author} {\bibinfo {author} {\bibfnamefont {Y.}~\bibnamefont
  {Aoki}} \emph {et~al.} (\bibinfo {collaboration} {RBC, UKQCD}),\ }\href
  {\doibase 10.1103/PhysRevD.83.074508} {\bibfield  {journal} {\bibinfo
  {journal} {Phys. Rev. D}\ }\textbf {\bibinfo {volume} {83}},\ \bibinfo
  {pages} {074508} (\bibinfo {year} {2011})},\ \Eprint
  {http://arxiv.org/abs/1011.0892} {arXiv:1011.0892 [hep-lat]} \BibitemShut
  {NoStop}%
\bibitem [{\citenamefont {Blum}\ \emph {et~al.}(2010)\citenamefont {Blum},
  \citenamefont {Zhou}, \citenamefont {Doi}, \citenamefont {Hayakawa},
  \citenamefont {Izubuchi}, \citenamefont {Uno},\ and\ \citenamefont
  {Yamada}}]{Blum:2010ym}%
  \BibitemOpen
  \bibfield  {author} {\bibinfo {author} {\bibfnamefont {T.}~\bibnamefont
  {Blum}}, \bibinfo {author} {\bibfnamefont {R.}~\bibnamefont {Zhou}}, \bibinfo
  {author} {\bibfnamefont {T.}~\bibnamefont {Doi}}, \bibinfo {author}
  {\bibfnamefont {M.}~\bibnamefont {Hayakawa}}, \bibinfo {author}
  {\bibfnamefont {T.}~\bibnamefont {Izubuchi}}, \bibinfo {author}
  {\bibfnamefont {S.}~\bibnamefont {Uno}}, \ and\ \bibinfo {author}
  {\bibfnamefont {N.}~\bibnamefont {Yamada}},\ }\href {\doibase
  10.1103/PhysRevD.82.094508} {\bibfield  {journal} {\bibinfo  {journal} {Phys.
  Rev. D}\ }\textbf {\bibinfo {volume} {82}},\ \bibinfo {pages} {094508}
  (\bibinfo {year} {2010})},\ \Eprint {http://arxiv.org/abs/1006.1311}
  {arXiv:1006.1311 [hep-lat]} \BibitemShut {NoStop}%
\bibitem [{\citenamefont {Arthur}\ \emph {et~al.}(2013)\citenamefont {Arthur}
  \emph {et~al.}}]{RBC:2012cbl}%
  \BibitemOpen
  \bibfield  {author} {\bibinfo {author} {\bibfnamefont {R.}~\bibnamefont
  {Arthur}} \emph {et~al.} (\bibinfo {collaboration} {RBC, UKQCD}),\ }\href
  {\doibase 10.1103/PhysRevD.87.094514} {\bibfield  {journal} {\bibinfo
  {journal} {Phys. Rev. D}\ }\textbf {\bibinfo {volume} {87}},\ \bibinfo
  {pages} {094514} (\bibinfo {year} {2013})},\ \Eprint
  {http://arxiv.org/abs/1208.4412} {arXiv:1208.4412 [hep-lat]} \BibitemShut
  {NoStop}%
\bibitem [{\citenamefont {Blum}\ \emph {et~al.}(2016)\citenamefont {Blum} \emph
  {et~al.}}]{RBC:2014ntl}%
  \BibitemOpen
  \bibfield  {author} {\bibinfo {author} {\bibfnamefont {T.}~\bibnamefont
  {Blum}} \emph {et~al.} (\bibinfo {collaboration} {RBC, UKQCD}),\ }\href
  {\doibase 10.1103/PhysRevD.93.074505} {\bibfield  {journal} {\bibinfo
  {journal} {Phys. Rev. D}\ }\textbf {\bibinfo {volume} {93}},\ \bibinfo
  {pages} {074505} (\bibinfo {year} {2016})},\ \Eprint
  {http://arxiv.org/abs/1411.7017} {arXiv:1411.7017 [hep-lat]} \BibitemShut
  {NoStop}%
\bibitem [{\citenamefont {Yang}\ \emph
  {et~al.}(2015{\natexlab{a}})\citenamefont {Yang} \emph
  {et~al.}}]{Yang:2014sea}%
  \BibitemOpen
  \bibfield  {author} {\bibinfo {author} {\bibfnamefont {Y.-B.}\ \bibnamefont
  {Yang}} \emph {et~al.},\ }\href {\doibase 10.1103/PhysRevD.92.034517}
  {\bibfield  {journal} {\bibinfo  {journal} {Phys. Rev.}\ }\textbf {\bibinfo
  {volume} {D92}},\ \bibinfo {pages} {034517} (\bibinfo {year}
  {2015}{\natexlab{a}})},\ \Eprint {http://arxiv.org/abs/1410.3343}
  {arXiv:1410.3343 [hep-lat]} \BibitemShut {NoStop}%
%%CITATION = ARXIV:1410.3343;%%
\bibitem [{\citenamefont {Ishikawa}\ \emph {et~al.}(2008)\citenamefont
  {Ishikawa} \emph {et~al.}}]{JLQCD:2007xff}%
  \BibitemOpen
  \bibfield  {author} {\bibinfo {author} {\bibfnamefont {T.}~\bibnamefont
  {Ishikawa}} \emph {et~al.} (\bibinfo {collaboration} {JLQCD}),\ }\href
  {\doibase 10.1103/PhysRevD.78.011502} {\bibfield  {journal} {\bibinfo
  {journal} {Phys. Rev. D}\ }\textbf {\bibinfo {volume} {78}},\ \bibinfo
  {pages} {011502} (\bibinfo {year} {2008})},\ \Eprint
  {http://arxiv.org/abs/0704.1937} {arXiv:0704.1937 [hep-lat]} \BibitemShut
  {NoStop}%
\bibitem [{\citenamefont {Sint}(2011)}]{Sint:2010eh}%
  \BibitemOpen
  \bibfield  {author} {\bibinfo {author} {\bibfnamefont {S.}~\bibnamefont
  {Sint}},\ }\href {\doibase 10.1016/j.nuclphysb.2011.02.002} {\bibfield
  {journal} {\bibinfo  {journal} {Nucl. Phys. B}\ }\textbf {\bibinfo {volume}
  {847}},\ \bibinfo {pages} {491} (\bibinfo {year} {2011})},\ \Eprint
  {http://arxiv.org/abs/1008.4857} {arXiv:1008.4857 [hep-lat]} \BibitemShut
  {NoStop}%
\bibitem [{\citenamefont {Aoki}\ \emph {et~al.}(2009)\citenamefont {Aoki} \emph
  {et~al.}}]{PACS-CS:2008bkb}%
  \BibitemOpen
  \bibfield  {author} {\bibinfo {author} {\bibfnamefont {S.}~\bibnamefont
  {Aoki}} \emph {et~al.} (\bibinfo {collaboration} {PACS-CS}),\ }\href
  {\doibase 10.1103/PhysRevD.79.034503} {\bibfield  {journal} {\bibinfo
  {journal} {Phys. Rev. D}\ }\textbf {\bibinfo {volume} {79}},\ \bibinfo
  {pages} {034503} (\bibinfo {year} {2009})},\ \Eprint
  {http://arxiv.org/abs/0807.1661} {arXiv:0807.1661 [hep-lat]} \BibitemShut
  {NoStop}%
\bibitem [{\citenamefont {Aoki}\ \emph
  {et~al.}(2010{\natexlab{a}})\citenamefont {Aoki} \emph
  {et~al.}}]{PACS-CS:2009sof}%
  \BibitemOpen
  \bibfield  {author} {\bibinfo {author} {\bibfnamefont {S.}~\bibnamefont
  {Aoki}} \emph {et~al.} (\bibinfo {collaboration} {PACS-CS}),\ }\href
  {\doibase 10.1103/PhysRevD.81.074503} {\bibfield  {journal} {\bibinfo
  {journal} {Phys. Rev. D}\ }\textbf {\bibinfo {volume} {81}},\ \bibinfo
  {pages} {074503} (\bibinfo {year} {2010}{\natexlab{a}})},\ \Eprint
  {http://arxiv.org/abs/0911.2561} {arXiv:0911.2561 [hep-lat]} \BibitemShut
  {NoStop}%
\bibitem [{\citenamefont {Aoki}\ \emph
  {et~al.}(2010{\natexlab{b}})\citenamefont {Aoki} \emph
  {et~al.}}]{PACS-CS:2010gyf}%
  \BibitemOpen
  \bibfield  {author} {\bibinfo {author} {\bibfnamefont {S.}~\bibnamefont
  {Aoki}} \emph {et~al.} (\bibinfo {collaboration} {PACS-CS}),\ }\href
  {\doibase 10.1007/JHEP08(2010)101} {\bibfield  {journal} {\bibinfo  {journal}
  {JHEP}\ }\textbf {\bibinfo {volume} {08}},\ \bibinfo {pages} {101} (\bibinfo
  {year} {2010}{\natexlab{b}})},\ \Eprint {http://arxiv.org/abs/1006.1164}
  {arXiv:1006.1164 [hep-lat]} \BibitemShut {NoStop}%
\bibitem [{\citenamefont {Aoki}\ \emph {et~al.}(2012)\citenamefont {Aoki} \emph
  {et~al.}}]{Aoki:2012st}%
  \BibitemOpen
  \bibfield  {author} {\bibinfo {author} {\bibfnamefont {S.}~\bibnamefont
  {Aoki}} \emph {et~al.},\ }\href {\doibase 10.1103/PhysRevD.86.034507}
  {\bibfield  {journal} {\bibinfo  {journal} {Phys. Rev. D}\ }\textbf {\bibinfo
  {volume} {86}},\ \bibinfo {pages} {034507} (\bibinfo {year} {2012})},\
  \Eprint {http://arxiv.org/abs/1205.2961} {arXiv:1205.2961 [hep-lat]}
  \BibitemShut {NoStop}%
\bibitem [{\citenamefont {Aoki}\ \emph {et~al.}(2022)\citenamefont {Aoki} \emph
  {et~al.}}]{FlavourLatticeAveragingGroupFLAG:2021npn}%
  \BibitemOpen
  \bibfield  {author} {\bibinfo {author} {\bibfnamefont {Y.}~\bibnamefont
  {Aoki}} \emph {et~al.} (\bibinfo {collaboration} {Flavour Lattice Averaging
  Group (FLAG)}),\ }\href {\doibase 10.1140/epjc/s10052-022-10536-1} {\bibfield
   {journal} {\bibinfo  {journal} {Eur. Phys. J. C}\ }\textbf {\bibinfo
  {volume} {82}},\ \bibinfo {pages} {869} (\bibinfo {year} {2022})},\ \Eprint
  {http://arxiv.org/abs/2111.09849} {arXiv:2111.09849 [hep-lat]} \BibitemShut
  {NoStop}%
\bibitem [{\citenamefont {Bruno}\ \emph {et~al.}(2020)\citenamefont {Bruno},
  \citenamefont {Campos}, \citenamefont {Fritzsch}, \citenamefont {Koponen},
  \citenamefont {Pena}, \citenamefont {Preti}, \citenamefont {Ramos},\ and\
  \citenamefont {Vladikas}}]{Bruno:2019vup}%
  \BibitemOpen
  \bibfield  {author} {\bibinfo {author} {\bibfnamefont {M.}~\bibnamefont
  {Bruno}}, \bibinfo {author} {\bibfnamefont {I.}~\bibnamefont {Campos}},
  \bibinfo {author} {\bibfnamefont {P.}~\bibnamefont {Fritzsch}}, \bibinfo
  {author} {\bibfnamefont {J.}~\bibnamefont {Koponen}}, \bibinfo {author}
  {\bibfnamefont {C.}~\bibnamefont {Pena}}, \bibinfo {author} {\bibfnamefont
  {D.}~\bibnamefont {Preti}}, \bibinfo {author} {\bibfnamefont
  {A.}~\bibnamefont {Ramos}}, \ and\ \bibinfo {author} {\bibfnamefont
  {A.}~\bibnamefont {Vladikas}} (\bibinfo {collaboration} {ALPHA}),\ }\href
  {\doibase 10.1140/epjc/s10052-020-7698-z} {\bibfield  {journal} {\bibinfo
  {journal} {Eur. Phys. J. C}\ }\textbf {\bibinfo {volume} {80}},\ \bibinfo
  {pages} {169} (\bibinfo {year} {2020})},\ \Eprint
  {http://arxiv.org/abs/1911.08025} {arXiv:1911.08025 [hep-lat]} \BibitemShut
  {NoStop}%
\bibitem [{\citenamefont {Durr}\ \emph {et~al.}(2011)\citenamefont {Durr},
  \citenamefont {Fodor}, \citenamefont {Hoelbling}, \citenamefont {Katz},
  \citenamefont {Krieg}, \citenamefont {Kurth}, \citenamefont {Lellouch},
  \citenamefont {Lippert}, \citenamefont {Szabo},\ and\ \citenamefont
  {Vulvert}}]{BMW:2010skj}%
  \BibitemOpen
  \bibfield  {author} {\bibinfo {author} {\bibfnamefont {S.}~\bibnamefont
  {Durr}}, \bibinfo {author} {\bibfnamefont {Z.}~\bibnamefont {Fodor}},
  \bibinfo {author} {\bibfnamefont {C.}~\bibnamefont {Hoelbling}}, \bibinfo
  {author} {\bibfnamefont {S.~D.}\ \bibnamefont {Katz}}, \bibinfo {author}
  {\bibfnamefont {S.}~\bibnamefont {Krieg}}, \bibinfo {author} {\bibfnamefont
  {T.}~\bibnamefont {Kurth}}, \bibinfo {author} {\bibfnamefont
  {L.}~\bibnamefont {Lellouch}}, \bibinfo {author} {\bibfnamefont
  {T.}~\bibnamefont {Lippert}}, \bibinfo {author} {\bibfnamefont {K.~K.}\
  \bibnamefont {Szabo}}, \ and\ \bibinfo {author} {\bibfnamefont
  {G.}~\bibnamefont {Vulvert}} (\bibinfo {collaboration} {BMW}),\ }\href
  {\doibase 10.1007/JHEP08(2011)148} {\bibfield  {journal} {\bibinfo  {journal}
  {JHEP}\ }\textbf {\bibinfo {volume} {08}},\ \bibinfo {pages} {148} (\bibinfo
  {year} {2011})},\ \Eprint {http://arxiv.org/abs/1011.2711} {arXiv:1011.2711
  [hep-lat]} \BibitemShut {NoStop}%
\bibitem [{\citenamefont {Martinelli}\ \emph {et~al.}(1995)\citenamefont
  {Martinelli}, \citenamefont {Pittori}, \citenamefont {Sachrajda},
  \citenamefont {Testa},\ and\ \citenamefont {Vladikas}}]{Martinelli:1994ty}%
  \BibitemOpen
  \bibfield  {author} {\bibinfo {author} {\bibfnamefont {G.}~\bibnamefont
  {Martinelli}}, \bibinfo {author} {\bibfnamefont {C.}~\bibnamefont {Pittori}},
  \bibinfo {author} {\bibfnamefont {C.~T.}\ \bibnamefont {Sachrajda}}, \bibinfo
  {author} {\bibfnamefont {M.}~\bibnamefont {Testa}}, \ and\ \bibinfo {author}
  {\bibfnamefont {A.}~\bibnamefont {Vladikas}},\ }\href {\doibase
  10.1016/0550-3213(95)00126-D} {\bibfield  {journal} {\bibinfo  {journal}
  {Nucl. Phys.}\ }\textbf {\bibinfo {volume} {B445}},\ \bibinfo {pages} {81}
  (\bibinfo {year} {1995})},\ \Eprint {http://arxiv.org/abs/hep-lat/9411010}
  {arXiv:hep-lat/9411010 [hep-lat]} \BibitemShut {NoStop}%
%%CITATION = HEP-LAT/9411010;%%
\bibitem [{\citenamefont {Aoki}\ \emph {et~al.}(2008)\citenamefont {Aoki} \emph
  {et~al.}}]{Aoki:2007xm}%
  \BibitemOpen
  \bibfield  {author} {\bibinfo {author} {\bibfnamefont {Y.}~\bibnamefont
  {Aoki}} \emph {et~al.},\ }\href {\doibase 10.1103/PhysRevD.78.054510}
  {\bibfield  {journal} {\bibinfo  {journal} {Phys. Rev.}\ }\textbf {\bibinfo
  {volume} {D78}},\ \bibinfo {pages} {054510} (\bibinfo {year} {2008})},\
  \Eprint {http://arxiv.org/abs/0712.1061} {arXiv:0712.1061 [hep-lat]}
  \BibitemShut {NoStop}%
%%CITATION = ARXIV:0712.1061;%%
\bibitem [{\citenamefont {Sturm}\ \emph {et~al.}(2009)\citenamefont {Sturm},
  \citenamefont {Aoki}, \citenamefont {Christ}, \citenamefont {Izubuchi},
  \citenamefont {Sachrajda},\ and\ \citenamefont {Soni}}]{Sturm:2009kb}%
  \BibitemOpen
  \bibfield  {author} {\bibinfo {author} {\bibfnamefont {C.}~\bibnamefont
  {Sturm}}, \bibinfo {author} {\bibfnamefont {Y.}~\bibnamefont {Aoki}},
  \bibinfo {author} {\bibfnamefont {N.~H.}\ \bibnamefont {Christ}}, \bibinfo
  {author} {\bibfnamefont {T.}~\bibnamefont {Izubuchi}}, \bibinfo {author}
  {\bibfnamefont {C.~T.~C.}\ \bibnamefont {Sachrajda}}, \ and\ \bibinfo
  {author} {\bibfnamefont {A.}~\bibnamefont {Soni}},\ }\href {\doibase
  10.1103/PhysRevD.80.014501} {\bibfield  {journal} {\bibinfo  {journal} {Phys.
  Rev.}\ }\textbf {\bibinfo {volume} {D80}},\ \bibinfo {pages} {014501}
  (\bibinfo {year} {2009})},\ \Eprint {http://arxiv.org/abs/0901.2599}
  {arXiv:0901.2599 [hep-ph]} \BibitemShut {NoStop}%
%%CITATION = ARXIV:0901.2599;%%
\bibitem [{\citenamefont {Hasan}\ \emph {et~al.}(2019)\citenamefont {Hasan},
  \citenamefont {Green}, \citenamefont {Meinel}, \citenamefont {Engelhardt},
  \citenamefont {Krieg}, \citenamefont {Negele}, \citenamefont {Pochinsky},\
  and\ \citenamefont {Syritsyn}}]{Hasan:2019noy}%
  \BibitemOpen
  \bibfield  {author} {\bibinfo {author} {\bibfnamefont {N.}~\bibnamefont
  {Hasan}}, \bibinfo {author} {\bibfnamefont {J.}~\bibnamefont {Green}},
  \bibinfo {author} {\bibfnamefont {S.}~\bibnamefont {Meinel}}, \bibinfo
  {author} {\bibfnamefont {M.}~\bibnamefont {Engelhardt}}, \bibinfo {author}
  {\bibfnamefont {S.}~\bibnamefont {Krieg}}, \bibinfo {author} {\bibfnamefont
  {J.}~\bibnamefont {Negele}}, \bibinfo {author} {\bibfnamefont
  {A.}~\bibnamefont {Pochinsky}}, \ and\ \bibinfo {author} {\bibfnamefont
  {S.}~\bibnamefont {Syritsyn}},\ }\href {\doibase 10.1103/PhysRevD.99.114505}
  {\bibfield  {journal} {\bibinfo  {journal} {Phys. Rev. D}\ }\textbf {\bibinfo
  {volume} {99}},\ \bibinfo {pages} {114505} (\bibinfo {year} {2019})},\
  \Eprint {http://arxiv.org/abs/1903.06487} {arXiv:1903.06487 [hep-lat]}
  \BibitemShut {NoStop}%
\bibitem [{\citenamefont {Carrasco}\ \emph {et~al.}(2014)\citenamefont
  {Carrasco} \emph {et~al.}}]{EuropeanTwistedMass:2014osg}%
  \BibitemOpen
  \bibfield  {author} {\bibinfo {author} {\bibfnamefont {N.}~\bibnamefont
  {Carrasco}} \emph {et~al.} (\bibinfo {collaboration} {European Twisted
  Mass}),\ }\href {\doibase 10.1016/j.nuclphysb.2014.07.025} {\bibfield
  {journal} {\bibinfo  {journal} {Nucl. Phys. B}\ }\textbf {\bibinfo {volume}
  {887}},\ \bibinfo {pages} {19} (\bibinfo {year} {2014})},\ \Eprint
  {http://arxiv.org/abs/1403.4504} {arXiv:1403.4504 [hep-lat]} \BibitemShut
  {NoStop}%
\bibitem [{\citenamefont {Giusti}\ \emph {et~al.}(2017)\citenamefont {Giusti},
  \citenamefont {Lubicz}, \citenamefont {Tarantino}, \citenamefont
  {Martinelli}, \citenamefont {Sanfilippo}, \citenamefont {Simula},\ and\
  \citenamefont {Tantalo}}]{Giusti:2017dmp}%
  \BibitemOpen
  \bibfield  {author} {\bibinfo {author} {\bibfnamefont {D.}~\bibnamefont
  {Giusti}}, \bibinfo {author} {\bibfnamefont {V.}~\bibnamefont {Lubicz}},
  \bibinfo {author} {\bibfnamefont {C.}~\bibnamefont {Tarantino}}, \bibinfo
  {author} {\bibfnamefont {G.}~\bibnamefont {Martinelli}}, \bibinfo {author}
  {\bibfnamefont {F.}~\bibnamefont {Sanfilippo}}, \bibinfo {author}
  {\bibfnamefont {S.}~\bibnamefont {Simula}}, \ and\ \bibinfo {author}
  {\bibfnamefont {N.}~\bibnamefont {Tantalo}},\ }\href {\doibase
  10.1103/PhysRevD.95.114504} {\bibfield  {journal} {\bibinfo  {journal} {Phys.
  Rev. D}\ }\textbf {\bibinfo {volume} {95}},\ \bibinfo {pages} {114504}
  (\bibinfo {year} {2017})},\ \Eprint {http://arxiv.org/abs/1704.06561}
  {arXiv:1704.06561 [hep-lat]} \BibitemShut {NoStop}%
\bibitem [{\citenamefont {Alexandrou}\ \emph {et~al.}(2021)\citenamefont
  {Alexandrou} \emph {et~al.}}]{ExtendedTwistedMass:2021gbo}%
  \BibitemOpen
  \bibfield  {author} {\bibinfo {author} {\bibfnamefont {C.}~\bibnamefont
  {Alexandrou}} \emph {et~al.} (\bibinfo {collaboration} {Extended Twisted
  Mass}),\ }\href {\doibase 10.1103/PhysRevD.104.074515} {\bibfield  {journal}
  {\bibinfo  {journal} {Phys. Rev. D}\ }\textbf {\bibinfo {volume} {104}},\
  \bibinfo {pages} {074515} (\bibinfo {year} {2021})},\ \Eprint
  {http://arxiv.org/abs/2104.13408} {arXiv:2104.13408 [hep-lat]} \BibitemShut
  {NoStop}%
\bibitem [{\citenamefont {L\"uscher}(2010)}]{Luscher:2010iy}%
  \BibitemOpen
  \bibfield  {author} {\bibinfo {author} {\bibfnamefont {M.}~\bibnamefont
  {L\"uscher}},\ }\href {\doibase 10.1007/JHEP08(2010)071} {\bibfield
  {journal} {\bibinfo  {journal} {JHEP}\ }\textbf {\bibinfo {volume} {08}},\
  \bibinfo {pages} {071} (\bibinfo {year} {2010})},\ \bibinfo {note} {[Erratum:
  JHEP 03, 092 (2014)]},\ \Eprint {http://arxiv.org/abs/1006.4518}
  {arXiv:1006.4518 [hep-lat]} \BibitemShut {NoStop}%
\bibitem [{\citenamefont {Borsanyi}\ \emph {et~al.}(2012)\citenamefont
  {Borsanyi} \emph {et~al.}}]{BMW:2012hcm}%
  \BibitemOpen
  \bibfield  {author} {\bibinfo {author} {\bibfnamefont {S.}~\bibnamefont
  {Borsanyi}} \emph {et~al.} (\bibinfo {collaboration} {BMW}),\ }\href
  {\doibase 10.1007/JHEP09(2012)010} {\bibfield  {journal} {\bibinfo  {journal}
  {JHEP}\ }\textbf {\bibinfo {volume} {09}},\ \bibinfo {pages} {010} (\bibinfo
  {year} {2012})},\ \Eprint {http://arxiv.org/abs/1203.4469} {arXiv:1203.4469
  [hep-lat]} \BibitemShut {NoStop}%
\bibitem [{ref()}]{ref_sm}%
  \BibitemOpen
  \href@noop {} {\bibinfo  {journal} {Supplementary materials}\ }\BibitemShut
  {NoStop}%
\bibitem [{\citenamefont {He}\ \emph {et~al.}(2022)\citenamefont {He},
  \citenamefont {Bi}, \citenamefont {Draper}, \citenamefont {Liu},
  \citenamefont {Liu},\ and\ \citenamefont {Yang}}]{He:2022lse}%
  \BibitemOpen
\bibfield  {journal} {  }\bibfield  {author} {\bibinfo {author} {\bibfnamefont
  {F.}~\bibnamefont {He}}, \bibinfo {author} {\bibfnamefont {Y.-J.}\
  \bibnamefont {Bi}}, \bibinfo {author} {\bibfnamefont {T.}~\bibnamefont
  {Draper}}, \bibinfo {author} {\bibfnamefont {K.-F.}\ \bibnamefont {Liu}},
  \bibinfo {author} {\bibfnamefont {Z.}~\bibnamefont {Liu}}, \ and\ \bibinfo
  {author} {\bibfnamefont {Y.-B.}\ \bibnamefont {Yang}} (\bibinfo
  {collaboration} {\ensuremath{\chi}QCD}),\ }\href {\doibase
  10.1103/PhysRevD.106.114506} {\bibfield  {journal} {\bibinfo  {journal}
  {Phys. Rev. D}\ }\textbf {\bibinfo {volume} {106}},\ \bibinfo {pages}
  {114506} (\bibinfo {year} {2022})},\ \Eprint
  {http://arxiv.org/abs/2204.09246} {arXiv:2204.09246 [hep-lat]} \BibitemShut
  {NoStop}%
\bibitem [{\citenamefont {Feng}\ \emph {et~al.}(2022)\citenamefont {Feng},
  \citenamefont {Jin},\ and\ \citenamefont {Riberdy}}]{Feng:2021zek}%
  \BibitemOpen
  \bibfield  {author} {\bibinfo {author} {\bibfnamefont {X.}~\bibnamefont
  {Feng}}, \bibinfo {author} {\bibfnamefont {L.}~\bibnamefont {Jin}}, \ and\
  \bibinfo {author} {\bibfnamefont {M.~J.}\ \bibnamefont {Riberdy}},\ }\href
  {\doibase 10.1103/PhysRevLett.128.052003} {\bibfield  {journal} {\bibinfo
  {journal} {Phys. Rev. Lett.}\ }\textbf {\bibinfo {volume} {128}},\ \bibinfo
  {pages} {052003} (\bibinfo {year} {2022})},\ \Eprint
  {http://arxiv.org/abs/2108.05311} {arXiv:2108.05311 [hep-lat]} \BibitemShut
  {NoStop}%
\bibitem [{\citenamefont {Zyla}\ \emph {et~al.}(2020)\citenamefont {Zyla} \emph
  {et~al.}}]{ParticleDataGroup:2020ssz}%
  \BibitemOpen
  \bibfield  {author} {\bibinfo {author} {\bibfnamefont {P.~A.}\ \bibnamefont
  {Zyla}} \emph {et~al.} (\bibinfo {collaboration} {Particle Data Group}),\
  }\href {\doibase 10.1093/ptep/ptaa104} {\bibfield  {journal} {\bibinfo
  {journal} {PTEP}\ }\textbf {\bibinfo {volume} {2020}},\ \bibinfo {pages}
  {083C01} (\bibinfo {year} {2020})}\BibitemShut {NoStop}%
\bibitem [{\citenamefont {Sharpe}(1997)}]{Sharpe:1997by}%
  \BibitemOpen
  \bibfield  {author} {\bibinfo {author} {\bibfnamefont {S.~R.}\ \bibnamefont
  {Sharpe}},\ }\href {\doibase 10.1103/PhysRevD.62.099901} {\bibfield
  {journal} {\bibinfo  {journal} {Phys. Rev. D}\ }\textbf {\bibinfo {volume}
  {56}},\ \bibinfo {pages} {7052} (\bibinfo {year} {1997})},\ \bibinfo {note}
  {[Erratum: Phys.Rev.D 62, 099901 (2000)]},\ \Eprint
  {http://arxiv.org/abs/hep-lat/9707018} {arXiv:hep-lat/9707018} \BibitemShut
  {NoStop}%
\bibitem [{\citenamefont {Della~Morte}\ \emph {et~al.}(2005)\citenamefont
  {Della~Morte}, \citenamefont {Hoffmann},\ and\ \citenamefont
  {Sommer}}]{DellaMorte:2005aqe}%
  \BibitemOpen
  \bibfield  {author} {\bibinfo {author} {\bibfnamefont {M.}~\bibnamefont
  {Della~Morte}}, \bibinfo {author} {\bibfnamefont {R.}~\bibnamefont
  {Hoffmann}}, \ and\ \bibinfo {author} {\bibfnamefont {R.}~\bibnamefont
  {Sommer}},\ }\href {\doibase 10.1088/1126-6708/2005/03/029} {\bibfield
  {journal} {\bibinfo  {journal} {JHEP}\ }\textbf {\bibinfo {volume} {03}},\
  \bibinfo {pages} {029} (\bibinfo {year} {2005})},\ \Eprint
  {http://arxiv.org/abs/hep-lat/0503003} {arXiv:hep-lat/0503003} \BibitemShut
  {NoStop}%
\bibitem [{\citenamefont {Edwards}\ and\ \citenamefont
  {Joo}(2005)}]{Edwards:2004sx}%
  \BibitemOpen
  \bibfield  {author} {\bibinfo {author} {\bibfnamefont {R.~G.}\ \bibnamefont
  {Edwards}}\ and\ \bibinfo {author} {\bibfnamefont {B.}~\bibnamefont {Joo}}
  (\bibinfo {collaboration} {SciDAC, LHPC, UKQCD}),\ }\bibfield  {booktitle}
  {\emph {\bibinfo {booktitle} {{Lattice field theory. Proceedings, 22nd
  International Symposium, Lattice 2004, Batavia, USA, June 21-26, 2004}}},\
  }\href {\doibase 10.1016/j.nuclphysbps.2004.11.254} {\bibfield  {journal}
  {\bibinfo  {journal} {Nucl. Phys. Proc. Suppl.}\ }\textbf {\bibinfo {volume}
  {140}},\ \bibinfo {pages} {832} (\bibinfo {year} {2005})},\ \bibinfo {note}
  {[,832(2004)]},\ \Eprint {http://arxiv.org/abs/hep-lat/0409003}
  {arXiv:hep-lat/0409003 [hep-lat]} \BibitemShut {NoStop}%
%%CITATION = HEP-LAT/0409003;%%
\bibitem [{\citenamefont {Clark}\ \emph {et~al.}(2010)\citenamefont {Clark},
  \citenamefont {Babich}, \citenamefont {Barros}, \citenamefont {Brower},\ and\
  \citenamefont {Rebbi}}]{Clark:2009wm}%
  \BibitemOpen
  \bibfield  {author} {\bibinfo {author} {\bibfnamefont {M.~A.}\ \bibnamefont
  {Clark}}, \bibinfo {author} {\bibfnamefont {R.}~\bibnamefont {Babich}},
  \bibinfo {author} {\bibfnamefont {K.}~\bibnamefont {Barros}}, \bibinfo
  {author} {\bibfnamefont {R.~C.}\ \bibnamefont {Brower}}, \ and\ \bibinfo
  {author} {\bibfnamefont {C.}~\bibnamefont {Rebbi}},\ }\href {\doibase
  10.1016/j.cpc.2010.05.002} {\bibfield  {journal} {\bibinfo  {journal}
  {Comput. Phys. Commun.}\ }\textbf {\bibinfo {volume} {181}},\ \bibinfo
  {pages} {1517} (\bibinfo {year} {2010})},\ \Eprint
  {http://arxiv.org/abs/0911.3191} {arXiv:0911.3191 [hep-lat]} \BibitemShut
  {NoStop}%
%%CITATION = ARXIV:0911.3191;%%
\bibitem [{\citenamefont {Babich}\ \emph {et~al.}(2011)\citenamefont {Babich},
  \citenamefont {Clark}, \citenamefont {Joo}, \citenamefont {Shi},
  \citenamefont {Brower},\ and\ \citenamefont {Gottlieb}}]{Babich:2011np}%
  \BibitemOpen
  \bibfield  {author} {\bibinfo {author} {\bibfnamefont {R.}~\bibnamefont
  {Babich}}, \bibinfo {author} {\bibfnamefont {M.~A.}\ \bibnamefont {Clark}},
  \bibinfo {author} {\bibfnamefont {B.}~\bibnamefont {Joo}}, \bibinfo {author}
  {\bibfnamefont {G.}~\bibnamefont {Shi}}, \bibinfo {author} {\bibfnamefont
  {R.~C.}\ \bibnamefont {Brower}}, \ and\ \bibinfo {author} {\bibfnamefont
  {S.}~\bibnamefont {Gottlieb}},\ }in\ \href {\doibase 10.1145/2063384.2063478}
  {\emph {\bibinfo {booktitle} {{SC11 International Conference for High
  Performance Computing, Networking, Storage and Analysis Seattle, Washington,
  November 12-18, 2011}}}}\ (\bibinfo {year} {2011})\ \Eprint
  {http://arxiv.org/abs/1109.2935} {arXiv:1109.2935 [hep-lat]} \BibitemShut
  {NoStop}%
%%CITATION = ARXIV:1109.2935;%%
\bibitem [{\citenamefont {Clark}\ \emph {et~al.}(2016)\citenamefont {Clark},
  \citenamefont {Jo}, \citenamefont {Strelchenko}, \citenamefont {Cheng},
  \citenamefont {Gambhir},\ and\ \citenamefont {Brower}}]{Clark:2016rdz}%
  \BibitemOpen
  \bibfield  {author} {\bibinfo {author} {\bibfnamefont {M.~A.}\ \bibnamefont
  {Clark}}, \bibinfo {author} {\bibfnamefont {B.}~\bibnamefont {Jo}}, \bibinfo
  {author} {\bibfnamefont {A.}~\bibnamefont {Strelchenko}}, \bibinfo {author}
  {\bibfnamefont {M.}~\bibnamefont {Cheng}}, \bibinfo {author} {\bibfnamefont
  {A.}~\bibnamefont {Gambhir}}, \ and\ \bibinfo {author} {\bibfnamefont
  {R.}~\bibnamefont {Brower}},\ }\href@noop {} {\  (\bibinfo {year} {2016})},\
  \Eprint {http://arxiv.org/abs/1612.07873} {arXiv:1612.07873 [hep-lat]}
  \BibitemShut {NoStop}%
%%CITATION = ARXIV:1612.07873;%%
\bibitem [{\citenamefont {Bi}\ \emph {et~al.}(2020)\citenamefont {Bi},
  \citenamefont {Xiao}, \citenamefont {Guo}, \citenamefont {Gong},
  \citenamefont {Sun}, \citenamefont {Xu},\ and\ \citenamefont
  {Yang}}]{Bi:2020wpt}%
  \BibitemOpen
  \bibfield  {author} {\bibinfo {author} {\bibfnamefont {Y.-J.}\ \bibnamefont
  {Bi}}, \bibinfo {author} {\bibfnamefont {Y.}~\bibnamefont {Xiao}}, \bibinfo
  {author} {\bibfnamefont {W.-Y.}\ \bibnamefont {Guo}}, \bibinfo {author}
  {\bibfnamefont {M.}~\bibnamefont {Gong}}, \bibinfo {author} {\bibfnamefont
  {P.}~\bibnamefont {Sun}}, \bibinfo {author} {\bibfnamefont {S.}~\bibnamefont
  {Xu}}, \ and\ \bibinfo {author} {\bibfnamefont {Y.-B.}\ \bibnamefont
  {Yang}},\ }\bibfield  {booktitle} {\emph {\bibinfo {booktitle} {{Proceedings,
  37th International Symposium on Lattice Field Theory (Lattice 2019): Wuhan,
  China, June 16-22 2019}}},\ }\href {\doibase 10.22323/1.363.0286} {\bibfield
  {journal} {\bibinfo  {journal} {PoS}\ }\textbf {\bibinfo {volume}
  {LATTICE2019}},\ \bibinfo {pages} {286} (\bibinfo {year} {2020})},\ \Eprint
  {http://arxiv.org/abs/2001.05706} {arXiv:2001.05706 [hep-lat]} \BibitemShut
  {NoStop}%
%%CITATION = ARXIV:2001.05706;%%
\bibitem [{\citenamefont {Borsanyi}\ \emph {et~al.}(2021)\citenamefont
  {Borsanyi} \emph {et~al.}}]{Borsanyi:2020mff}%
  \BibitemOpen
  \bibfield  {author} {\bibinfo {author} {\bibfnamefont {S.}~\bibnamefont
  {Borsanyi}} \emph {et~al.},\ }\href {\doibase 10.1038/s41586-021-03418-1}
  {\bibfield  {journal} {\bibinfo  {journal} {Nature}\ }\textbf {\bibinfo
  {volume} {593}},\ \bibinfo {pages} {51} (\bibinfo {year} {2021})},\ \Eprint
  {http://arxiv.org/abs/2002.12347} {arXiv:2002.12347 [hep-lat]} \BibitemShut
  {NoStop}%
\bibitem [{\citenamefont {Gracey}(2023)}]{Gracey:2022vjc}%
  \BibitemOpen
  \bibfield  {author} {\bibinfo {author} {\bibfnamefont {J.~A.}\ \bibnamefont
  {Gracey}},\ }\href {\doibase 10.1140/epjc/s10052-022-11088-0} {\bibfield
  {journal} {\bibinfo  {journal} {Eur. Phys. J. C}\ }\textbf {\bibinfo {volume}
  {83}},\ \bibinfo {pages} {181} (\bibinfo {year} {2023})},\ \Eprint
  {http://arxiv.org/abs/2210.12420} {arXiv:2210.12420 [hep-ph]} \BibitemShut
  {NoStop}%
\bibitem [{\citenamefont {Zhang}\ \emph {et~al.}(2021)\citenamefont {Zhang},
  \citenamefont {Li}, \citenamefont {Huo}, \citenamefont {Sch\"afer},
  \citenamefont {Sun},\ and\ \citenamefont {Yang}}]{Zhang:2020rsx}%
  \BibitemOpen
  \bibfield  {author} {\bibinfo {author} {\bibfnamefont {K.}~\bibnamefont
  {Zhang}}, \bibinfo {author} {\bibfnamefont {Y.-Y.}\ \bibnamefont {Li}},
  \bibinfo {author} {\bibfnamefont {Y.-K.}\ \bibnamefont {Huo}}, \bibinfo
  {author} {\bibfnamefont {A.}~\bibnamefont {Sch\"afer}}, \bibinfo {author}
  {\bibfnamefont {P.}~\bibnamefont {Sun}}, \ and\ \bibinfo {author}
  {\bibfnamefont {Y.-B.}\ \bibnamefont {Yang}} (\bibinfo {collaboration}
  {\ensuremath{\chi}QCD}),\ }\href {\doibase 10.1103/PhysRevD.104.074501}
  {\bibfield  {journal} {\bibinfo  {journal} {Phys. Rev. D}\ }\textbf {\bibinfo
  {volume} {104}},\ \bibinfo {pages} {074501} (\bibinfo {year} {2021})},\
  \Eprint {http://arxiv.org/abs/2012.05448} {arXiv:2012.05448 [hep-lat]}
  \BibitemShut {NoStop}%
\bibitem [{\citenamefont {Chang}\ \emph {et~al.}(2021)\citenamefont {Chang},
  \citenamefont {Liu}, \citenamefont {Raya}, \citenamefont
  {Rodr\'\i{}guez-Quintero},\ and\ \citenamefont {Yang}}]{Chang:2021vvx}%
  \BibitemOpen
  \bibfield  {author} {\bibinfo {author} {\bibfnamefont {L.}~\bibnamefont
  {Chang}}, \bibinfo {author} {\bibfnamefont {Y.-B.}\ \bibnamefont {Liu}},
  \bibinfo {author} {\bibfnamefont {K.}~\bibnamefont {Raya}}, \bibinfo {author}
  {\bibfnamefont {J.}~\bibnamefont {Rodr\'\i{}guez-Quintero}}, \ and\ \bibinfo
  {author} {\bibfnamefont {Y.-B.}\ \bibnamefont {Yang}},\ }\href {\doibase
  10.1103/PhysRevD.104.094509} {\bibfield  {journal} {\bibinfo  {journal}
  {Phys. Rev. D}\ }\textbf {\bibinfo {volume} {104}},\ \bibinfo {pages}
  {094509} (\bibinfo {year} {2021})},\ \Eprint
  {http://arxiv.org/abs/2105.06596} {arXiv:2105.06596 [hep-lat]} \BibitemShut
  {NoStop}%
\bibitem [{\citenamefont {Gasser}\ and\ \citenamefont
  {Leutwyler}(1985)}]{Gasser:1984gg}%
  \BibitemOpen
  \bibfield  {author} {\bibinfo {author} {\bibfnamefont {J.}~\bibnamefont
  {Gasser}}\ and\ \bibinfo {author} {\bibfnamefont {H.}~\bibnamefont
  {Leutwyler}},\ }\href {\doibase 10.1016/0550-3213(85)90492-4} {\bibfield
  {journal} {\bibinfo  {journal} {Nucl. Phys. B}\ }\textbf {\bibinfo {volume}
  {250}},\ \bibinfo {pages} {465} (\bibinfo {year} {1985})}\BibitemShut
  {NoStop}%
\bibitem [{\citenamefont {Boyle}\ \emph {et~al.}(2016)\citenamefont {Boyle}
  \emph {et~al.}}]{Boyle:2015exm}%
  \BibitemOpen
  \bibfield  {author} {\bibinfo {author} {\bibfnamefont {P.}~\bibnamefont
  {Boyle}} \emph {et~al.},\ }\href {\doibase 10.1103/PhysRevD.93.054502}
  {\bibfield  {journal} {\bibinfo  {journal} {Phys. Rev. D}\ }\textbf {\bibinfo
  {volume} {93}},\ \bibinfo {pages} {054502} (\bibinfo {year} {2016})},\
  \Eprint {http://arxiv.org/abs/1511.01950} {arXiv:1511.01950 [hep-lat]}
  \BibitemShut {NoStop}%
\bibitem [{\citenamefont {Yang}\ \emph
  {et~al.}(2015{\natexlab{b}})\citenamefont {Yang}, \citenamefont {Chen},
  \citenamefont {Draper}, \citenamefont {Gong}, \citenamefont {Liu},
  \citenamefont {Liu},\ and\ \citenamefont {Ma}}]{Yang:2014xsa}%
  \BibitemOpen
  \bibfield  {author} {\bibinfo {author} {\bibfnamefont {Y.-B.}\ \bibnamefont
  {Yang}}, \bibinfo {author} {\bibfnamefont {Y.}~\bibnamefont {Chen}}, \bibinfo
  {author} {\bibfnamefont {T.}~\bibnamefont {Draper}}, \bibinfo {author}
  {\bibfnamefont {M.}~\bibnamefont {Gong}}, \bibinfo {author} {\bibfnamefont
  {K.-F.}\ \bibnamefont {Liu}}, \bibinfo {author} {\bibfnamefont
  {Z.}~\bibnamefont {Liu}}, \ and\ \bibinfo {author} {\bibfnamefont {J.-P.}\
  \bibnamefont {Ma}},\ }\href {\doibase 10.1103/PhysRevD.91.074516} {\bibfield
  {journal} {\bibinfo  {journal} {Phys. Rev.}\ }\textbf {\bibinfo {volume}
  {D91}},\ \bibinfo {pages} {074516} (\bibinfo {year} {2015}{\natexlab{b}})},\
  \Eprint {http://arxiv.org/abs/1405.4440} {arXiv:1405.4440 [hep-ph]}
  \BibitemShut {NoStop}%
%%CITATION = ARXIV:1405.4440;%%
\end{thebibliography}
\end{document}